\documentclass[twocolumn,showkeys,showpacs,preprintnumbers,prd,superscriptaddress,nofootinbib]{revtex4-1}
\usepackage{graphicx}
\usepackage{epsf}
\usepackage{bm}
\usepackage{float}
\usepackage{amsmath}
\usepackage{amsfonts}
\usepackage{amssymb}
\usepackage{epstopdf}
\usepackage{color}
\usepackage[dvipsnames]{xcolor}
\usepackage{verbatim}
\usepackage{multirow}
\usepackage{bm}
\usepackage[colorlinks = true,
            linkcolor = blue,
            urlcolor  = blue,
            citecolor = blue,
            anchorcolor = blue]{hyperref}
\usepackage[capitalize]{cleveref}
\usepackage{enumitem}

\usepackage{soul}

\usepackage{braket}

\makeatletter\let\expandableinput\@@input\makeatother

\begin{document}

\title{Exploring new physics in the late Universe's expansion\\ through non-parametric inference}

\author{Miguel A. Sabogal}
\email{miguel.sabogal@ufrgs.br}
\affiliation{Instituto de F\'{i}sica, Universidade Federal do Rio Grande do Sul, 91501-970 Porto Alegre RS, Brazil}

\author{\"{O}zg\"{u}r Akarsu}
\email{akarsuo@itu.edu.tr}
\affiliation{Department of Physics, Istanbul Technical University, Maslak 34469 Istanbul, Turkey}

\author{Alexander Bonilla}
\email{alexanderbonilla@on.br}
\affiliation{Observat\'orio Nacional, Rua General Jos\'e Cristino 77, S\~{a}o Crist\'ov\~{a}o, 20921-400 Rio de Janeiro, RJ, Brazil}

\author{Eleonora Di Valentino}
\email{e.divalentino@sheffield.ac.uk}
\affiliation{School of Mathematics and Statistics, University of Sheffield, Hounsfield Road, Sheffield S3 7RH, United Kingdom}

\author{Rafael C. Nunes}
\email{rafadcnunes@gmail.com}
\affiliation{Instituto de F\'{i}sica, Universidade Federal do Rio Grande do Sul, 91501-970 Porto Alegre RS, Brazil}
\affiliation{Divis\~{a}o de Astrof\'{i}sica, Instituto Nacional de Pesquisas Espaciais, Avenida dos Astronautas 1758, S\~{a}o Jos\'{e} dos Campos, 12227-010, S\~{a}o Paulo, Brazil}

\begin{abstract}
In this study, we investigate deviations from the Planck-$\Lambda$CDM model in the late universe ($z \lesssim 2.5$) using the Gaussian Processes method, with minimal assumptions. Our goal is to understand where exploring new physics in the late universe is most relevant. We analyze recent Cosmic Chronometers (CC), Type Ia Supernovae (SN), and Baryon Acoustic Oscillations (BAO) data. By examining reconstructions of the dimensionless parameter $\delta(z)$, which measures deviations of the Hubble parameter from the Planck-$\Lambda$CDM predictions, we identify intriguing features at low ($z \lesssim 0.5$) and high ($z \gtrsim 2$) redshifts. Deviations from the Planck-$\Lambda$CDM model were not significant between $0.5\lesssim z \lesssim2$. Using the combined CC+SN+BAO dataset, we gain insights into dark energy (DE) dynamics, resembling characteristics of omnipotent DE, extending beyond quintessence and phantom models. DE exhibits n-quintessence traits for $z\gtrsim2$, transitioning with a singularity around $z\sim2$ to usual phantom traits in $1\lesssim z\lesssim2$. DE characteristics differ between scenarios ($H_0$-SH0ES and $H_0$-$\Lambda$\&CMB), with $H_0$-SH0ES leaning towards phantom traits and $H_0$-$\Lambda$\&CMB towards quintessence. We suggest exploring new physics at $z\lesssim0.5$ and $1.5\lesssim z\lesssim2.5$, particularly around $z = 2$, to understand cosmological tensions such as $H_0$ and $S_8$.
\end{abstract}



\maketitle

\section{Introduction}
\label{sec:introduction}

In the realm of cosmology, cosmological parameters play a pivotal role, shedding light on the universe's structure, composition, and dynamics. Derived from meticulous analysis of diverse astronomical observations, these parameters are crucial for our understanding of the cosmos. Among them, the Hubble Constant ($H_0$), signifying the current expansion rate of the universe, stands out as particularly challenging to ascertain. Other key parameters include the dark energy (DE) equation of state (EoS) parameter ($w_{\rm DE}$), the present-day density parameters of baryons ($\Omega_{\rm b}$) and cold dark matter (CDM) ($\Omega_{\rm cdm}$), and the weighted amplitude of matter fluctuations ($S_8= \sigma_8 \sqrt{\Omega_{\rm m}/0.3}$, where $\Omega_{\rm m}= \Omega_{\rm b}+\Omega_{\rm cdm}$ and $\sigma_8$ is the amplitude of mass fluctuations on scales of $8h^{-1}\,\mathrm{Mpc}$), each contributing uniquely to our understanding of cosmic phenomena~\cite{Planck:2018vyg,eBOSS:2020yzd,DES:2021wwk,KiDS:2020suj,freedman2023progress,Escamilla:2023oce,DiValentino:2020vvd}.

The `Hubble Constant tension' presents a significant challenge in modern cosmology~\cite{Verde:2019ivm,Knox:2019rjx,Riess:2019qba,DiValentino:2020zio,Kamionkowski:2022pkx}. This tension, a discrepancy in $H_0$ measurements, emerges when contrasting the Planck-CMB estimate~\citep{Planck:2018vyg}, based on the standard $\Lambda$CDM model (or its simple canonical extensions), with the local distance ladder measurements by the SH0ES team~\citep{Riess:2021jrx, Riess_2022, murakami2023leveraging}, showing a significant difference exceeding 5$\sigma$. Late-time measurements generally support a higher $H_0$ value (refer to the discussion in~\cite{DiValentino:2021izs}). This discrepancy has sparked intense interest in exploring new physics that might explain these variations in our understanding of the universe~\cite{Abdalla:2022yfr,DiValentino:2021izs,Perivolaropoulos_2022,akarsu2024lambda}.

Analyzing cosmological data in a model-independent manner is a challenging yet crucial pursuit, as it enables the exploration of the underlying dynamics of the universe without reliance on specific theoretical models like the widely accepted $\Lambda$CDM model. Model-independent approaches are proven to be particularly valuable for testing the robustness of cosmological parameters and identifying potential deviations from our current understanding of the cosmos. Cosmographic approaches~\cite{Catto_n_2008,Capozziello_2018,D_Agostino_2023}, a subset of model-independent methods, aim to describe the universe's expansion history and geometry without being tethered to specific theoretical models. They involve conducting a series expansion of a cosmological observable around present-day universe (around redshift $z=0$), using the data to constrain the kinematic parameters such as Hubble parameter, deceleration parameter. Despite their wide usage, cosmographic methods face challenges at higher $z$ values. An excellent alternative to these methods is the use of Gaussian Processes (\texttt{GPs}) for reconstructing cosmological parameters in a model-independent manner. As powerful and versatile tools in machine learning and statistical analysis~\cite{rasmussen2006gaussian,williams2005gaussian}, GPs are particularly adept at modeling and analyzing complex, non-linear, and noisy data. Being a form of non-parametric Bayesian modeling, GPs find applications in regression, classification, optimization, and uncertainty quantification. In the field of cosmology, GPs have been successfully employed for reconstructing the dynamics of DE, modified gravity, cosmic curvature, estimates of the Hubble constant, and various other perspectives~\cite{2010PhRvD..82j3502H,2010PhRvL.105x1302H,seikel2012reconstruction, Shafieloo_2012, Sahni_2014, Belgacem_2020,bonilla2021measurements,Wang_2017,Bengaly_2020,G_mez_Valent_2018,Bonilla_2022,Dinda_2023,Avila_2022,Renzi_2022,Rodrigues_2022,Sun_2021,Keeley:2020aym,Escamilla:2023shf}.

In this work, our primary goal is to utilize Gaussian Processes for model-independent inference, leveraging measurements of the universe's expansion rate from Cosmic Chronometers (CC), Type Ia Supernovae (SN), and Baryon Acoustic Oscillations (BAO) samples. We aim to explore potential deviations from the standard $\Lambda$CDM model, particularly at low redshifts, to gain a deeper understanding of the late Universe's expansion dynamics and to contribute to the ongoing debate surrounding the Hubble tension. Our analysis encompasses two main objectives: first, to determine the degree of deviation from the $\Lambda$CDM model at the background level, based exclusively on data pertaining to the expansion rate of the universe. Secondly, our analysis aims to explore the dynamic nature of DE, particularly the intriguing possibility of encountering negative energy density values in the Universe's late stages, thereby challenging conventional extensions of the standard $\Lambda$CDM model~\cite{Wang:2018fng,Escamilla:2021uoj, Escamilla:2023shf,Dutta:2018vmq,Visinelli:2019qqu,Akarsu:2019hmw,Akarsu:2021fol,Akarsu:2022typ,Akarsu:2023mfb,Sen:2021wld,Adil:2023exv}.
In particular, we highlight Ref.~\cite{Wang:2018fng}, which reports a 3.7$\sigma$ preference for an evolving effective DE density that can assume negative values for $z \gtrsim 2.3$.  We also refer to Refs.~\cite{Dutta:2018vmq,Colgain:2021pmf,Raveri_2023,pogosian2022imprints} for similar conclusions.

This paper is structured as follows: Section \ref{GP} introduces the Gaussian Processes (GPs) methodology and dataset used in this study. Section \ref{Results} details our key findings. Finally, Section \ref{Conclusion} discusses the conclusions drawn from our analysis and outlines future research directions.

\section{Gaussian Processes Regression and Data Set}
\label{GP}

Gaussian processes (GPs) are a fully Bayesian technique, characterizing distributions over functions and serving as an extension of Gaussian distributions into function space~\cite{rasmussen2006gaussian,seikel2012reconstruction}. This approach allows for the reconstruction of a function, indicated as $f(x)$, using observational data $\left\{\left(x_{i}, y_{i}\right) \mid i=1, \ldots, N\right\}$, without requiring a predetermined specific functional parameterization.

A Gaussian process (GP) can be expressed as
\begin{equation}
f(x) \sim \mathcal{G} \mathcal{P} \left( \mu(x), k(x, \tilde{x}) \right) \hspace{0.05cm}, 
\end{equation}
where the value of $f(x)$, when evaluated at a point $x$, is a Gaussian random variable with a mean of $\mu(x)$ and is correlated with the value at another point $\tilde{x}$ through the covariance function $\operatorname{cov}(f(x), f(\tilde{x}))=k(x, \tilde{x})$. A wide range of covariance functions exists~\cite{williams2005gaussian}. The most well-known and commonly used are the Mat\'{e}rn class of kernels ($M_{\nu}$), and the Squared-Exponential (SE) covariance function, defined as $k_{\rm SE}(x, \tilde{x})=\sigma_{f}^{2} \exp \left(- |x-\tilde{x}|^{2} / 2 \ell^{2} \right)$. The SE kernels, known for their smoothness (infinitely differentiable), are ideal for reconstructing function derivatives and depend on only two hyper-parameters, $\sigma_{f}$ and $\ell$, which characterize the function's smoothness. The characteristic length scale $\ell$ indicates the displacement in the $x$-direction required for a significant change in $f(x)$, while the signal variance $\sigma_{f}$ represents the typical change in the $y$-direction. However, due to their inherent smoothness, SE kernels predominantly capture global characteristics, not local ones. To address this limitation, the Mat\'{e}rn class kernels are advantageous. The general form of these kernels is given by
\begin{equation}
k_{_{M_\nu}}(x^\prime)=\sigma_f^2 \frac{2^{1-\nu}}{\Gamma(\nu)}\left(\frac{\sqrt{2 \nu} x^\prime}{\ell}\right)^\nu K_\nu\left(\frac{\sqrt{2 \nu} x^\prime}{\ell}\right) \hspace{0.05cm},
\end{equation}
where $x^\prime = |x-\tilde{x}|$, $K_\nu$ is the modified Bessel function of the second kind, $\Gamma(\nu)$ is the standard Gamma function, and $\nu$ is a strictly positive parameter. The $M_\nu$ kernel converges to the SE kernel as $\nu \rightarrow \infty$ and takes an explicit analytic form for half-integer values of $\nu$, i.e., $\left\{ \nu = n/2 \mid n=1,3,5, \ldots \right\}$. We focus primarily on $\nu = 7/2$ and $\nu=9/2$, as they correspond to smooth functions with a high predictability of higher-order derivatives~\cite{bonilla2021measurements}. 

On the other hand, the selection between a smooth or wavy function also depends on the distribution of the data. In our case, we have a sufficiently uniform distribution that allows us to use the aforementioned kernels with confidence. Additionally, the primary features in the reconstruction of the functions and their derivatives enable us to estimate the parameters of interest with high precision, particularly at low redshifts (to be discussed in~\cref{rho_DE_plots}). Another possibility, when the data has high dispersion, is to bin the data, allowing the application of a smooth function. However, we must exercise caution with this technique. If applied excessively, it may remove correlations between data points that could have particular characteristics of interest, leaving only the global characteristics (see, e.g., Ref.~\cite{colgain2023mcmc}).

The hyper-parameters are optimized for the observed data $f\left(x_{i}\right) + \sigma_{i}$ by minimizing the log marginal likelihood function, defined as
\begin{equation} \label{loglike}
\begin{split}
\ln \mathcal{L} & =-\frac{1}{2}(\boldsymbol{y}-\boldsymbol{\mu})^{T}[K(\boldsymbol{X}, \boldsymbol{X})+C]^{-1}(\boldsymbol{y}-\boldsymbol{\mu}) \\ 
& \hspace{0.38cm} -\frac{1}{2} \ln |K(\boldsymbol{X}, \boldsymbol{X})+C|-\frac{N}{2} \ln 2 \pi \hspace{0.1cm}.
\end{split}
\end{equation}
Here, $K(\boldsymbol{X}, \boldsymbol{X})$ is the covariance matrix with components $k(x_{i}, x_{j})$, and $C$ is the covariance matrix for the set of $N$ observations. $\boldsymbol{\mu}$ represents the assumed mean (for our model-independent analysis, $\boldsymbol{\mu} = 0$), and $\boldsymbol{y}$ is the data vector. It is noteworthy that $\mathcal{L} = p\left(\boldsymbol{y} \mid \boldsymbol{X}, \sigma_{f}, \ell\right)$ depends only on the observation locations $\boldsymbol{X}$, not on the points $\boldsymbol{X}^{*}$ where we seek to reconstruct the function. The reconstruction $\boldsymbol{f}^{*}$ is achieved through the posterior distribution (as detailed in the Appendix of Ref.~\cite{seikel2012reconstruction}),
\begin{equation}
\boldsymbol{f}^{*} \mid \boldsymbol{X}^{*}, \boldsymbol{X}, \boldsymbol{y} \sim \mathcal{N}\left(\overline{\boldsymbol{f}^{*}}, \operatorname{cov}\left(\boldsymbol{f}^{*}\right)\right) \hspace{0.1cm}
\label{postirior}
\end{equation}
with the mean $\overline{\boldsymbol{f}^{*}}$ given by
\begin{equation}
\overline{\boldsymbol{f}^{*}}=\boldsymbol{\mu}^{*}+K\left(\boldsymbol{X}^{*}, \boldsymbol{X}\right)[K(\boldsymbol{X}, \boldsymbol{X})+C]^{-1}(\boldsymbol{y}-\boldsymbol{\mu}), 
\end{equation}
and the covariance defined as
\begin{equation}
\begin{split}
\operatorname{cov}\left(\boldsymbol{f}^{*}\right) & = K\left(\boldsymbol{X}^{*}, \boldsymbol{X}^{*}\right)
\\
& - K\left(\boldsymbol{X}^{*}, \boldsymbol{X}\right)[K(\boldsymbol{X}, \boldsymbol{X})+C]^{-1} K\left(\boldsymbol{X}, \boldsymbol{X}^{*}\right).
\end{split}
\end{equation}

GPs are a highly effective method, as they can also be utilized to reconstruct the derivatives of $f(x)$; this is due to the fact that the derivative of a Gaussian process is itself a GP~\cite{williams2005gaussian},
\begin{equation}
f'(x) \sim {\mathcal{G}} {\mathcal{P}}\left(\mu'(x), \frac{\partial^{2} k(x, \tilde{x})}{\partial x \partial \tilde{x}}\right) \hspace{0.1cm}, 
\end{equation}
albeit the effectiveness of this procedure is contingent upon the differentiability of the chosen kernel. It is important to note that the hyper-parameters are trained in the same way as for the reconstruction of $f(x)$, since the marginal likelihood $\mathcal{L}$ is dependent solely on the observations and not on the function we aim to reconstruct.\\

We summarize below the datasets used in our analysis:
\begin{itemize}[wide]
\item \textit{Cosmic Chronometers (CC):} The CC approach, a powerful method for tracing the history of cosmic expansion, measures the Hubble parameter, $H(z)$, at various redshifts. In our analysis, we utilize the compilation of $H(z)$ measurements provided in Table I of Ref.~\cite{Moresco_2022}, which includes 32 measurements distributed over a redshift range of $0 < z < 2$.

\item \textit{Type Ia Supernovae (SN):} We incorporate SN distance moduli measurements from the Pantheon sample, consisting of 1048 SNeIa within the redshift range $0.01 < z < 2.3$~\cite{Scolnic_2018}. These measurements are utilized to constrain the normalized expansion rate, defined as $E(z) = H(z)/H_0$~\cite{Riess_2018}. We consider the six data points reported in~\cite{haridasu2018improved}. To convert $E(z)$ estimates into $H(z)$, a value for $H_0$ is required. Given the tension between the Planck CMB base-$\Lambda$CDM estimate and SH0ES team measurements, we rescale these data set by the corresponding values of $H_0$ estimated/measured from/by each probe. Therefore, when using SN data, we consider two distinct scenarios:
\begin{enumerate}[label=(\roman*)]
\item \textbf{$H_0$-SH0ES}: We rescaled the SN data using the Hubble estimate as ${H_0 = 73.04 \pm 1.04}\,{\rm \,km\, s^{-1}\, Mpc^{-1}}$, based on the SH0ES distance ladder measurement of $H_0$~\cite{Riess:2021jrx}.
\item \textbf{$H_0$-$\Lambda$\&CMB}: We rescaled the SN data using the Hubble estimate as ${H_0 = 67.4 \pm 0.5}\,{\rm \,km\, s^{-1}\, Mpc^{-1}}$, based on the Planck 2018 CMB base-$\Lambda$CDM best-fit estimation~\cite{Planck:2018vyg}.
\end{enumerate}

\item \textit{Baryon Acoustic Oscillations (BAO):} BAO data provide another crucial cosmological probe. The expanding spherical wave, generated by baryonic perturbations from acoustic oscillations during the recombination epoch, is traceable through the correlation function of large-scale structures. This information is then used to measure geometrical distances and the rate of expansion of the universe at various redshifts. We incorporate $H(z)$ measurements from various astronomical surveys, as compiled in Table III of Ref.~\cite{Mukherjee_2021}. As demonstrated in Ref.~\cite{bonilla2021measurements}, we have confirmed that different $r_{\rm d}$ (sound horizon at drag epoch) input values do not affect the GP analysis.
\end{itemize}

\begin{figure*}[ht!!]
    \centering
    \includegraphics[width=5.9cm]{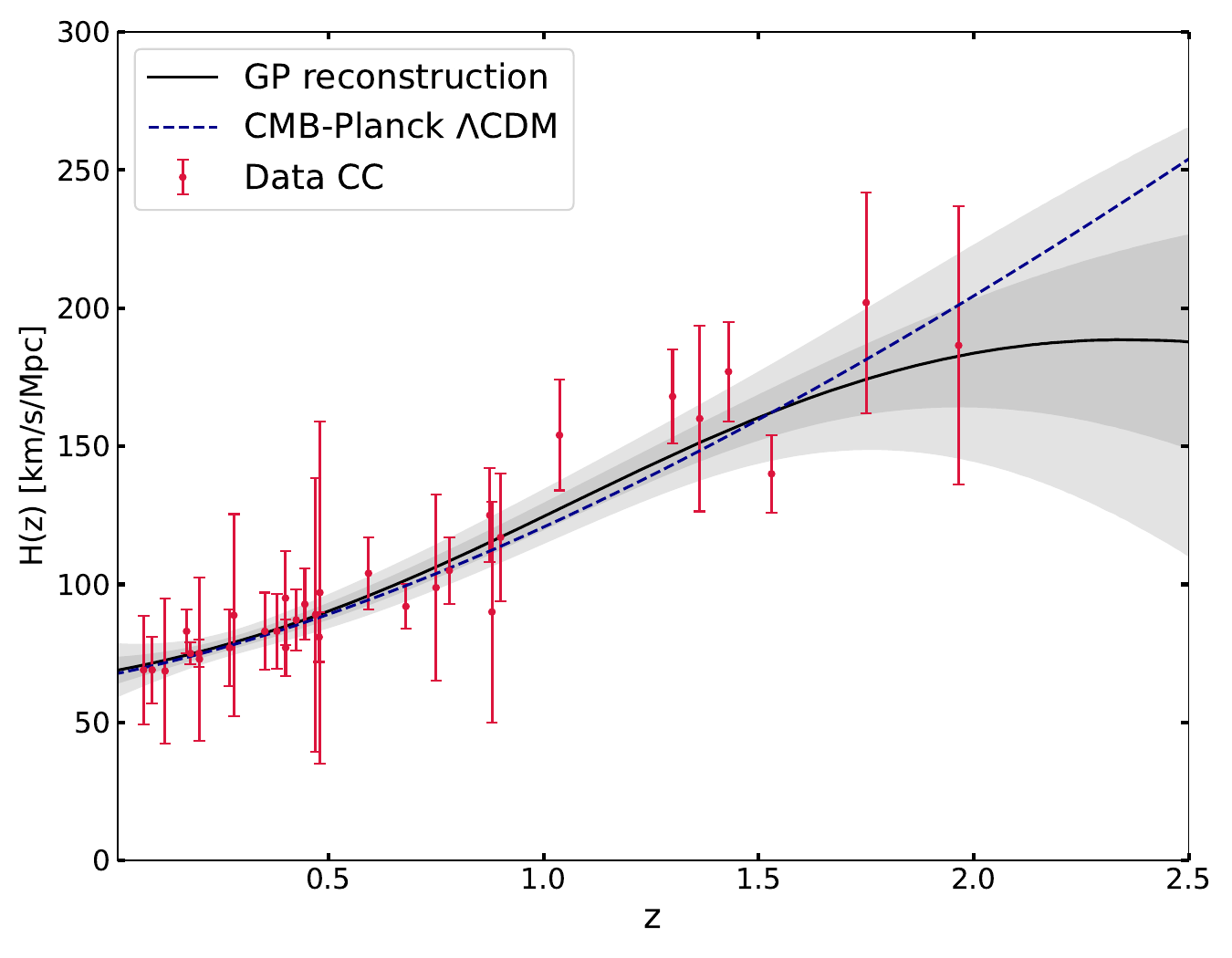} 
    \includegraphics[width=5.9cm]{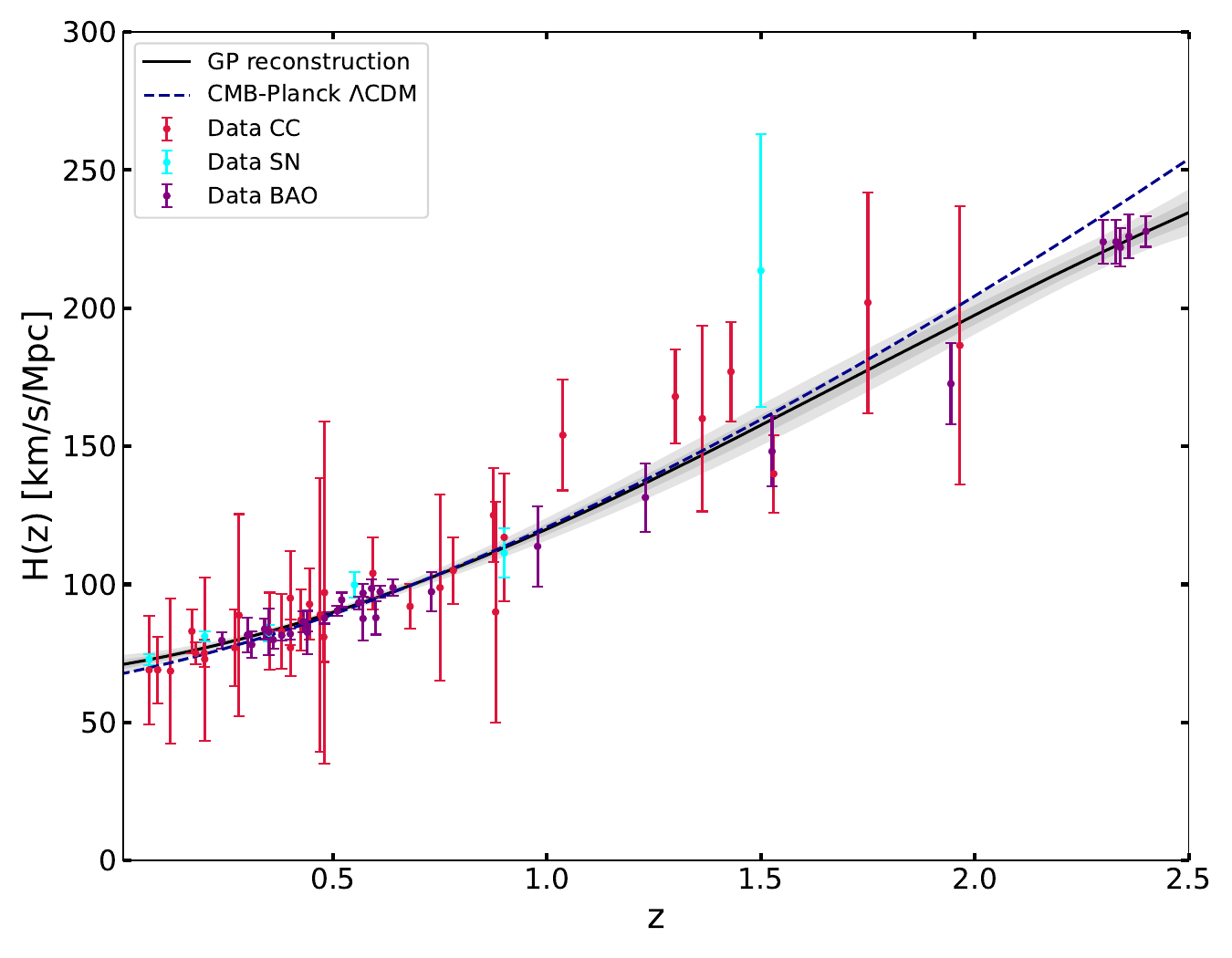}
    \includegraphics[width=5.9cm]{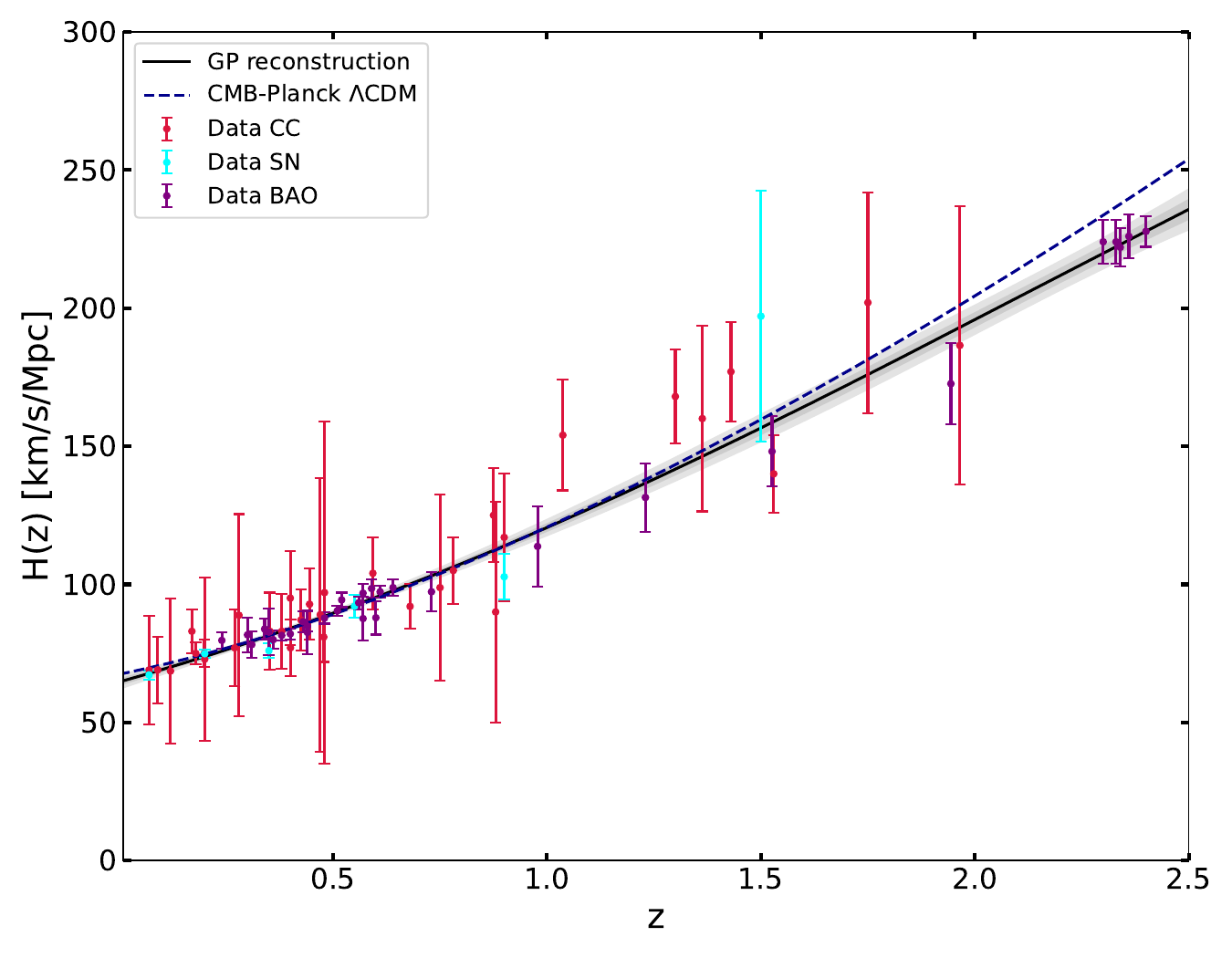}
    \caption{Left Panel: Reconstruction of the expansion rate of the Universe from the CC sample. Middle Panel: Reconstruction of the expansion rate of the Universe using CC+SN+BAO for the case of $H_0$-SH0ES. Right Panel: Same as in the middle panel, but with $H_0$-$\Lambda$\&CMB. In all panels, the blue dashed line represents the $\Lambda$CDM model with parameters fixed at the best-fit values predicted by CMB-Planck~\cite{Planck:2018vyg}.}
    \label{Hz_sample}
\end{figure*}

Expanding upon the methodology and datasets described earlier, Figure \ref{Hz_sample} showcases our reconstruction of the Universe's expansion rate, $H(z)$, utilizing both CC only data and the combined CC+SN+BAO data. For comparison, we display the $\Lambda$CDM model with parameters fixed at the best-fit values predicted by CMB-Planck~\cite{Planck:2018vyg}. In the CC analysis (left panel), the reconstruction yields $H_0 = 68.6 \pm 5.06\,{\rm \,km\, s^{-1}\, Mpc^{-1}}$, at 1$\sigma$ CL, when evaluating the $H(z)$ function at $z=0$. For the combined CC+SN+BAO data, reconstruction yields $H_0 = 70.75 \pm 1.77\,{\rm \,km\, s^{-1}\, Mpc^{-1}}$ for the $H_0$-SH0ES scenario (middle panel) and $H_0 = 64.64 \pm 1.43\,{\rm \,km\, s^{-1}\, Mpc^{-1}}$ for the $H_0$-$\Lambda$\&CMB scenario (right panel), both at 1$\sigma$ CL. It is noted that the CC-only analysis does not provide robust constraints on the expansion rate for $z > 1$, but including the SN and BAO data significantly improves the observational constraints on $H(z)$ reconstruction. The joint CC+SN+BAO analysis allows us to infer and evaluate the expansion rate functions of the universe with good/reasonable accuracy up to $z \simeq 2.5$. One immediate observation is that the high redshift (viz., $z>2.0$) BAO data points yield $H(z)$ values lower than those predicted by the Planck-$\Lambda$CDM model ($H(z)_{\Lambda\rm CDM}$), so predictably, our reconstructed $H(z)$ values at those redshifts are also lower. A more detailed exploration and analysis of these findings will be presented in the subsequent section, where we delve deeper into the implications and nuances of our results.

\section{Results}
\label{Results}

We aim to explore possible deviations from the standard $\Lambda$CDM model at low redshifts (specifically, $z\lesssim2.5$, which is the range covered by our data), while adhering to a framework of minimal assumptions. To quantitatively assess these deviations, we define the dimensionless quantity $\delta(z)$, representing the relative difference in the Hubble parameter $H(z)$ from its predicted value by the $\Lambda$CDM model, $H(z)_{\Lambda \rm CDM}$, at a given redshift $z$~\cite{Akarsu:2022lhx}:
\begin{equation}
\label{delta}
\delta(z) = \frac{H(z)-H(z)_{\Lambda \rm CDM}}{H(z)_{\Lambda \rm CDM}}.
\end{equation}
This formulation enables quantification of deviations in the background cosmological evolution compared to the $\Lambda$CDM model. In this context, $H(z)$ is reconstructed using GP (Gaussian Process) numerical routines provided by the publicly available GaPP (Gaussian Processes in Python) code~\cite{seikel2012reconstruction}, while $H(z)_{\Lambda \rm CDM}$ is determined using the best-fit values derived from the Planck-CMB data~\cite{Planck:2018vyg}.

The primary physical characteristic of the $\delta(z)$ function is straightforward: ${\delta(z) > 0}$ implies deviations toward a faster-expanding universe than Planck-$\Lambda$CDM predicts, while ${\delta(z) < 0}$ indicates a slower expansion in comparison, and a value of $\delta(z) = 0$ aligns with Planck-$\Lambda$CDM expectations. Considering that an ever-expanding universe requires $H(z)>0$, it follows that $\delta(z)>-1$ would always hold. On the other hand, despite the $\Lambda$CDM model grappling with various cosmological tensions, in the era of high-precision observational cosmology, and theoretical challenges, such as the cosmological constant problem~\cite{Weinberg:1988cp,Weinberg:2000yb,Peebles:2002gy}, it remains the simplest model that explains most cosmological data with remarkable accuracy~\cite{Planck:2018vyg,eBOSS:2020yzd,KiDS:2020suj,DES:2021wwk}. Thus, we anticipate only minor deviations from the Planck-$\Lambda$CDM model, specifically $|\delta(z)|\ll1$ for $z\lesssim2.5$, the redshift range under investigation. The largest discrepancies known, such as the SH0ES $H_0$ measurement at $H_0=73.04 \pm 1.04$ km s${}^{-1}$ Mpc${}^{-1}$~\cite{Riess:2021jrx} and the $H(2.33)=224\pm8$ km s${}^{-1}$ Mpc${}^{-1}$ Ly-$\alpha$-quasar data~\cite{eBOSS:2020yzd}, correspond to $\delta(z=0)\sim0.08$ and $\delta(z=2.33)\sim-0.05$, respectively. Nonetheless, small but statistically significant deviations from $\delta(z)=0$ and their signs can offer valuable insights into the universe's late-time kinematics and dynamics, under the assumption of gravity theory such as general relativity. Accordingly, in what follows, we will first discuss our results on the kinematics, namely $\delta(z)$, and then the corresponding dynamics, viz., the dynamics of dark energy, by assuming a spatially flat FRW (Friedmann-Robertson-Walker) background and the general theory of relativity (GR).

\begin{figure}[ht!]
    \centering
    \includegraphics[width=7cm]{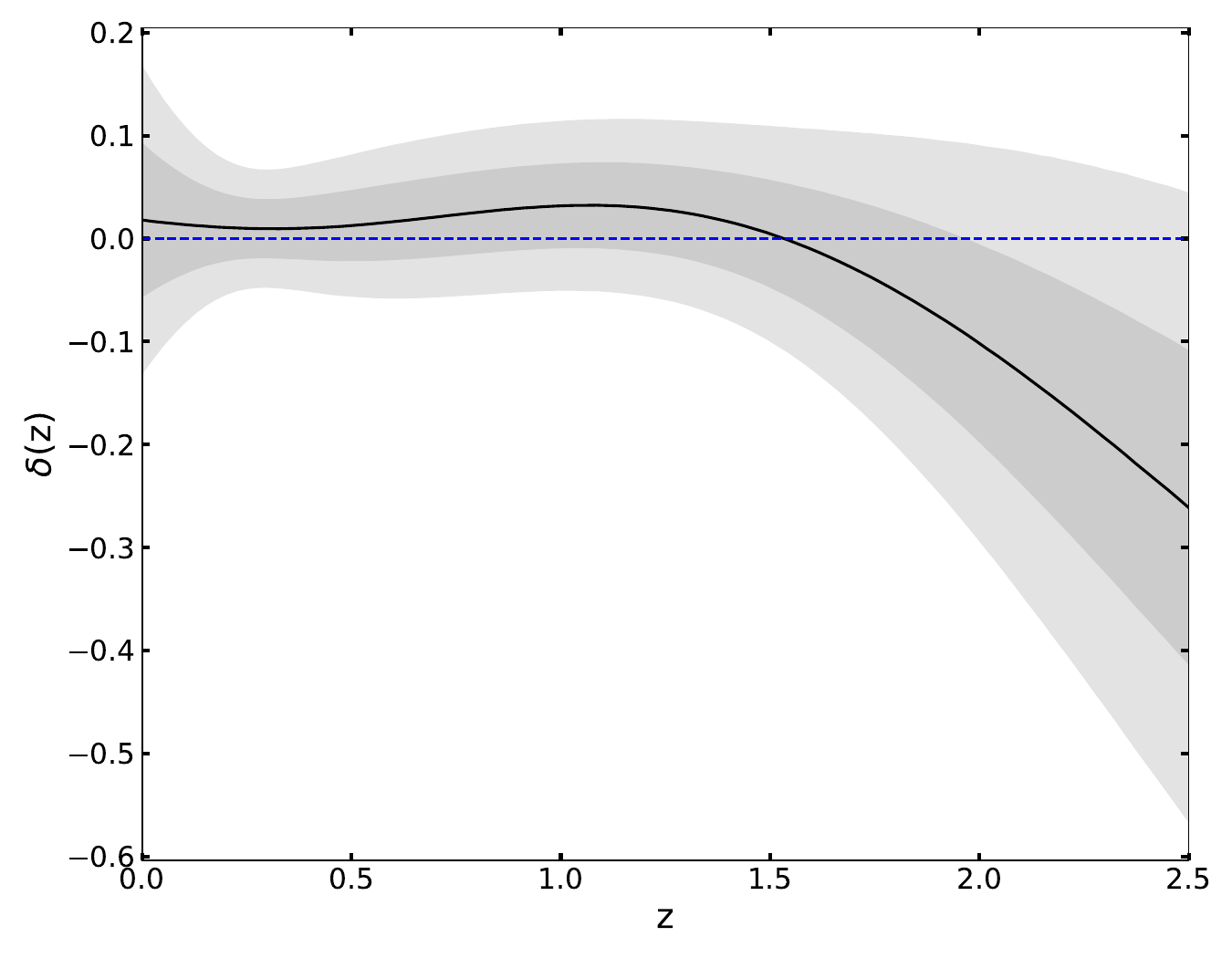}
    \caption{Reconstruction at 1$\sigma$ and 2$\sigma$ confidence levels of the $\delta(z)$ function using the CC sample. The blue-dashed line is the null hypothesis.}
    \label{delta_CC}
\end{figure}

Our reconstruction of the $\delta(z)$ function from the CC sample alone is depicted in Figure \ref{delta_CC}. In this analysis, $\delta(z)$ consistently aligns with the null hypothesis across the studied redshift range, implying $\delta(z) = 0$ at a 2$\sigma$ confidence level. On the other hand, it is noteworthy that for redshifts greater than approximately $1.5$ ($z \gtrsim 1.5$), there is a discernible trend of $\delta(z) < 0$, observed at a 1$\sigma$ confidence level. When evaluated at present, the $\delta(z=0)$ value is found to be $\delta(z=0)=0.018 \pm 0.075$ at a 1$\sigma$ confidence level.

\begin{figure}
    \centering
    \includegraphics[width=7cm]{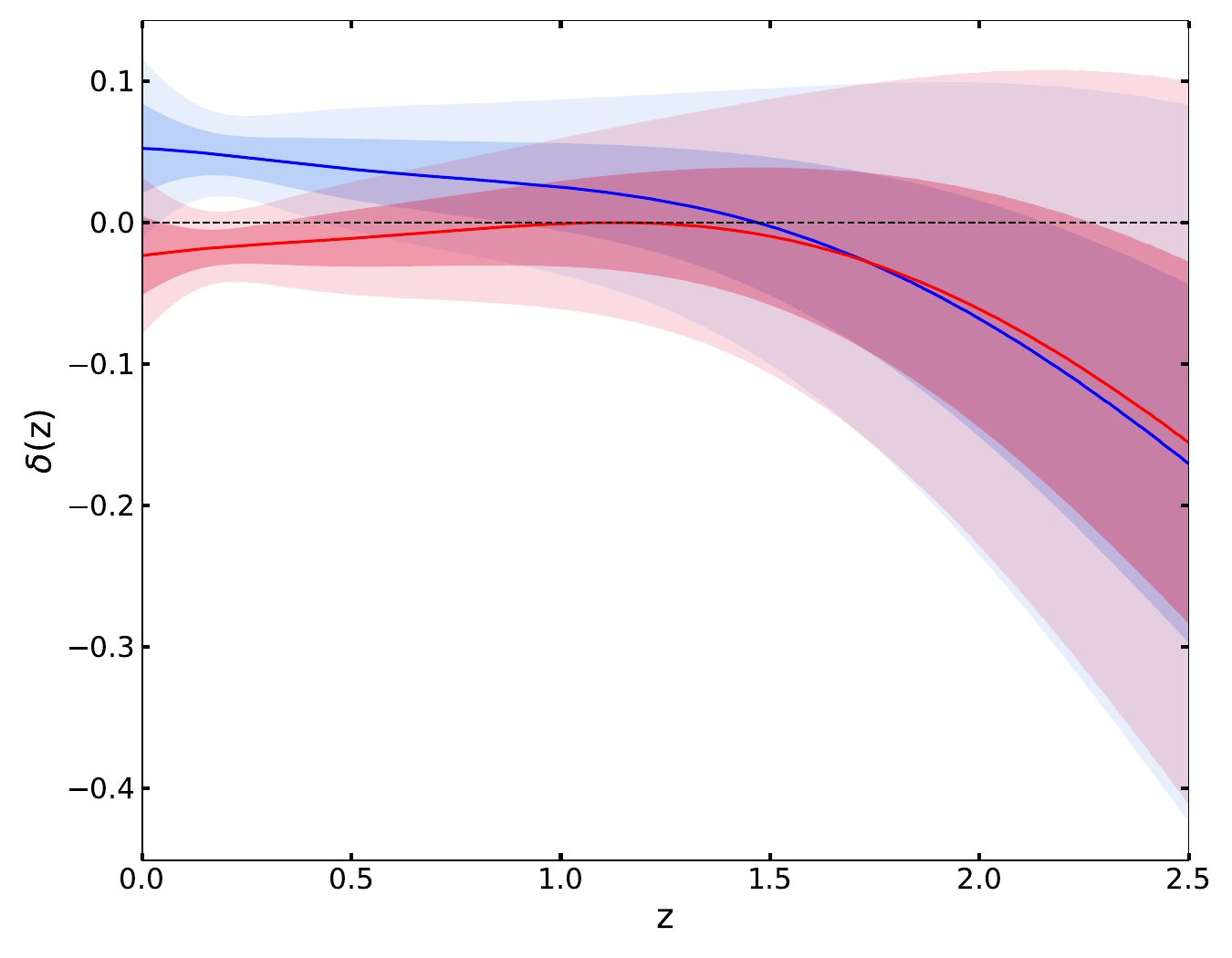} 
    \includegraphics[width=7cm]{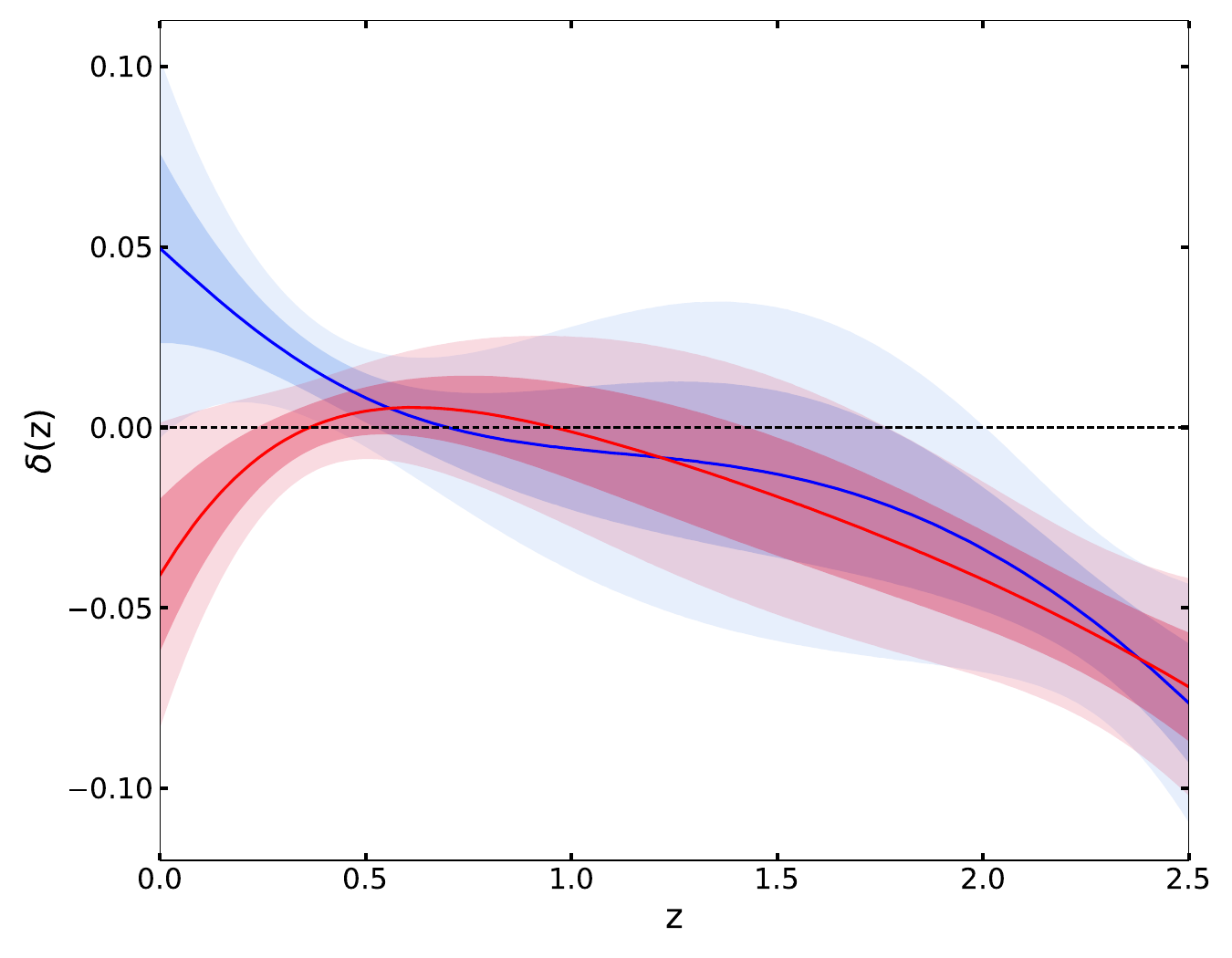} 
    \caption{Upper panel: Reconstruction at 1$\sigma$ and 2$\sigma$ confidence levels of the $\delta(z)$ function using the CC+SN sample in the cases of $H_0$-SH0ES (blue) and $H_0$-$\Lambda$\&CMB (red). Lower panel: Same as in the upper panel, but now including BAO data in the analysis, i.e, CC+SN+BAO. For both cases, the black-dashed line is the null hypothesis.}
    \label{delta_CC_SN_BAO}
\end{figure}

In the upper panel of Figure \ref{delta_CC_SN_BAO}, we present the reconstruction of the $\delta(z)$ function from the CC+SN joint analysis considering both the $H_0$-SH0ES and $H_0$-$\Lambda$\&CMB scenarios. Notably, for redshifts $z \gtrsim 0.5$, the reconstructions from both scenarios show statistical equivalence, suggesting that the background evolution of the universe is independent of the chosen scenario for cosmic distances beyond $z \gtrsim 0.5$, even though the $H_0$ values used---to convert $E(z)$
estimates into $H(z)$ estimates---in these two scenarios are in significant (5$\sigma$) tension, which may be due to the fact that SN data are very sparse beyond $z=1$ and that when combined with CC data, they have no impact at high redshifts, but they do show an impact at low redshifts (for a similar observation, see, e.g., Ref.~\cite{colgain2023mcmc}). Similarly, regardless of the scenario chosen, for $z\gtrsim1.5$, we observe a trend of $\delta(z) < 0$ at a 1$\sigma$ confidence level, in line with the CC only analysis. On the other hand, for $z \lesssim 0.5$, we notice divergent behaviors in the $\delta(z)$ evolution between the two scenarios, particularly pronounced for $z\lesssim0.2$. Namely, in the $H_0$-$\Lambda$\&CMB scenario, the reconstructed $\delta(z)$ aligns with the null hypothesis, while in the $H_0$-SH0ES scenario, $\delta(z) > 0$ at a 1$\sigma$ confidence level for $z < 0.2$. And, evaluating $\delta(z)$ at the present time, we find $\delta(z=0) = -0.023 \pm 0.028$ (1$\sigma$ CL) for the $H_0$-$\Lambda$\&CMB scenario and $\delta(z=0) = 0.052 \pm 0.032$ (1$\sigma$ CL) for the $H_0$-SH0ES scenario, indicating a 1.8$\sigma$ discrepancy. This discrepancy exceeds 2$\sigma$ level at $z\sim0.2$. These findings may offer valuable insights into the so-called $H_0$ tension, suggesting that deviations from the Planck-$\Lambda$CDM predicted background evolution of the universe are more pronounced for $z \lesssim 0.2$.

\begin{figure}[ht!]
    \centering
     \includegraphics[width=7cm]{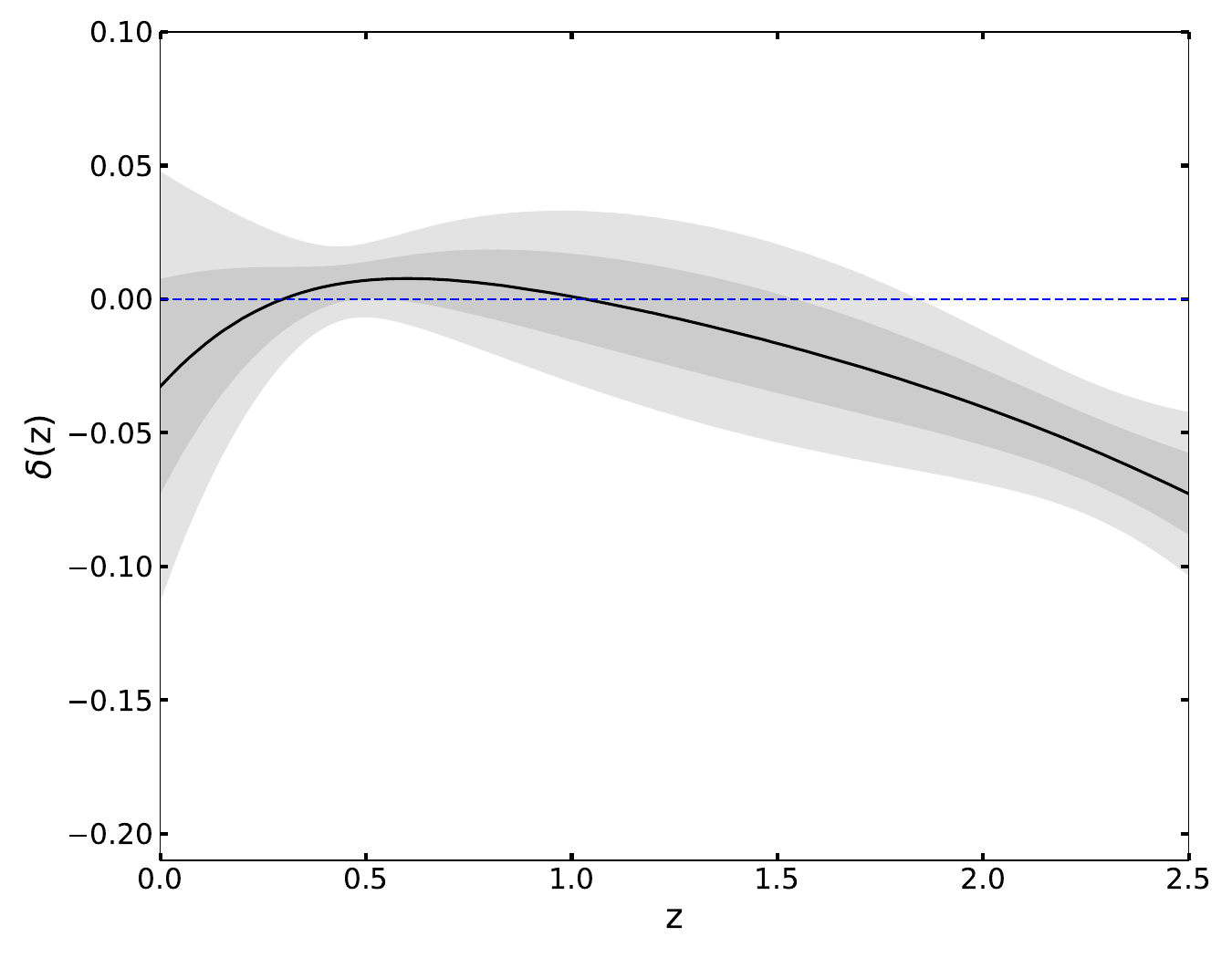}
    \caption{Reconstruction at 1$\sigma$ and 2$\sigma$ confidence levels of the $\delta(z)$ function using the CC+BAO sample. The blue dashed line represents the null hypothesis.}
    \label{delta_CC_BAO}
\end{figure}

In the lower panel of Figure~\ref{delta_CC_SN_BAO}, the reconstruction of the $\delta(z)$ function with the inclusion of the BAO dataset (with the combined CC+SN+BAO dataset) is presented. Comparing this with the upper panel (noting the different scales of the $\delta(z)$ axis in the panels), the deviations from the null hypothesis, corresponding to the Planck-$\Lambda$CDM, are either nearly the same or reduced in magnitude but exhibit increased statistically significance. Moreover, this enhanced statistical significance leads to a more pronounced distinction between the $H_0$-SH0ES and $H_0$-$\Lambda$\&CMB scenarios. However, the general nature of these deviations, more emphasized now, remains similar to that observed in the CC+SN joint analysis. Particularly for $z \lesssim 0.2$, the deviation between the two scenarios is more evident.  For instance, at present ($z=0$), we find $\delta(z=0) = 0.050 \pm 0.026$ for the $H_0$-SH0ES scenario and $\delta(z=0) =  -0.041 \pm 0.021$ for the $H_0$-$\Lambda$\&CMB scenario, showing an increased tension of $2.7\sigma$ between the two, compared to the one found in the CC+SN analysis. This analysis suggests that the chosen $H_0$ scaling value influences $\delta(z)$, leading to preferences for $\delta(z) < 0$ and $\delta(z) > 0$ for $z \lesssim 0.2$ for the $H_0$-$\Lambda$\&CMB and $H_0$-SH0ES scenarios, respectively. Notably, the preference for $\delta(z=0)<0$ in the $H_0$-$\Lambda$\&CMB scenario at about a 1$\sigma$ confidence level is interesting, as a negative $\delta(z=0)$ implies a slower expansion rate of the present-day universe than predicted by Planck-$\Lambda$CDM, thus exacerbating the $H_0$ tension associated with Planck-$\Lambda$CDM.

Another notable finding from the analysis using the combined CC+SN+BAO dataset is that for $z\gtrsim1.5$, the deviations of $\delta(z)$ from the null hypothesis are smaller in both scenarios. However, independent of whether the $H_0$-SH0ES or $H_0$-$\Lambda$\&CMB scenario is chosen, the trend for $\delta(z)<0$ begins at slightly lower redshifts. This trend, observed at a 1$\sigma$ level in the CC and CC+SN cases, attains a significance level of over 2$\sigma$ for $z\gtrsim2$. Thus, regardless of variations in the $H(z)$ function beyond the Planck-$\Lambda$CDM model, the deviations in $H(z)$ for $z \gtrsim 1.5$ consistently indicate a universe expanding at a rate slower than predicted by Planck-$\Lambda$CDM. For $z \lesssim 0.5$, however, the expansion rate hinges on the ability of the considered cosmological model to predict the $H_0$ parameter, even allowing for $\delta(z) < 0$ when the BAO dataset is included in the analysis.

In~\cref{delta_CC_BAO}, we assess the influence on the $\delta(z)$ function by excluding the SN data. Consequently, this scenario does not require consideration of the $H_0$-SH0ES and $H_0$-$\Lambda$\&CMB estimates. Notably, for low redshifts, we observe complete statistical agreement with $\delta(z) = 0$, which remains robust up to $z$ $\sim$ 1.8. Conversely, at higher redshifts (exceeding $1.5$), a similar trend to previously discussed results is observed: $\delta(z)$ becomes increasingly negative with increasing $z$. An interesting relevant discussion can be found in Ref.~\cite{colgain2023mcmc} (in particular, see \textit{Section~6: Concluding remarks} of the reference), where the authors discuss the use of CC and BAO data in estimating the value of $H_0$ and other derivatives related to $\Omega_{\rm m}$.

In~\cref{B} we quantify the impact of choosing a zero mean function and various kernel choices on the reconstructions within the $\delta(z)$ function. Our findings reveal that all reconstructions in the function $\delta(z)$, and consequently all our main results, remains unchanged within these perspectives. Since all other quantities assessed in this study are reconstructions derived from the $\delta(z)$ function, we can extend the same conclusions to all other functions evaluated throughout this study. In conclusion, our findings remain consistent even when assessing the impact of statistical decisions.

We note that our analysis thus far has implicitly assumed a spatially maximally symmetric spacetime, viz., the Robertson-Walker (RW) spacetime metric, without committing to a specific gravity theory. Accordingly, we have discussed only the kinematical deviations from the Planck-$\Lambda$CDM scenario, represented by $\delta(z)$. Introducing a gravity theory (relating the kinematics and the physical ingredient of the universe with each other) would enable us to extend our discussions to dynamics, though this would entail some compromise on our minimal assumption approach. Choosing a modest yet reasonable step forward, we attribute all the kinematical deviations from the Planck-$\Lambda$CDM model to the deviations of a minimally interacting DE from the cosmological constant.\footnote{One could consider other options, such as attributing the kinematical deviations from the Planck-$\Lambda$CDM model to modifications in the theory of gravity, non-minimal interaction between DE and CDM, or a modified dark matter equation of state. However, these options do not align with the minimal approach we intend to follow in the current paper. \label{footnote1}} This allows us to define a parameter that can be directly derived from the $\delta(z)$ function to characterize the dynamics of DE~\cite{Akarsu:2022lhx}. In particular, we assume that general relativity (GR) governs the expansion dynamics of the universe, and given that in the late universe—at redshifts relevant to the data we use in our analyses—radiation is negligible, meaning only pressureless matter (baryons and CDM) and dark energy (DE) are relevant, and that DE interacts only gravitationally. This parameter, denoted as $\Omega_{\rm \Delta DE}(z)$, quantifies the deviation of the DE density parameter from that of the cosmological constant in the Planck-$\Lambda$CDM scenario as a quadratic function of $\delta(z)$ as follows:
\begin{equation}
\label{OrhoDE}
\Omega_{\rm \Delta DE}(z) \equiv \frac{\Delta \rho_{\rm DE}(z)}{3 H^2_{\Lambda \rm CDM}(z)} = \delta(z)[2+\delta(z)]\approx2\delta(z),
\end{equation}
where ${\Delta\rho_{\rm DE}(z)\equiv\rho_{\rm DE}(z)-\rho_{\Lambda}}$ (with ${\rho_\Lambda=\rho_{\rm DE0}}$). For small deviations from Planck-$\Lambda$CDM, i.e., $|\delta(z)|\ll1$, this simplifies to approximately twice the $\delta(z)$ function: $\Omega_{\rm \Delta DE}(z)\approx2\delta(z)$. The advantage of reconstructing this parameter lies in its alignment with our minimal assumption approach, making our compromise minimal, as it allows direct derivation from the reconstructions of $\delta(z)$, without needing additional information beyond what we have used for the reconstructions of the $\delta(z)$ function.

\begin{figure}
    \centering
    \includegraphics[width=7cm]{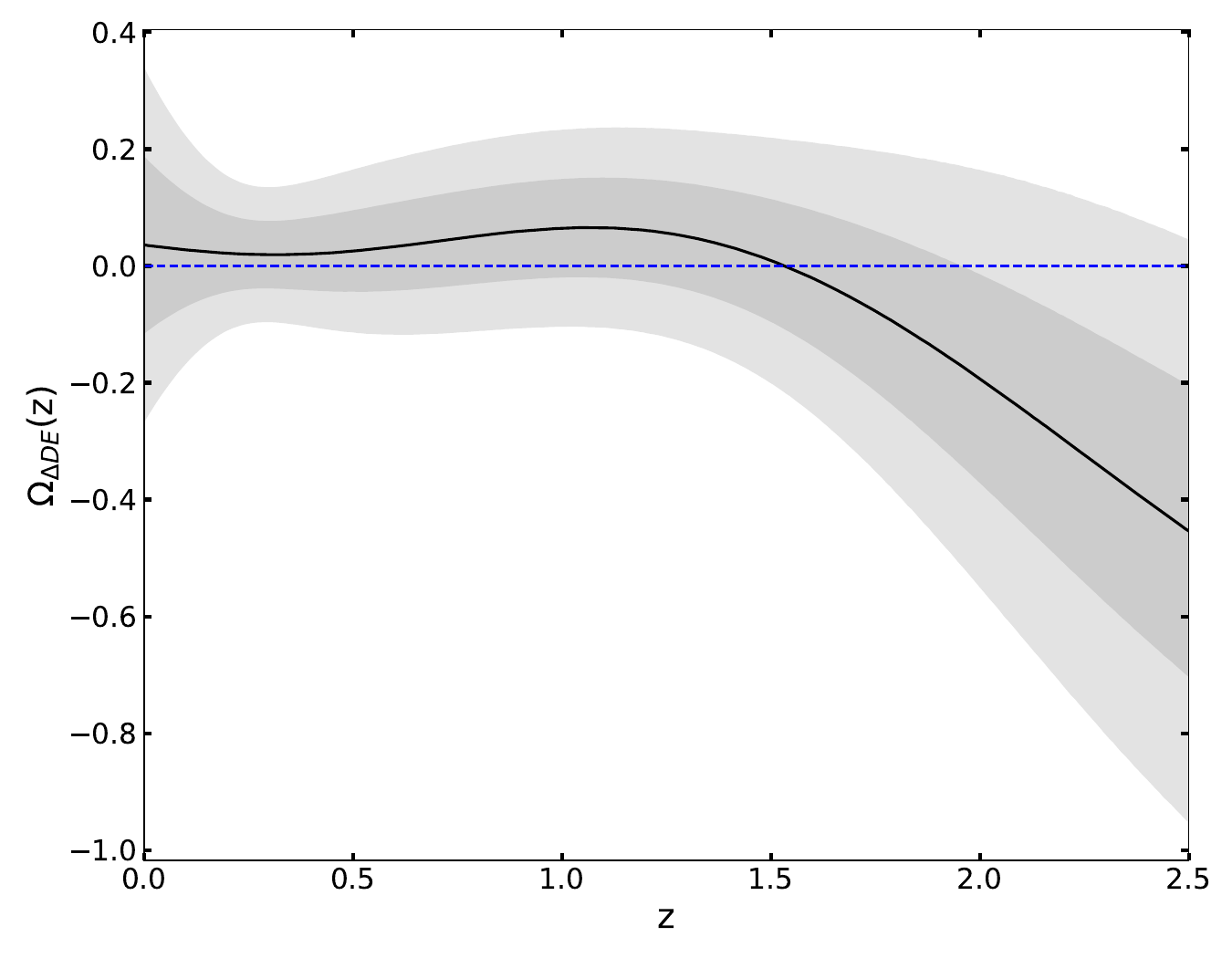}
    \caption{Reconstruction at 1$\sigma$ and 2$\sigma$ confidence levels of the $\Omega_{\rm \Delta DE}(z)$ function for the CC sample. The blue-dashed line is the null hypothesis.}
    \label{fig_DE_CC}
\end{figure}

\begin{figure}
    \centering
    \includegraphics[width=7cm]{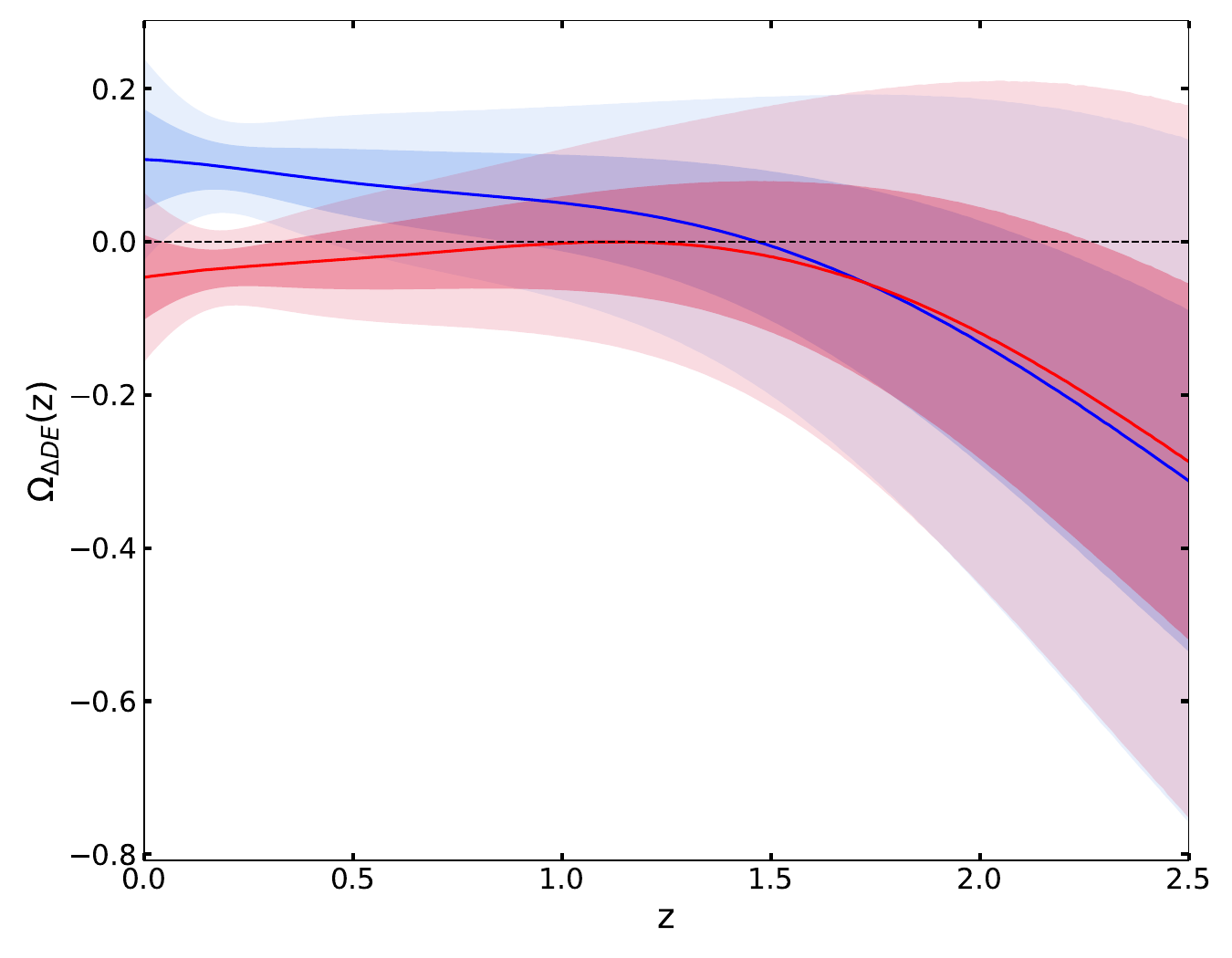}
    \includegraphics[width=7cm]{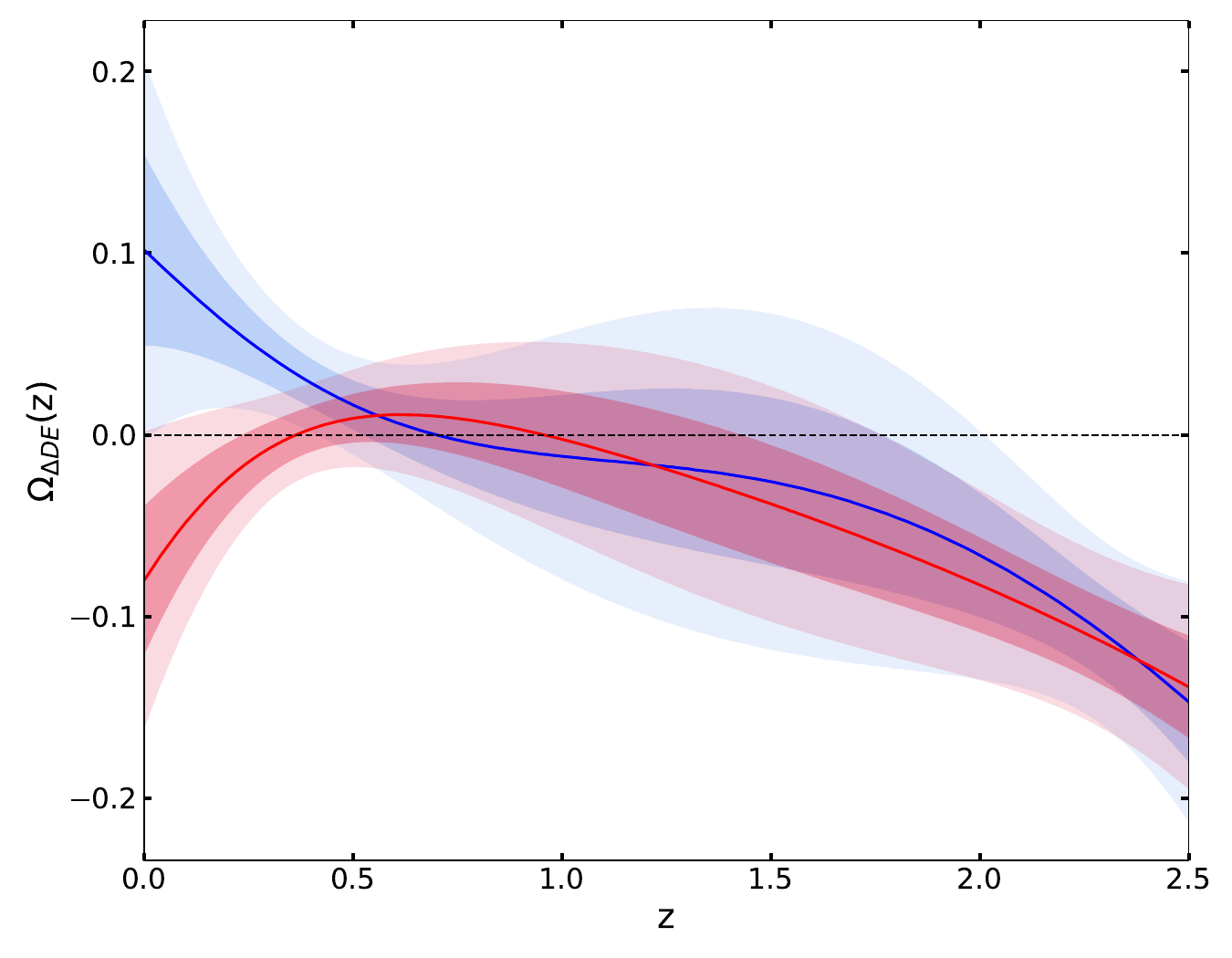} 
    \caption{Upper Panel: Reconstruction at 1$\sigma$ and 2$\sigma$ confidence levels of the $\Omega_{\rm \Delta DE}(z)$ function for the CC+SN sample in the case of $H_0$-SH0ES (blue) and $H_0$-$\Lambda$\&CMB (red) with scaling SN data. Lower Panel: Same as the upper panel, but now including BAO data in the analysis (CC+SN+BAO). For both cases, the black-dashed line is the null hypothesis.}
    \label{fig_DE}
\end{figure}

In our reconstructions of the $\delta(z)$ function, we observe deviations from the null hypothesis, $\delta(z)=0$, are small, $|\delta(z)|\lesssim0.1$ at a 2$\sigma$ CL, for $z\lesssim1.5$, regardless of the sample used in the analysis, whether CC, CC+SN, or CC+SN+BAO. The largest deviations occur at higher redshifts ($z\gtrsim2$), with $\delta(z)$ reaching approximately $-0.6 $ for CC and approximately $-0.4$ for CC+SN, both at a 2$\sigma$ CL, at $z=2.5$. These large negative deviations at $z\gtrsim2$ cause the quadratic term, $\delta(z)^2$, in Eq.\eqref{OrhoDE} to push $\Omega_{\rm \Delta DE}(z)$ towards the null hypothesis ($\Omega_{\rm \Delta DE}(z)=0$ corresponding to $\delta(z)=0$), contrary to the linear term $2\delta(z)$ that pushes $\Omega_{\rm \Delta DE}(z)$ towards negative values. This phenomenon seems to suggest that increasing $\delta(z)$ does not necessarily lead to a larger deviation from the cosmological constant, yet this is unlikely due to the condition $\delta(z)>-1$ ensuring $H(z)>0$. This phenomenon is most evident in the CC sample, as seen when comparing Fig.~\ref{delta_CC} and Fig.~\ref{fig_DE_CC}, where $\Omega_{\rm \Delta DE}(z)\approx2\delta(z)$ for $z\lesssim1.5$, while $-0.6\lesssim\delta(z)\lesssim0.05$ and $-0.8\lesssim\Omega_{\rm \Delta DE}(z)\lesssim0.1$ at $z=2.5$. A similar situation, albeit less pronounced, is observed in the CC+SN analysis; compare the upper panels of Figs.~\ref{delta_CC_SN_BAO} and \ref{fig_DE}. With the inclusion of BAO data (CC+SN+BAO) in the analysis, deviations confined in $|\delta(z)|\lesssim0.1$ across the entire redshift range from $0$ to $2.5$. Therefore, $\Omega_{\rm \Delta DE}(z)\approx2\delta(z)$ is generally a good approximation for the CC+SN+BAO sample, suggesting that reconstructions of $\Omega_{\rm \Delta DE}(z)$ and $\delta(z)$ will exhibit almost the same pattern, as can be seen when comparing the lower panels of Figs.~\ref{delta_CC_SN_BAO} and \ref{fig_DE}. 

Consequently, one can straightforwardly map our discussions from the $\delta(z)$ reconstructions to those of $\Omega_{\rm \Delta DE}(z)$. For instance, in the CC+SN analysis, we have $\Omega_{\rm \Delta DE}(z=0)=0.107 \pm 0.066$ for the $H_0$-SH0ES scenario and $\Omega_{\rm \Delta DE}(z=0)=-0.046 \pm 0.055$ for the $H_0$-$\Lambda$\&CMB scenario. In the CC+SN+BAO analysis, we find $\Omega_{\rm \Delta DE}(z=0)=-0.102 \pm 0.053$ for the $H_0$-SH0ES scenario and $\Omega_{\rm \Delta DE}(z=0)=-0.080 \pm 0.041$ for the $H_0$-$\Lambda$\&CMB scenario. These results align with the approximation $\Omega_{\rm \Delta DE}(z)\approx2\delta(z)$, as compared to their corresponding $\delta(z=0)$ estimates. Similarly, regardless of the sample used---CC, CC+SN, or CC+SN+BAO---for redshifts $z \gtrsim 1.5$, the best-fit expectations for $\Omega_{\rm \Delta DE}(z)$ tend to be negative. From the statistical perspective on the other hand, we observe that while $\Omega_{\rm \Delta DE}(z)$ is still compatible with the null-hypothesis---$\Omega_{\rm \Delta DE}(z) = 0$, implying no deviation from the cosmological constant of the Planck-$\Lambda$CDM---within a 2$\sigma$ confidence level for the CC and CC+SN analyses, in the case of CC+SN+BAO, the best-fit expectation of $\Omega_{\rm \Delta DE}(z)$ passes below zero at lower redshifts ($z\sim1$) compared to the CC and CC+SN analyses, and the significance of $\Omega_{\rm \Delta DE}(z) < 0$ exceeds the 2$\sigma$ level for $z\gtrsim2$. Given this straightforward relation between $\delta(z)$ and $\Omega_{\rm \Delta DE}(z)$, can we still glean additional insights beyond what we have learned from $\delta(z)$? Notably, the fact that the energy density of the pressureless matter component increases as $(1+z)^3$ and the parameter $\Omega_{\rm \Delta DE}(z)$ increasingly assumes/tends to assume negative values for $z\gtrsim2$, surpassing the 2$\sigma$ significance level for the CC+SN+BAO sample, suggests that the DE density might also assume negative values at high-redshifts ($z\gtrsim2$). To investigate this possibility further, we can explore the parameters characterizing the DE based on our $\delta(z)$ reconstructions. This will require another departure from our minimal assumption approach, yet could still yield informative results.

\begin{figure*}
    \centering
    \includegraphics[width=5.5cm]{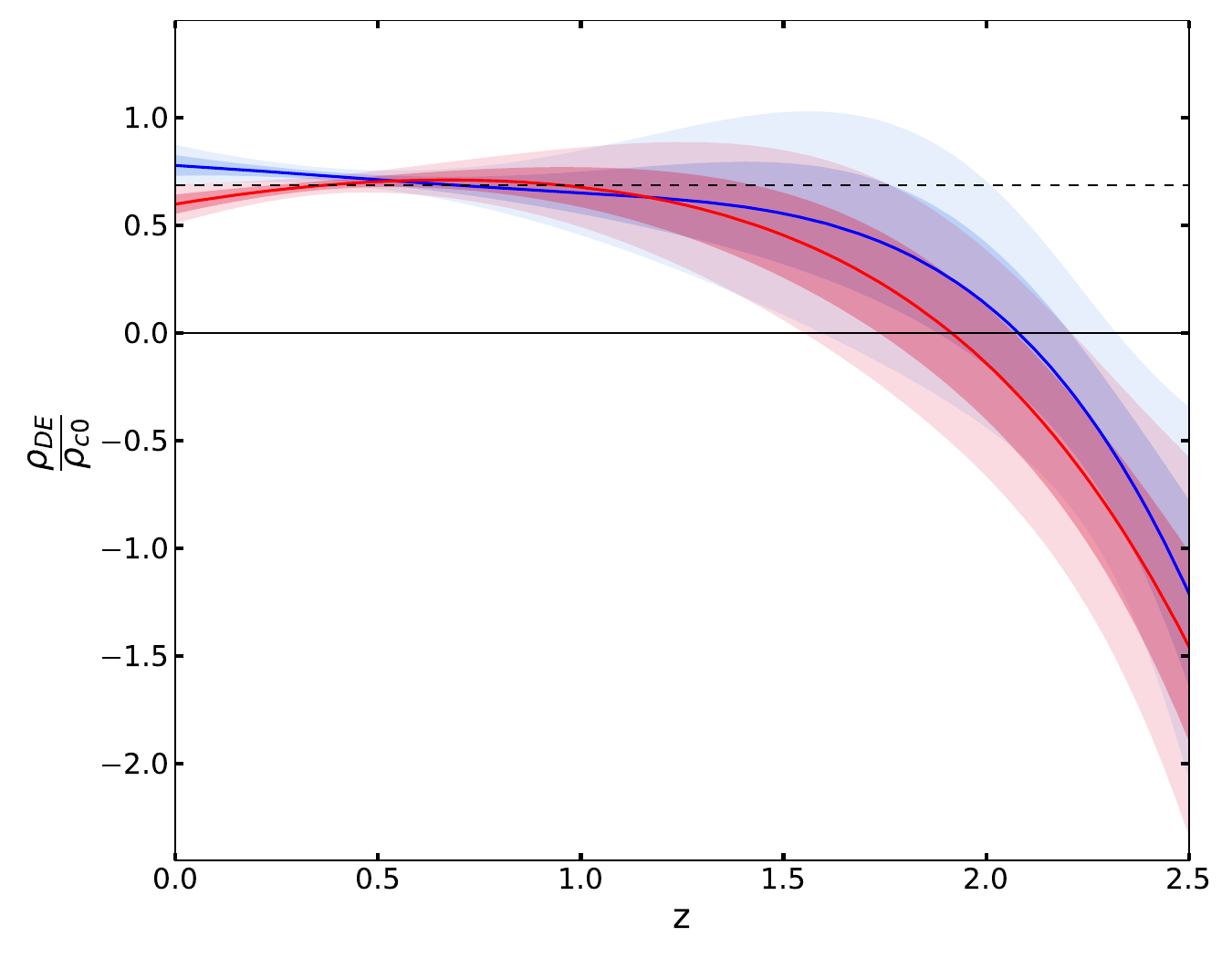} \,\,\,\, 
    \includegraphics[width=5.5cm]{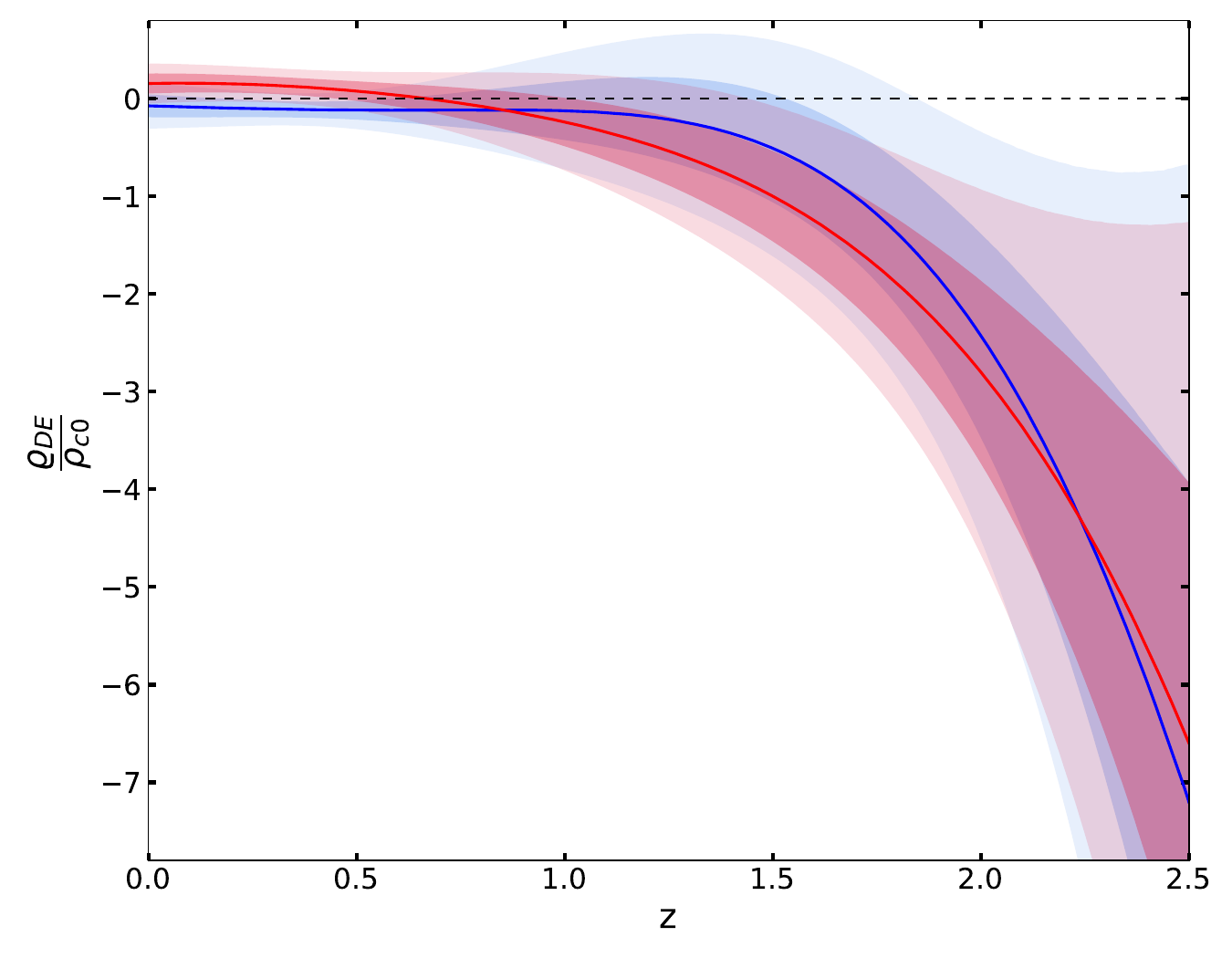} \,\,\,\, 
    \includegraphics[width=5.5cm]{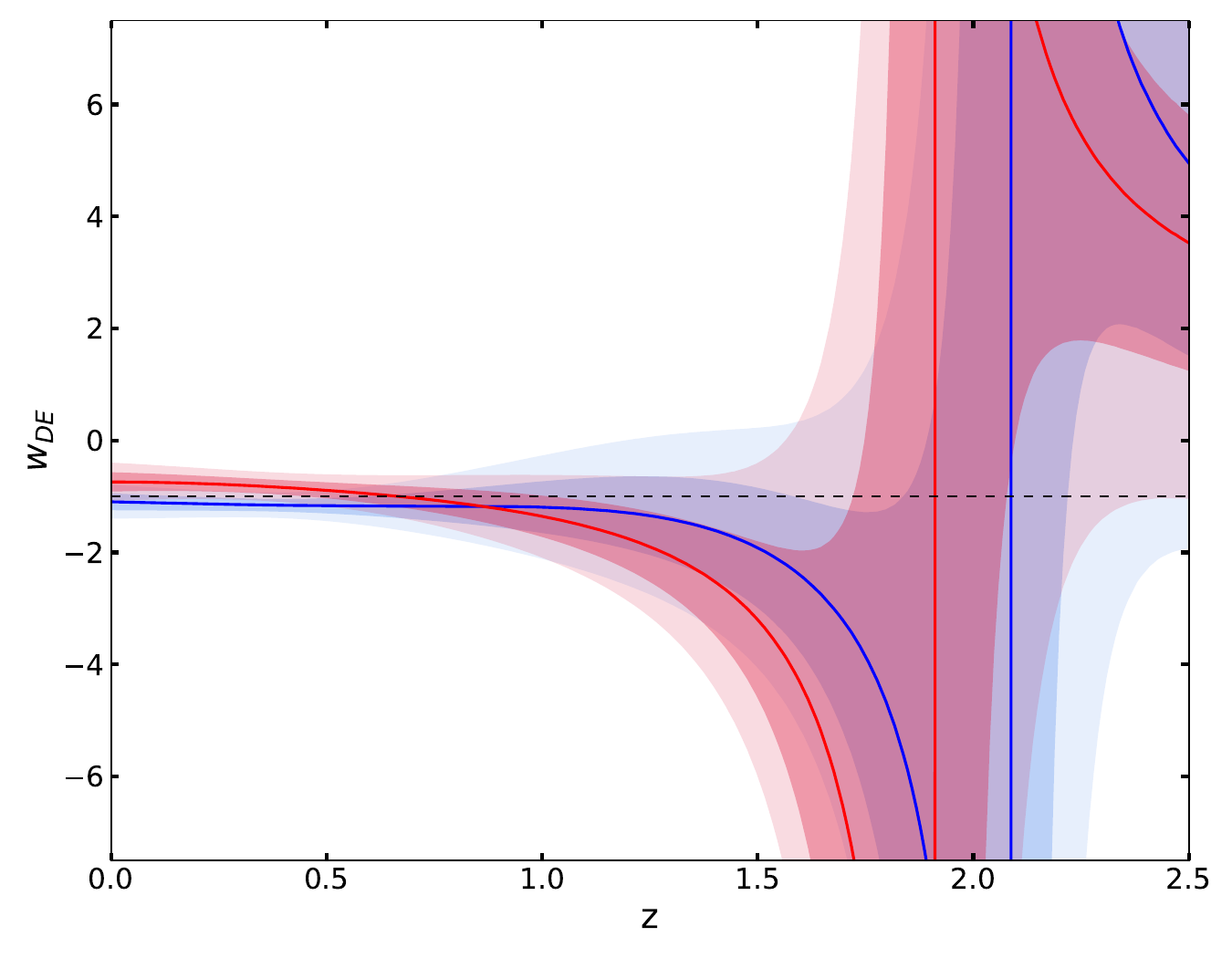}
    \caption{Left panel: Reconstruction at 1$\sigma$ and 2$\sigma$ confidence levels of the $\rho_{\rm DE}(z)/\rho_{\rm c0}$ function using the CC+SN+BAO for the $H_0$-SH0ES (blue) and $H_0$-$\Lambda$\&CMB cases (red), where the black-solid line reference the zero value. Middle Panel: Similar to the left panel, but for the DE inertial mass density scaled with the present-day critical density, $\varrho_{\rm DE}(z) / \rho_{\rm c0}$. Right Panel: Similar to the left panel, but for the DE equation of state parameter $w_{\rm DE}(z)$. The dashed lines indicate the parameters corresponding to the cosmological constant of the Planck-$\Lambda$CDM model, namely, $\frac{\rho_{\rm DE}}{\rho_{\rm c0}}=0.684$ (best fit value~\cite{Planck:2018vyg}), $\varrho_{\rm DE}=0$, and $w_{\rm DE}=-1$.}
    \label{rho_DE_plots}
\end{figure*}

Compromising slightly further from our minimal assumption approach, we can reconstruct the dynamics of dark energy based on our $\delta(z)$ function reconstructions. We adopt the methodology outlined in Ref.~\cite{Akarsu:2022lhx} to establish the relationship between DE dynamics and $\delta(z)$. The evolution of the DE density is then described as follows:
\begin{equation}
     \rho_{\rm DE}(z)
     =3H_0^2\Omega_{\rm DE0}+3H_{\Lambda \rm CDM}^2(z)\delta(z)[2+\delta(z)]
     \label{rho_DE}
\end{equation}
where $\Omega_{\rm DE0}\equiv\Omega_{\rm DE}(z=0)$ is the present-day density parameter of the DE. The continuity equation, $\dot\rho_{\rm DE}(z)+3H(z)\varrho_{\rm DE}(z)=0$, leads to $\varrho_{\rm DE}(z)\equiv\rho_{\rm DE}(z)+p_{\rm DE}(z)=\frac{1+z}{3} \rho_{\rm DE}^{\prime}(z)$ for the inertial mass density, and $w_{\rm DE}(z)\equiv-1+\varrho_{\rm DE}(z)/\rho_{\rm DE}(z)=-1+\frac{1+z}{3} \rho_{\rm DE}^{\prime}(z)/\rho_{\rm DE}(z)$ for the EoS parameter, where $\rho_{\rm DE}(z)$ and $p_{\rm DE}(z)$ are respectively the DE density and pressure, and $^{\prime}\equiv d/dz$. Accordingly, we have:
\begin{equation}
\label{eqn:varrho}
\begin{aligned}
      \varrho_{\rm DE}&=2(1+z)H^2_{\Lambda\rm CDM}\\
      &\quad\times \left[\frac{H'_{\Lambda\rm CDM}}{H_{\Lambda\rm CDM}}\delta(\delta+2)+\delta'(\delta+1)\right],
\end{aligned}
\end{equation}
\begin{equation}
\begin{aligned}\label{eq:eos}
    w_{\rm DE}&=-1+\frac{2(1+z)H_{\Lambda\rm CDM}^2}{3H_0^2\Omega_{\rm DE0}+3H_{\Lambda \rm CDM}^2\delta(2+\delta)}\\
    &\quad\quad\quad\quad\times \left[\frac{H'_{\Lambda\rm CDM}}{H_{\Lambda\rm CDM}}\delta(\delta+2)+\delta'(\delta+1)\right].
\end{aligned}
\end{equation}
For small deviations from the cosmological constant of the Planck-$\Lambda$CDM model, i.e., $\delta(z)\ll1$, these parameters approximately are:
\begin{align}
   \rho_{\rm DE}&\approx 3H_0^2 \Omega_{\rm DE0}+6H_{\Lambda \rm CDM}^2\delta,\\
     \varrho_{\rm DE}&\approx2(1+z)H^2_{\Lambda\rm CDM}\left[\frac{2H'_{\Lambda\rm CDM}}{H_{\Lambda\rm CDM}}\delta+\delta'\right],\\
      w_{\rm DE}&\approx-1+\frac{2(1+z)H^2_{\Lambda\rm CDM}\left[2\frac{H'_{\Lambda\rm CDM}}{H_{\Lambda\rm CDM}}\delta+\delta'\right]}{3H_0^2\Omega_{\rm DE0}+6H_{\Lambda \rm CDM}^2\delta}.
\end{align}

Note that for $\delta=0$, these parameters accurately describe a cosmological constant, with $\rho_{\rm DE}=3H_0^2 \Omega_{\rm DE0}$, $\varrho_{\rm DE}=0$, $w_{\rm DE}=-1$. We observe that the DE density (which is positive today) changes sign in the past, if $\delta(z)$ reaches a large enough negative value:
\begin{equation}
    \rho_{\rm DE}(z)<0\quad\textnormal{for}\quad\delta(z)<-1+\sqrt{1-\frac{\Omega_{\rm DE0}H_0^2}{H^2_{\Lambda \rm CDM}}},
\end{equation}
which can be approximate to $\delta\lessapprox-H_0^2 \Omega_{\rm DE0}/2H_{\Lambda \rm CDM}^2$ for $\delta\ll1$.

Our reconstructions indicate that the best-fit of $\delta(z)$ remains relatively constant (thus $\delta'(z)\sim 0$) around $z=1$, and $\delta(z)\lesssim 0$ with $\delta'(z)< 0$ for $z\gtrsim1.5$, becoming statistically more pronounced for $z\gtrsim2.0$. Using this information and the above equations, we reach several conclusions. \textbf{(i)} Firstly, the inertial mass density of DE is expected to be negative for $z\gtrsim1.5$. \textbf{(ii)} Secondly, it is likely that, with increasing $z$, the DE density transitions from its late-time positive values to negative values at a redshift higher than $\sim1.5$. \textbf{(iii)} Finally, given that $\rho_{\rm DE}(z=0)>0$ and $w_{\rm DE}(z=0)\sim-1$, this transition should be accompanied by an EoS parameter diverging to $-\infty$ just before the transition redshift, exhibiting a singularity at the transition redshift, and then returning from $+\infty$ to approach a finite value of EoS at redshifts larger than the transitions redshift (see Ref.~\cite{Akarsu:2019hmw,Akarsu:2022lhx,Ozulker:2022slu} for further discussions). In light of these intriguing possibilities, we explored the reconstruction of these parameters from our $\delta(z)$ reconstructions for the complete CC+SN+BAO dataset. However, it should be noted that these equations incorporate dependencies on the Planck-$\Lambda$CDM input in the $H_{\Lambda \rm CDM}^2(z)$ function and the $\Omega_{\rm DE0}$ value. For both, we have used the best-fit values from the Planck-$\Lambda$CDM~\cite{Planck:2018vyg}.

In \cref{rho_DE_plots}, we present reconstructions of DE density (left panel) and inertial mass density (middle panel), both scaled by the present-day critical energy density of the universe ($\rho_{\rm c0}=3H_0^2$), alongside the EoS parameter (right panel). It is important to emphasize that in the Planck-$\Lambda$CDM model, $\frac{\rho_{\text{DE}}(z)}{\rho_{\text{c0}}} = 0.684$ (best fit value~\cite{Planck:2018vyg}) remains constant as a function of $z$. In the left panel of this figure, the dashed line represents this value, while the black-solid line represents the vanishing dark energy density, to visually emphasize the potential for negative density values of DE at high $z$ in our reconstructions. In both the $H_0$-$\Lambda$\&CMB and $H_0$-SH0ES scenarios, for $z \lesssim 1.5$, the DE remains consistent with the null hypothesis---namely, a positive cosmological constant with the Planck-$\Lambda$CDM best-fit value of $\rho_{\rm DE}/\rho_{\rm c0} = 0.684$~\cite{Planck:2018vyg}, a null inertial mass density, $\varrho_{\rm DE} = 0$, and equation of state parameter $w_{\rm DE} = -1$, which describe the cosmological constant---within a 2$\sigma$ confidence level. However, analyzing the behavior of DE within a 1$\sigma$ confidence level, or considering the best-fits, we note in both scenarios that DE exhibits phantom region characteristics with an EoS parameter yielding large negative values for $1.5\lesssim z \lesssim2$ and gradually approaching minus unity with decreasing redshift. On the other hand, while DE in the $H_0$-SH0ES scenario consistently aligns with a positive cosmological constant, showing a slight preference towards phantom behavior, in the $H_0$-$\Lambda$\&CMB scenario it surpasses the phantom divide line ($w=-1$) at $z\sim1.0$, subsequently demonstrating quintessence-like behavior with increasing deviation from the cosmological constant as redshift decreases. This deviation is most pronounced at the lowest redshifts in the $H_0$-$\Lambda$\&CMB scenario, exemplified by $w_{\rm DE}(z=0) = -0.74 \pm 0.17$ and $\varrho_{\rm DE}(z=0) / \rho_{\rm c0} = -0.07 \pm 0.12$, indicating a $1.5\sigma$ tension with a positive cosmological constant. Note the decreasing trend of the DE density with decreasing redshift for $z\lesssim0.5$, reaching $\rho_{\rm DE}(z=0) / \rho_{\rm c0} = 0.60 \pm 0.04$ at $z=0$. Conversely, in the $H_0$-SH0ES scenario, there is a slight increase in DE density for $z\lesssim0.5$, reaching $\rho_{\rm DE}(z=0) / \rho_{\rm c0} =0.78 \pm 0.05$ at $z=0$. The deviation from a positive cosmological constant remains stable for $z\lesssim1$, with $w_{\rm DE}(z=0) = -1.09 \pm 0.15$ and $\varrho_{\rm DE}(z=0) / \rho_{\rm c0} = -0.07 \pm 0.12$ at $z=0$, marking only a 0.6$\sigma$ significance level discrepancy at $z=0$. We refer the reader to Ref.~\cite{Escamilla:2023oce} for a comprehensive examination of the current state of constraints on the present-day EoS parameter of DE. One may observe that these findings are in line with our results from the $\delta(z)$ reconstruction for the CC+SN+BAO sample. Notably, there is an increasing trend of $\delta(z)$ with decreasing $z$---consistent with $\delta(z)\sim0$ in the range $0.5\lesssim z\lesssim2.5$ and with best-fits approaching $\delta(z)\sim0$ at $z\sim0.7$---for both scenarios in the range $0.5\lesssim z<2.5$. For $z\lesssim0.5$, there is an enhanced increasing trend of $\delta(z)$, achieving positive values within a 1$\sigma$ confidence level in the $H_0$-SH0ES scenario, while a decreasing trend of $\delta(z)$ develops, reaching negative values within a 1$\sigma$ confidence level in the $H_0$-$\Lambda$\&CMB scenario.

These results provide valuable insights into the dynamics of DE, revealing behaviors that are conventionally unexpected. Among our most intriguing findings, we note that for $z \lesssim 0.5$, the consistency with the cosmological constant is more robust, within a $1\sigma$ confidence level, in the $H_0$-SH0ES scenario---using the SH0ES $H_0$ measurement, which is in significant (5$\sigma$) tension with the Planck-$\Lambda$CDM predicted $H_0$---than in the $H_0$-$\Lambda$\&CMB scenario, which adopts the Planck-$\Lambda$CDM predicted $H_0$ and thus readily assumes a cosmological constant as the DE component. This implies that, for $z \lesssim 0.5$, the data prefer an $H_0$ larger than the Planck-$\Lambda$CDM prediction but still favor a cosmological constant as DE. Another intriguing observation emerges when considering the best-fit of the reconstructions for $z \lesssim 0.5$. It is not surprising to find $\delta(z) > 0$, indicative of phantom-like DE behavior, in the $H_0$-SH0ES scenario, as this assumes the SH0ES $H_0$ measurement, which is larger than the Planck-$\Lambda$CDM prediction at a 5$\sigma$ significance level. However, what is surprising is that in the $H_0$-$\Lambda$\&CMB scenario, $\delta(z) < 0$ suggesting a quintessence-like DE behavior. This implies an $H_0$ smaller than the Planck-$\Lambda$CDM prediction, i.e., a trend that exacerbates the existing $H_0$ tension within the standard $\Lambda$CDM model, which the $H_0$-$\Lambda$\&CMB scenario is based on.

An observation that may be more intriguing than all these arises when relating the DE dynamics to the fact that, in both scenarios, the increasing (decreasing) trend of $\delta(z)$ with decreasing (increasing) redshift extends to the highest redshift of $z=2.5$, and moreover $\delta(z)$ becomes negative with increasing significance, beyond a 2$\sigma$ confidence level, for $z \gtrsim 2$. This can be explained by the presence of DE that changes the sign of its energy density around $z \sim 2$, as seen in the left panel of Figure~\ref{rho_DE_plots}. To better understand this phenomenon, let us focus on the region beyond $z\sim1.5$. Firstly, we observe in the middle panel of Figure~\ref{rho_DE_plots} that the inertial mass density of the DE is negative, $\varrho_{\rm DE}(z)<0$, for $z\gtrsim1.5$ and becomes increasingly negative with increasing redshift. This behavior is reminiscent of phantom DE (described by $\rho_{\rm DE}>0$ along with $w_{\rm DE}<-1$), which typically has negative inertial mass density ($\rho_{\rm DE}+p_{\rm DE}<0$). However, unlike typical phantom DE, here the DE density does not have a minimum at $\rho_{\rm DE}=0$. Instead, it becomes zero at a certain redshift $z_{\rm p}\sim2$, i.e., $\rho_{\rm DE}(z=z_{\rm p})=0$, and assumes negative values for $z\gtrsim2$. Specifically, it transitions to negative energy density values at $z_{\rm p}=1.91^{+0.16}_{-0.18}$ ($1\sigma$ CL) in the $H_0$-$\Lambda$\&CMB scenario and at $z_{\rm p}=2.08^{+0.12}_{-0.20}$ ($1\sigma$ CL) in the $H_0$-SH0ES scenario. We see from the continuity equation, $w_{\rm DE}(z)=-1+\frac{1+z}{3} \rho_{\rm DE}'(z)/\rho_{\rm DE}(z)$, that a negative energy density decreasing with increasing redshift (namely, $\rho_{\rm DE}(z)<0$ and $\rho_{\rm DE}'(z)<0$) implies a quintessence-like EoS parameter, i.e, $w_{\rm DE}>-1$. We also observe that since the DE density decreases with increasing redshift (i.e., $\rho_{\rm DE}'(z)<0$) for the entire range of $z\gtrsim1.5$, its equation of state (EoS) parameter must diverge to minus infinity as the positive $\rho_{\rm DE}(z)$ approaches zero (for $1.5\gtrsim z > z_{\rm p}$). It exhibits a singularity when the DE density reaches zero (at $z=z_{\rm p}$), and then at redshifts beyond $z_{\rm p}$, as the DE density changes sign to become negative, the EoS parameter starts decreasing from positive infinity and approaching a finite value larger than minus unity, see the left and right panels of Figure~\ref{rho_DE_plots}. Such a singularity (pole), represented as ${\lim_{z\to z_{\rm p}^\pm}w_{\rm DE}(z)=\pm\infty}$, is necessary for a minimally interacting DE that changes the sign of its energy density, becoming positive in the late universe~\cite{Ozulker:2022slu}.

Our reconstructions of DE dynamics, utilizing our $\delta(z)$ reconstructions, closely align with most properties of Omnipotent Dark Energy~\cite{Adil:2023exv}, which offers an extended description/classification of DE---to meet a need that arises when DE is allowed to assume negative energy densities dynamically, going beyond traditional DE models like quintessence, phantom, and quintom~\cite{Copeland:2006wr}. Specifically, when tracing the history from the past to the present in both scenarios, we observe DE exhibiting n-quintessence characteristics ($\rho_{\rm DE}<0$ and $w_{\rm DE}>-1$) at high redshifts ($z\gtrsim2$). It then undergoes a transition in its energy density sign, accompanied by a singularity in its EoS parameter, around $z\sim2$, subsequently yielding p-phantom characteristics in the redshift range of $1\lesssim z\lesssim2$ in both scenarios. At lower redshifts, the DE characteristics differ in the two scenarios; though consistent with a positive cosmological constant at a 2$\sigma$ confidence level for $z\lesssim1.5$, in the $H_0$-SH0ES scenario, the DE tends to remain in the p-phantom region until today. In contrast, in the $H_0$-$\Lambda$\&CMB scenario, it leans towards the p-quintessence region ($\rho_{\rm DE}>0$ and $w_{\rm DE}>-1$).

We note that our reconstructed DE dynamics from the combined CC+SN+BAO dataset can accommodate certain DE models that can  support large $H_0$ values, thus alleviating the $H_0$ tension when these models are constrained by combining the Planck CMB data along with the data from the late-universe observations. Such models include Graduated Dark Energy (gDE)~\cite{Akarsu:2019hmw}, the $\Lambda_{\rm s}$CDM model~\cite{Akarsu:2021fol,Akarsu:2022typ,Akarsu:2023mfb} (which replaces the standard $\Lambda$CDM model's cosmological constant with a  rapidly sign-switching cosmological constant), models considering dynamical DE with positive energy density on top of an AdS background \cite{Visinelli:2019qqu,Dutta:2018vmq,Sen:2021wld}, and the DMS20 Omnipotent Dark Energy model~\cite{DiValentino:2020naf,Adil:2023exv}. Besides these, Interacting Dark Energy (IDE) models~\cite{Kumar:2017dnp,DiValentino:2017iww,Yang:2018uae,Pan:2019gop,Kumar:2019wfs,DiValentino:2019ffd,DiValentino:2019jae,Lucca:2020zjb,Gomez-Valent:2020mqn,Kumar:2021eev,Nunes:2022bhn,Bernui:2023byc} also suggest a late-time solution to the $H_0$ tension. However, recent model-independent reconstructions of the IDE kernel, utilizing binned and Gaussian process methods, suggest a sign-change in DE density around $z\sim2$, similar to our findings~\cite{Escamilla:2023shf}. This indicates that IDE models do not account for the sign-change in the DE density at $z\sim2$; that is, they do not eliminate this phenomenon and guarantee that DE density is always positive~\cite{Escamilla:2023shf}. Our analysis highlights the necessity of exploring new physics within the redshift range of $1.5\lesssim z\lesssim2.5$, with a particular focus around $z = 2$. This targeted approach in the late universe is not only crucial for a deeper understanding of the dynamics in this specific range but also offers the potential to address the cosmological tensions, such as the $H_0$ and $S_8$ tensions. Our findings thus emphasize the importance of this redshift window in the ongoing quest for answers in cosmology.

It is worth emphasizing before closing this section that the above discussion regarding the possibility of having negative DE densities at high redshifts ($1.5 \lesssim z \lesssim 2.5$) is based on our minimal assumption approach, which led us to relate the kinematical deviations encoded in the $\delta$ reconstructions from the Planck-$\Lambda$CDM model to the deviations of a minimally interacting DE from the cosmological constant. As noted in~\cref{footnote1}, one could consider attributing these kinematical deviations to different types of new physics, if not to some undiscovered systematics in the data. One reasonable alternative could be allowing the dynamics of DM to deviate from its standard $\rho_{\rm dm} \propto (1+z)^3$ evolution (while keeping DE as a cosmological constant), particularly since DM is traditionally the dominant constituent of the universe at redshifts related to our negative DE findings. This could be done by defining $\Delta\rho_{\rm dm} \equiv \rho_{\rm dm}(z) - \rho_{\rm dm0}(1+z)^3$ (where $\rho_{\rm dm0}$ represents the present-day energy density of DM in the standard $\Lambda$CDM model) and relating this to $\delta(z)$. In this case, $\delta(z) > 0$ and $\delta(z) < 0$ from our reconstructions would be interpreted as DM density values larger and smaller than those predicted by the Planck-$\Lambda$CDM model at various redshifts, respectively, implying DM energy density deviating from $\rho_{\rm dm} \propto (1+z)^3$ and, possibly (necessarily if DM is assumed to be minimally interacting), an EoS parameter deviating from $w_{\rm dm} = 0$. We do not delve into this interesting possibility in the current work and postpone it to future research, but we refer to a few studies that might provide insights regarding such an option. Ref.~\cite{Malekjani:2023ple} suggests a preference for negative DE densities for $z > 1$ using Pantheon-plus data, but also discusses that this finding might be indicating a need to question the assumption of pressureless matter energy density scaling as $(1+z)^3$ in the late universe. Ref.~\cite{Escamilla:2023shf}, studying the model-independent reconstruction (using binned and Gaussian process) of DE and DM that are allowed to interact non-minimally with each other (so that DM energy density evolution is allowed to deviate from its standard behavior), finds altered dynamics for DM density but does not eliminate the possibility of negative DE densities at high redshifts. Ref.~\cite{Wen:2023wes} introduces a model to address the $H_0$ tension, which involves a new term in the Friedmann equation that behaves like a cosmological constant today and a phantom at low redshifts, but behaves like DM with negative energy densities ($\rho_{\rm dm} < 0$ and $w_{\rm dm} \sim 0$) at high redshifts in addition to the standard CDM matter. Finally, our findings on the possibility of negative DE densities at high redshifts could also be interpreted as an indication of possible new physics that can be effectively modeled as a minimally interacting DE that assumes negative density values at high redshifts.

\subsection{Reconstructing a phenomenological model using the information from $\delta(z)$ reconstructions}

\begin{table*}[ ]
    \renewcommand{\arraystretch}{1.5}
    \setlength{\tabcolsep}{8pt}
    \centering
\begin{tabular}{c|c|cc|cc}
\hline \hline  \multicolumn{2}{c|}{ } &  \multicolumn{2}{c|}{$H_0$-SH0ES scenario} & \multicolumn{2}{c}{$H_0$-$\Lambda$\&CMB scenario}\\
\hline
Parameters & CC & CC +SN & CC+SN+BAO & CC +SN & CC+SN+BAO \\
\hline
$\Omega_{\mathrm{m0}}$ & $0.320^{+0.044}_{-0.033}$  & $ 0.338^{+0.036}_{-0.014}$ & $0.346^{+0.029}_{-0.008}$ & $0.314 \pm 0.032$  & $0.342^{+0.032}_{-0.011}$  \\
$ C $ & $-0.2 \pm 1.3$ & $-0.34 \pm 0.73$ & $-0.37^{+0.39}_{-0.55}$ & $0.0^{+1.3}_{-1.0}$  & $-0.72^{+0.44}_{-0.87}$ \\
$ D $ & $0.2 \pm 1.4$ & $0.28 \pm 0.77$ & $0.54^{+0.67}_{-0.41}$  & $-0.01^{+0.96}_{-1.3}$  & $0.27 \pm 0.51$ \\
$H_{0}\,[{\rm \,km\, s^{-1}\, Mpc^{-1}}]$ & $68.5 \pm 4.0$ & $70.7^{+2.2}_{-1.9}$ & $70.9^{+1.8}_{-2.0}$ & $65.7 \pm 2.2$ &  $64.8 \pm 1.3$ \\
\hline \hline
\end{tabular}
\caption{Marginalized constraints, expressed as mean values with a 68\% confidence level, are provided for the free parameters of the test $H_{\rm th}$ theoretical model. These constraints are derived from various data-set combinations considered in this study.}
    \label{tab:H_model}
\end{table*}

\begin{figure}
    \centering
    \includegraphics[width=8cm]{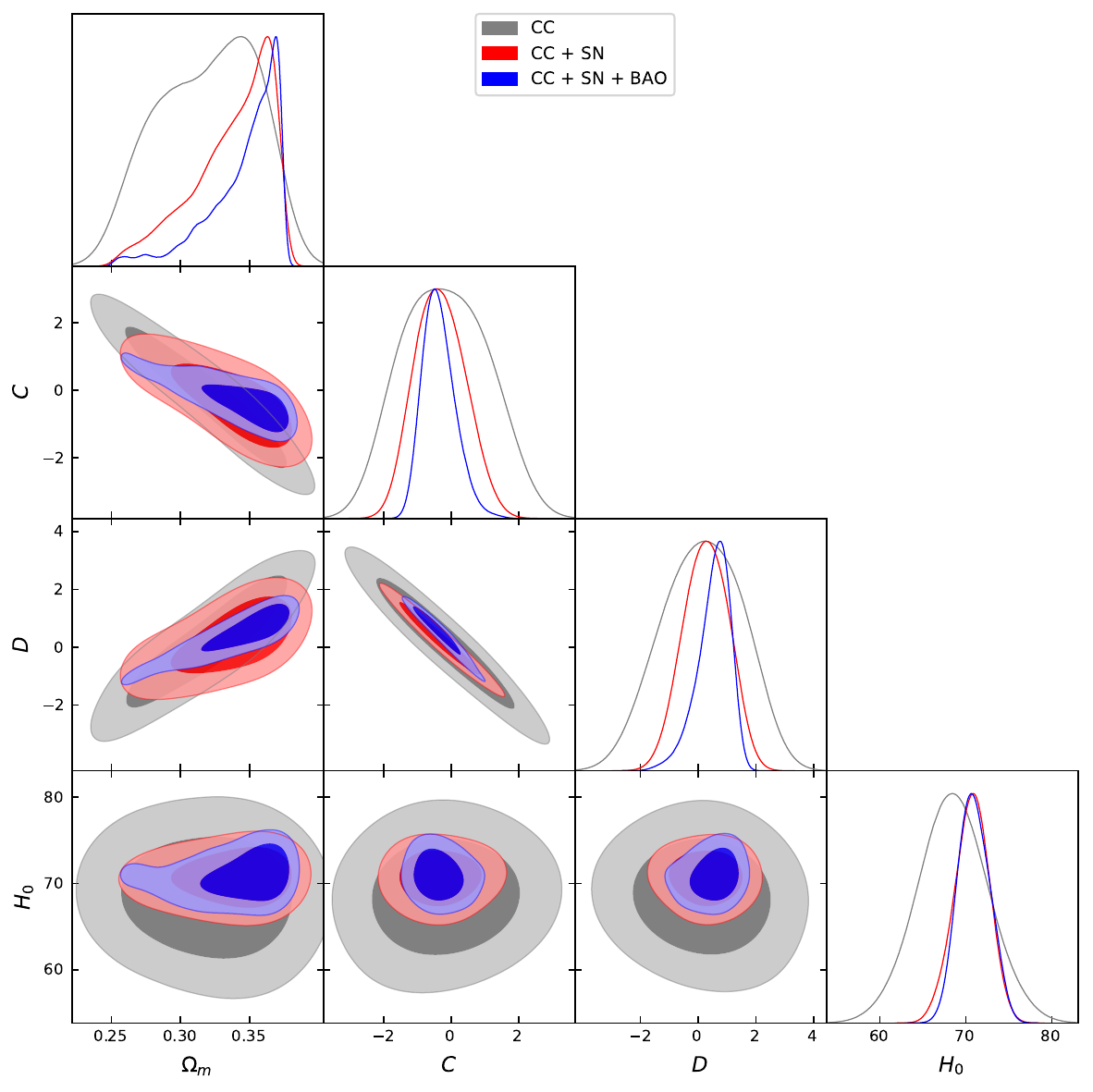}
    \includegraphics[width=8cm]{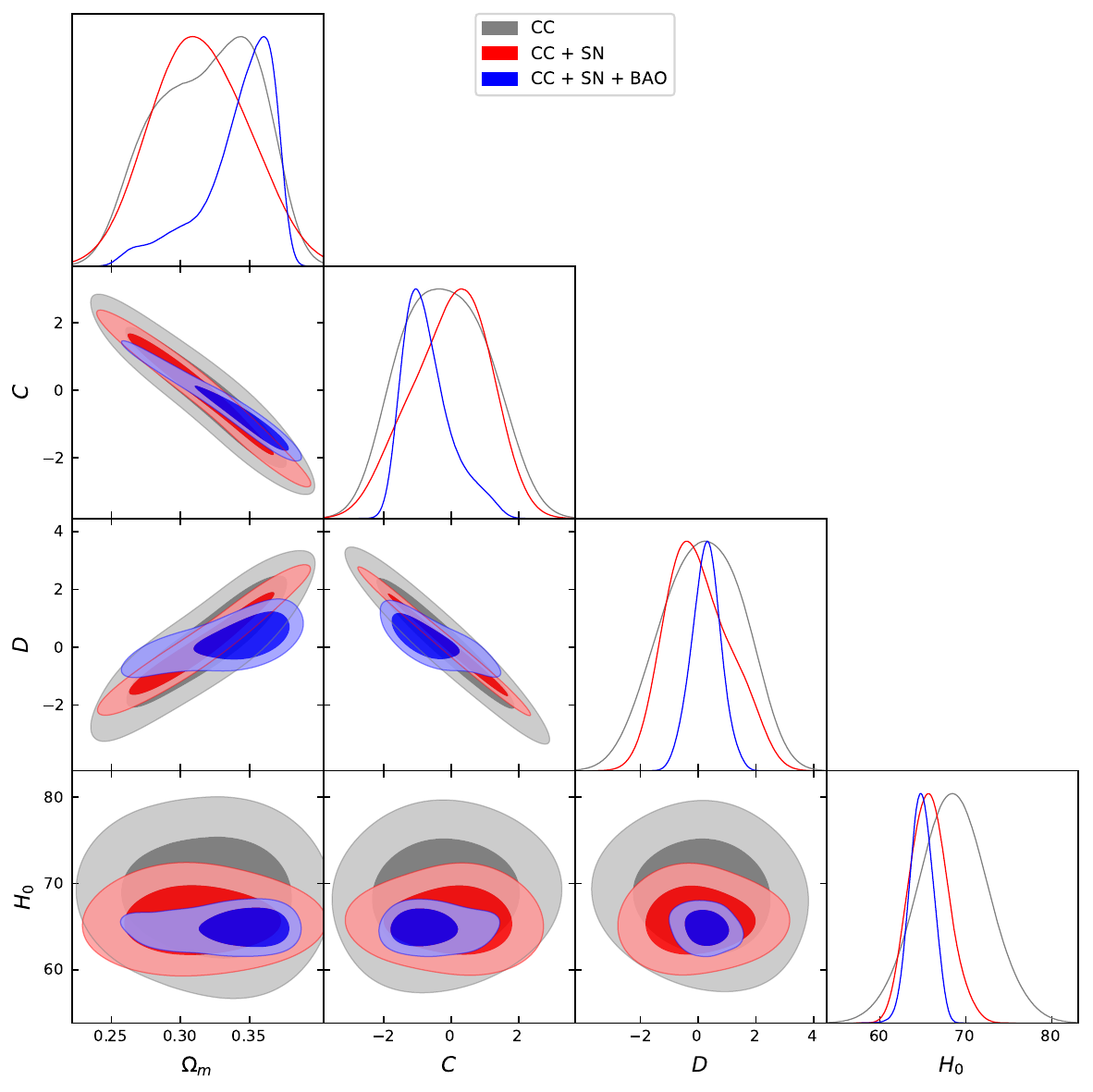}
    \caption{Upper panel: 1D posterior distribution and 2D contour plots for the model in equation (\ref{H_th}) using the CC+SN+BAO sample in the $H_0$-SH0ES case. Lower panel: Similar plots for the $H_0$-$\Lambda$\&CMB case. }
    \label{fig_MCMC}
\end{figure}

After discussing all the GP reconstructions, our objective in this section is to create an interface for analyzing and reconstructing a cosmological $H(z)$ model, utilizing the information from the $\delta(z)$ function. To achieve this, we define the covariance of the reconstruction function as follows:
\begin{equation}
\label{cov}
  {\rm Cov}_{\delta} = \sum_l^i \sum_k^j {\rm Cov}(\delta(z_k), \delta(z_l)) + \sigma_{kl}^{2} \eta_{kl} ,
\end{equation}
where $\delta(z)$ information is derived from Eq.~\eqref{delta}, and $\eta_{kl}$ represents the Kronecker delta.

To maximize the logarithmic likelihood function through Markov Chain Monte Carlo (MCMC) statistical analysis, we employ the MontePython code~\cite{Brinckmann:2018cvx, Audren:2012wb}. This process involves sampling the parameter space to perform a comprehensive exploration. The log-likelihood function is expressed as:
\begin{equation}
\ln \mathcal{L} = - \frac{1}{2} \Delta \delta^T {\rm Cov}_{\delta}^{-1} \Delta \delta,
\end{equation}
where
\begin{equation}
 \Delta \delta = \delta_{\rm theory}(\theta,z) - \delta_{\rm GP}(z),
\end{equation}
and ${\rm Cov}_{\delta}$ is the corresponding total covariance matrix as defined in Eq.~\eqref{cov}. In this context, $\theta$ represents a vector of free parameters in a given theory, encapsulating the $\delta_{\rm theory}$ function that is under test. The equation captures the difference between the theoretical predictions, $\delta_{\rm theory}$, and the GP prediction, $\delta_{\rm GP}$, using the inverse covariance matrix ${\rm Cov}_{\delta}^{-1}$. This formulation is integral to the MCMC analysis carried out using the MontePython code.

To incorporate a generalized theoretical function for the Hubble parameter, $H(z)$, we adopt the model proposed in~\cite{lemos2019model, Efstathiou_2021}:
\begin{equation}
\label{H_th}
 \frac{H_{\rm th}^2(z)}{H_0^2} = \Omega_{\rm m0}(1+z)^3 + (1-\Omega_{\rm m0}) [1 + C z +  D \ln(1 + z) ],
\end{equation}
where $C$ and $D$ are dimensionless free parameters. This model offers a flexible representation of the evolution of the Hubble parameter, capturing the dynamics of both the matter-dominated and dark energy-dominated regimes through the incorporation of the $C$ and $D$ terms.

In the course of our analysis and chain processing, we employ the Python package GetDist.\footnote{\url{https://github.com/cmbant/getdist}} This package offers a suite of valuable tools and functionalities essential for managing and interpreting parameter chains, thereby significantly enhancing the efficiency and reliability of our data analysis workflows.

Table~\ref{tab:H_model} summarizes our statistical results for the free parameters of the model at 68\% CL. Figure~\ref{fig_MCMC} displays the 2D joint and 1D marginalized posterior probability distributions for the baseline parameters of the theory. During our statistical analysis, we apply uniform priors as follows: $\Omega_{\rm m0} \in [0.2, 0.4]$, $C \in [-4, 4]$, and $D \in [-4, 4]$. These priors have been selected based on a combination of theoretical considerations and statistical reasoning, as detailed by the authors in~\cite{lemos2019model}. While for $H_{0}$ a Gaussian prior was applied, with mean and standard deviation $\sigma$ coming from the GP reconstruction in each scenario.  

Upon examination of the results, it becomes apparent that an analysis solely based on CC does not effectively constrain the parameter space of the model. However, with the inclusion of reconstructed information from SN and BAO, a significant enhancement in the model's parameter space is observed. In the CC+SN+BAO analysis, the dynamics of $H_{\rm th}$ can be recovered with good precision, and the parameters C and D are well-constrained.

Considering all statistical information for $\delta(z)$ with $z \in [0, 2.5]$, we do not observe significant deviations in the dynamics compared to the $\Lambda$CDM model. The methodology presented here can be applied to analyze any phenomenological $H_{\rm th}(z)$ function, showcasing its versatility in probing the underlying cosmological dynamics.

\section{Conclusion}
\label{Conclusion}

In conclusion, our investigation into the expansion dynamics of the late Universe, using Gaussian Processes (GP) for model-independent inference, has provided valuable insights. The reconstructions of the $\delta(z)$ function~\cite{Akarsu:2022lhx}—a parameter measuring the deviation from the expansion of the Universe as described by the Planck-$\Lambda$CDM model, with $\delta(z)=0$ implying no deviation—and consequently the $H(z)$ behaviors, have been achieved by adhering to a framework of minimal assumptions. Extending this approach to the $\Omega_{\rm \Delta DE}(z)$ functions and ultimately to the dynamics of dark energy, namely, $\rho_{\rm DE}(z)$, $\varrho_{\rm DE}(z)$, and $w_{\rm DE}(z)$, we utilized data from the late universe extending up to a redshift of $\sim2.5$, including Cosmic Chronometers (CC), Type Ia Supernovae (SN), and Baryon Acoustic Oscillations (BAO) samples. This analysis has shed light on potential kinematic deviations from the Planck-$\Lambda$CDM model and subsequent dynamical deviations of dark energy from cosmological constant.

Intriguing and unexpected features in the evolution of the $\delta(z)$ function were revealed, particularly at low redshifts ($z \lesssim 0.5$), where the impact of the current $H_0$ tension is more pronounced, and at high redshifts ($z \gtrsim2$), while significant deviations from the Planck-$\Lambda$CDM model were not observed in the range $0.5\lesssim z \lesssim2$. Variations in the behavior of $\delta(z)$ and $\Omega_{\rm \Delta DE}(z)$ (which was shown to be approximately $2\delta(z)$, as data mostly favor $|\delta(z)|\ll1$), depending on whether the $H_0$-SH0ES or $H_0$-$\Lambda$\&CMB scenario is chosen, underscore the sensitivity of these functions to the choice of the Hubble constant. Moreover, the inclusion of BAO data has further refined our constraints, emphasizing the role of additional cosmological probes in enhancing observational precision. The joint analysis of CC+SN+BAO not only improved the robustness of our constraints but also extended our ability to infer expansion dynamics up to $z \simeq 2.5$. The derived functions provide a comprehensive view of the late Universe, revealing nuanced deviations that warrant further theoretical exploration.

Overall, our results contribute to the ongoing debate regarding the Hubble tension and its potential implications for new physics at late times or any problem in the measurement estimates \cite{Moresco_2023}. The model-independent approach adopted in this study offers a valuable framework for understanding the intricacies of cosmic expansion, paving the way for more refined investigations into the fundamental nature of the Universe.

Utilizing our $\delta(z)$ reconstructions from the combined CC+SN+BAO dataset, we've gained insights into the dynamics of dark energy (DE), which align with the properties of Omnipotent Dark Energy~\cite{Adil:2023exv} and extend beyond conventional DE models like quintessence, phantom, and quintom~\cite{Copeland:2006wr}. In both scenarios, we observe DE exhibiting n-quintessence characteristics at high redshifts ($z\gtrsim2$). It then undergoes a transition in its energy density sign, accompanied by a singularity in its EoS parameter, around $z\sim2$, subsequently yielding p-phantom characteristics in the redshift range of $1\lesssim z\lesssim2$. At lower redshifts, the DE characteristics differ in the two scenarios; though consistent with $\Lambda>0$ at a 2$\sigma$ confidence level for $z\lesssim1.5$, in the $H_0$-SH0ES scenario, DE tends to remain in the p-phantom region until today. In contrast, in the $H_0$-$\Lambda$\&CMB scenario, it leans towards the p-quintessence region. Our reconstructed DE dynamics are compatible with a class of models addressing the $H_0$ tension, including the gDE model~\cite{Akarsu:2019hmw}, the $\Lambda_{\rm s}$CDM model~\cite{Akarsu:2021fol,Akarsu:2022typ,Akarsu:2023mfb}, models considering dynamical DE with positive energy density on top of an AdS background \cite{Visinelli:2019qqu,Dutta:2018vmq,Sen:2021wld}, and the DMS20 Omnipotent Dark Energy model~\cite{DiValentino:2020naf,Adil:2023exv}. Furthermore, IDE models~\cite{Kumar:2017dnp,DiValentino:2017iww,Yang:2018uae,Pan:2019gop,Kumar:2019wfs,DiValentino:2019ffd,DiValentino:2019jae,Lucca:2020zjb,Gomez-Valent:2020mqn,Kumar:2021eev,Nunes:2022bhn,Bernui:2023byc} also propose a late-time solution to the $H_0$ tension. However, recent model-independent reconstructions of the IDE kernel~\cite{Escamilla:2023shf}, using Gaussian process methods, suggest a sign-switch in DE density around $z\sim2$, mirroring our findings. This indicates that IDE models may not fully account for the sign-switching in DE density at $z\sim2$.

Our analysis underlines the need to explore new physics at very low redshifts ($z\lesssim0.5$) and within the redshift range of $1.5\lesssim z\lesssim2.5$, especially around $z = 2$. This targeted approach in the late universe is crucial for a deeper understanding of the dynamics in this specific range and offers potential solutions to cosmological tensions, such as the $H_0$ and $S_8$ tensions. Thus, our findings emphasize the significance of these redshift windows in the ongoing quest for answers in cosmology.

\appendix 
\section{An independent test of $H_0$-$\Lambda$\&CMB and local}

To conduct the analysis outlined in the main text utilizing the SN sample, it is necessary to estimate a value for $H_0$ in order to rescale the SN data as an input for statistical reconstructions. Given the existing tension in determining $H_0$, the most reasonable choices are the values adopted in the $H_0$-SH0ES and $H_0$-$\Lambda$\&CMB contexts. As an alternative approach, we propose the following test:
\begin{enumerate}[nosep]
\item Initially, we reconstruct the $H(z)$ function solely based on the CC sample.
\item Subsequently, we assess $H(z=0)$ at a 1$\sigma$ confidence level.
\item Notably, the value obtained in step (ii) is independent of any other dataset, whether it be from CMB or local observations. Consequently, we utilize this independently derived value as the input to rescale the SN sample.
\end{enumerate}

\begin{figure}[ht!]
    \centering
    \includegraphics[width=7cm]{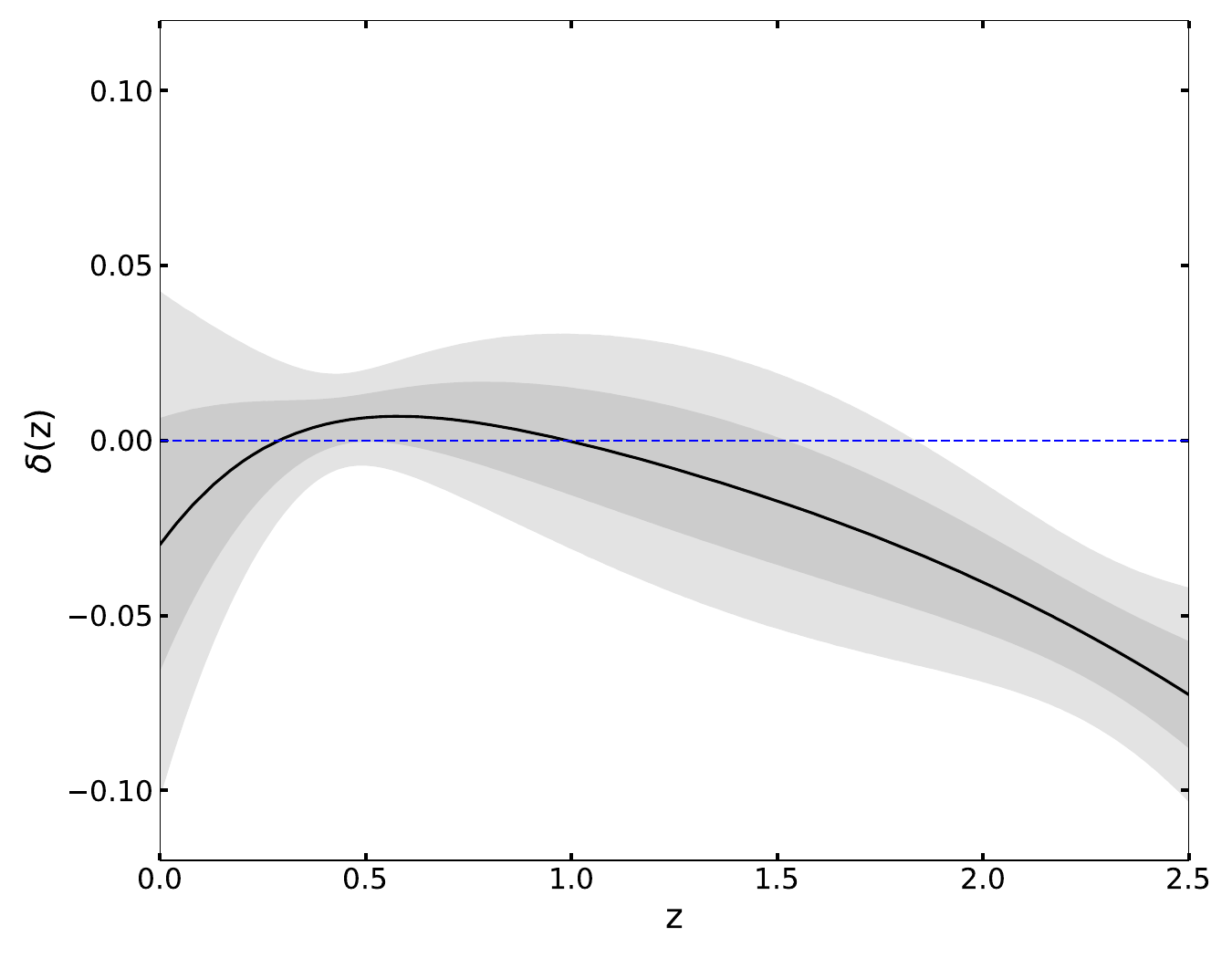} 
    \caption{Reconstruction at 1$\sigma$ and 2$\sigma$ CL of $\delta(z)$ function using the CC+SN+BAO sample, assuming $H_0$ input constraints from CC reconstruction (Figure \ref{Hz_sample} left panel) for scaling SN data. The blue-dashed line is the null hypothesis.}
    \label{delta_CC_SN_CC_BAO}
\end{figure}

\cref{delta_CC_SN_CC_BAO} depicts the reconstruction of the $\delta(z)$ function based on the considerations outlined above. Notably, the reconstructed $\delta(z=0)$ and its values up to $z = 0.5$ are entirely consistent with the null hypothesis. However, for $z > 2$, there persists a pronounced preference at a 2$\sigma$ confidence level for $\delta(z) < 0$. The disparities observed at low redshifts when compared to the $H_0$-SH0ES and $H_0$-$\Lambda$\&CMB scenarios, particularly for $z < 0.5$, can be attributed to the substantial error associated with the $H_0$ value derived solely from CC samples. This higher uncertainty in the input value propagates into larger error bars for all $\delta(z)$ reconstructions. In summary, in the spirit of the methodology presented here, we conclude that only measurements of $H_0$ with an accuracy of approximately 2\% or less may reveal potential inadequacies that could prompt the exploration of new physics at late times.

\section{Assessing the influence of statistical decisions}
\label{B}

Gaussian processes offer a regression approach that does not rely on a predefined parametric model to extract insights from observational data. However, it is necessary to specify a functional form for the kernel and an `initial guess', typically represented by the mean function. While much research focuses on exploring the impact of different kernels on outcomes, the significance of the mean function is often overlooked, though there are a few exceptions~\cite{Holsclaw_2011,Shafieloo_2012,Hwang_2023}.

In this appendix, we present results regarding the reconstruction of the $\delta(z)$ function. We investigate four distinct kernel choices and extend our analysis to include assumptions about the mean function beyond the null choice.

\begin{figure}[ht!]
    \centering
    \includegraphics[width=7cm]{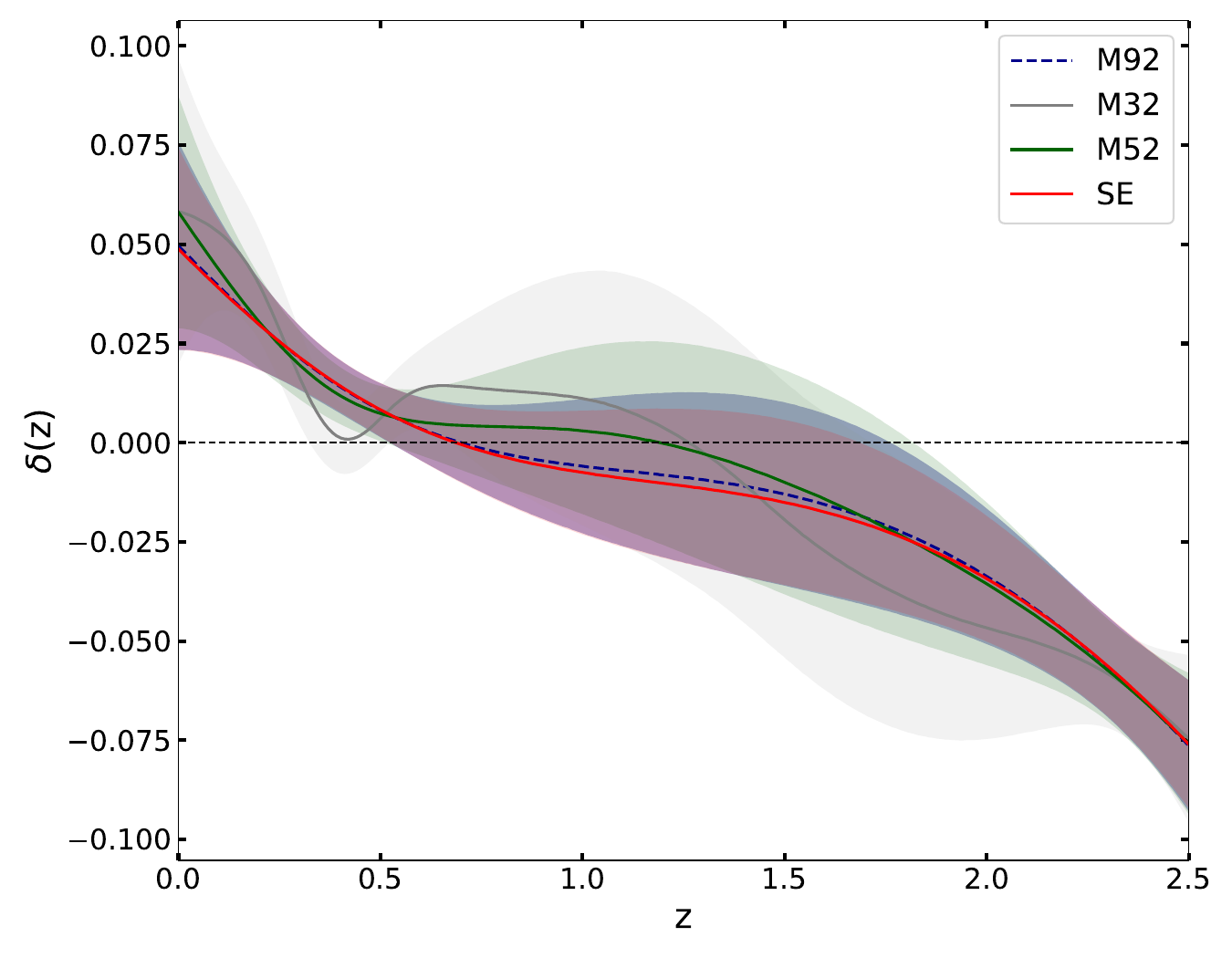} 
    \includegraphics[width=7cm]{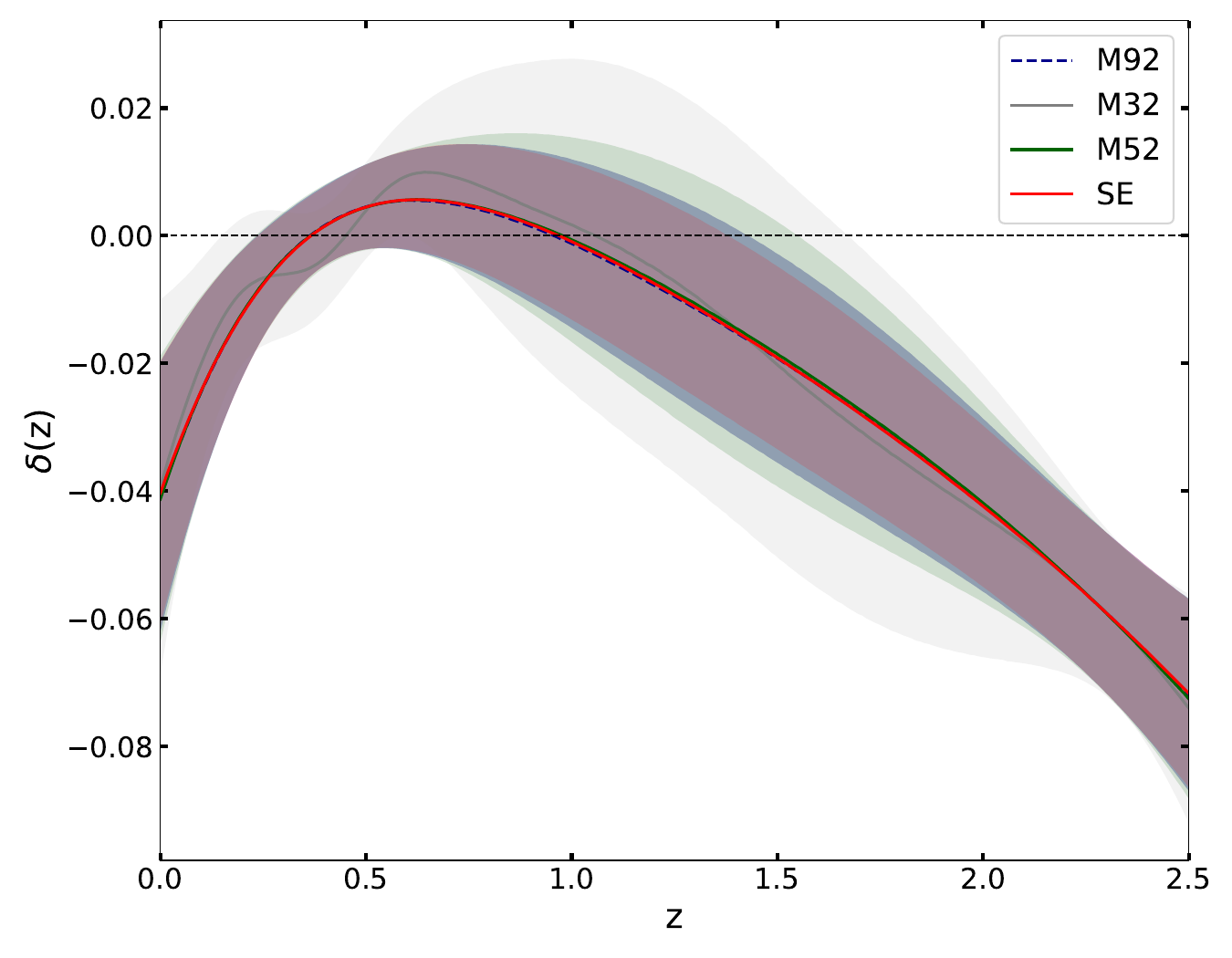} 
    \caption{Upper Panel: Reconstruction of the $\delta(z)$ function at 1$\sigma$ CL for the combined CC+SN+BAO sample with $H_0$-SH0ES, incorporating scaled SN data. Lower Panel: Similar reconstruction for the $H_0$-$\Lambda$\&CMB scenario. In both instances, the black dashed line represents the null hypothesis. In the legend, the notations M92, M32, M52, and SE represent the Mat\'{e}rn class kernels with $\nu = 9/2,\, 3/2,\, 5/2$, and the Squared-Exponential (SE) kernel, respectively.}
    \label{delta_kernel}
\end{figure}

\cref{delta_kernel} shows, in the upper panel, the reconstruction of the $\delta(z)$ function at 1$\sigma$ CL from the analysis of the combined CC+SN+BAO sample within the $H_0$-SH0ES framework. Four different kernel assumptions were employed: the Mat\'{e}rn class kernel with $\nu =9/2,\, 3/2,\, 5/2$, and the SE kernel. It is evident that all reconstructions exhibit a high degree of statistical compatibility with each other. In the lower panel, we depict the reconstruction for the $H_0$-$\Lambda$\&CMB scenario, where we observe a similar trend. Consequently, we can infer that all analyses and conclusions drawn in the main text remain robust and independent of the choice of kernel.

\begin{figure}[ht!]
    \centering
    \includegraphics[width=7cm]{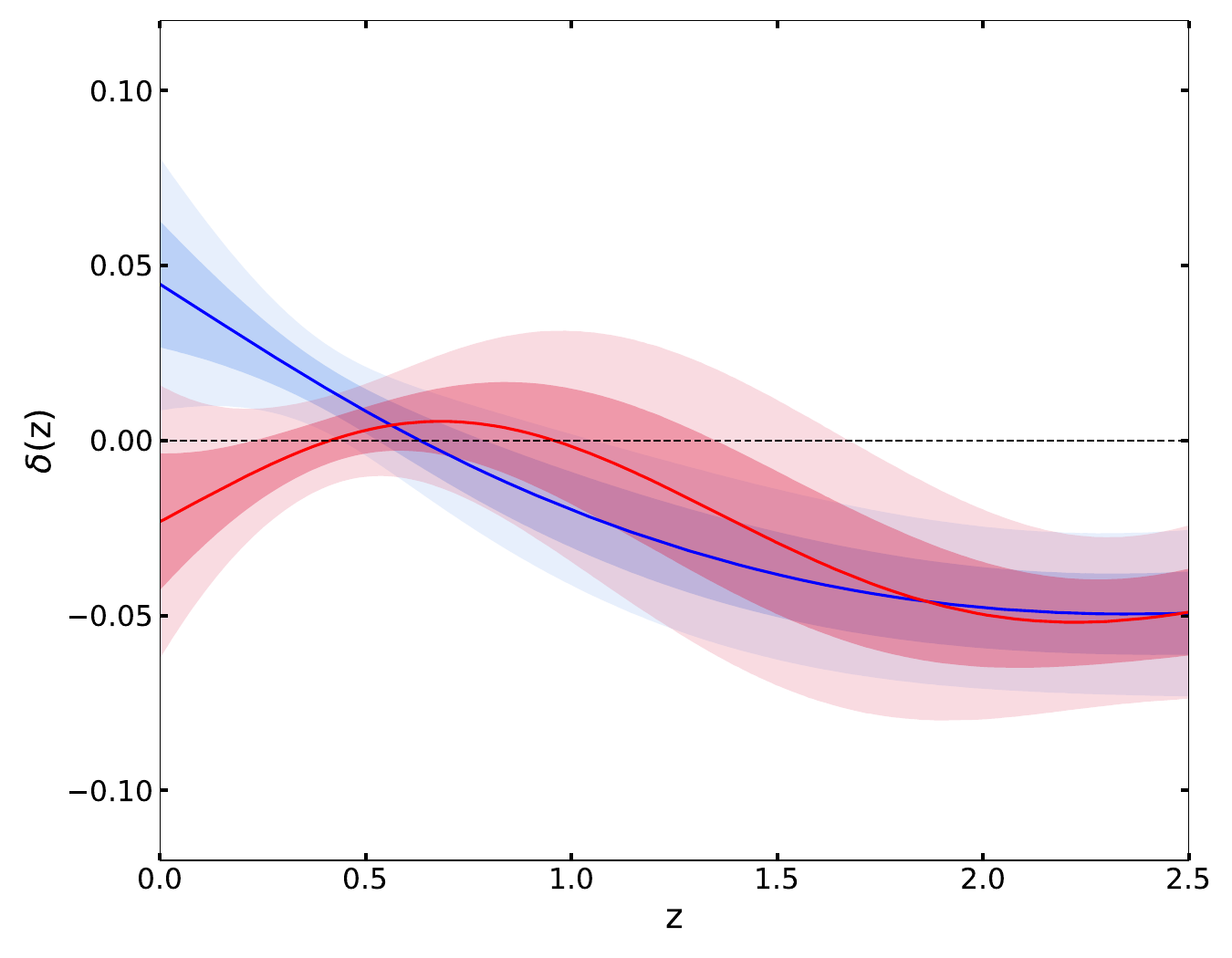} 
    \caption{Reconstruction at $1\sigma$ and $2\sigma$ CL of the $\delta(z)$ function using the combined CC+SN+BAO sample for the cases of $H_0$-SH0ES (blue) and $H_0$-$\Lambda$\&CMB (red). For both cases, the $\Lambda$CDM framework has been assumed for the mean function.} 
    \label{delta_mean_function}
\end{figure}

A recent study conducted by~\cite{Hwang_2023} examined the influence of the mean function. The findings highlighted that assuming a zero mean function led to an inability to accurately capture the reference model used for generating the data. Following the methodology outlined in~\cite{Hwang_2023}, we selected a mean function expected to reasonably describe the data, such as the $\Lambda$CDM model from CMB-Planck data best-fit values. \cref{delta_mean_function} illustrates the resulting reconstructions for the function $\delta(z)$ for both cases of interest in our main results. We can observe that the results are in perfect statistical agreement with the main results presented in the lower panel of~\cref{delta_CC_SN_BAO}. Additionally, we can observe that the reconstruction at high $z$ becomes more robust, although this does not change any of the main conclusions in our results. Thus, we conclude that our main results are invariant under the choice of the mean function.

\begin{acknowledgments}
The authors express their gratitude to the referees for their valuable comments and suggestions, which have greatly enhanced the overall quality of the work. M.A.S. received support from the CAPES scholarship. \"{O}.A. acknowledges the support of the Turkish Academy of Sciences in the scheme of the Outstanding Young Scientist Award (T\"{U}BA-GEB\.{I}P). A. B. acknowledges a fellowship (44.291/2018-
0) of the PCI Program - MCTI and CNPq. E.D.V acknowledges support from the Royal Society through a Royal Society Dorothy Hodgkin Research Fellowship. R.C.N. thanks the financial support from the Conselho Nacional de Desenvolvimento Cient\'{i}fico e Tecnologico (CNPq, National Council for Scientific and Technological Development) under the project No. 304306/2022-3, and the Fundação de Amparo à pesquisa do Estado do RS (FAPERGS, Research Support Foundation of the State of RS) for partial financial support under the project No. 23/2551-0000848-3. This article is based upon work from the COST Action CA21136 ``Addressing observational tensions in cosmology with systematics and fundamental physics (CosmoVerse), supported by COST (European Cooperation in Science and Technology).
\end{acknowledgments}

\bibliographystyle{apsrev4-1}
\bibliography{main.bib}

\begin{thebibliography}{89}%
\makeatletter
\providecommand \@ifxundefined [1]{%
 \@ifx{#1\undefined}
}%
\providecommand \@ifnum [1]{%
 \ifnum #1\expandafter \@firstoftwo
 \else \expandafter \@secondoftwo
 \fi
}%
\providecommand \@ifx [1]{%
 \ifx #1\expandafter \@firstoftwo
 \else \expandafter \@secondoftwo
 \fi
}%
\providecommand \natexlab [1]{#1}%
\providecommand \enquote  [1]{``#1''}%
\providecommand \bibnamefont  [1]{#1}%
\providecommand \bibfnamefont [1]{#1}%
\providecommand \citenamefont [1]{#1}%
\providecommand \href@noop [0]{\@secondoftwo}%
\providecommand \href [0]{\begingroup \@sanitize@url \@href}%
\providecommand \@href[1]{\@@startlink{#1}\@@href}%
\providecommand \@@href[1]{\endgroup#1\@@endlink}%
\providecommand \@sanitize@url [0]{\catcode `\\12\catcode `\$12\catcode `\&12\catcode `\#12\catcode `\^12\catcode `\_12\catcode `\%12\relax}%
\providecommand \@@startlink[1]{}%
\providecommand \@@endlink[0]{}%
\providecommand \url  [0]{\begingroup\@sanitize@url \@url }%
\providecommand \@url [1]{\endgroup\@href {#1}{\urlprefix }}%
\providecommand \urlprefix  [0]{URL }%
\providecommand \Eprint [0]{\href }%
\providecommand \doibase [0]{http://dx.doi.org/}%
\providecommand \selectlanguage [0]{\@gobble}%
\providecommand \bibinfo  [0]{\@secondoftwo}%
\providecommand \bibfield  [0]{\@secondoftwo}%
\providecommand \translation [1]{[#1]}%
\providecommand \BibitemOpen [0]{}%
\providecommand \bibitemStop [0]{}%
\providecommand \bibitemNoStop [0]{.\EOS\space}%
\providecommand \EOS [0]{\spacefactor3000\relax}%
\providecommand \BibitemShut  [1]{\csname bibitem#1\endcsname}%
\let\auto@bib@innerbib\@empty
\bibitem [{\citenamefont {Aghanim}\ \emph {et~al.}(2020)\citenamefont {Aghanim} \emph {et~al.}}]{Planck:2018vyg}%
  \BibitemOpen
  \bibfield  {author} {\bibinfo {author} {\bibfnamefont {N.}~\bibnamefont {Aghanim}} \emph {et~al.} (\bibinfo {collaboration} {Planck}),\ }\href {\doibase 10.1051/0004-6361/201833910} {\bibfield  {journal} {\bibinfo  {journal} {Astron. Astrophys.}\ }\textbf {\bibinfo {volume} {641}},\ \bibinfo {pages} {A6} (\bibinfo {year} {2020})},\ \bibinfo {note} {[Erratum: Astron.Astrophys. 652, C4 (2021)]},\ \Eprint {http://arxiv.org/abs/1807.06209} {arXiv:1807.06209 [astro-ph.CO]} \BibitemShut {NoStop}%
\bibitem [{\citenamefont {Alam}\ \emph {et~al.}(2021)\citenamefont {Alam} \emph {et~al.}}]{eBOSS:2020yzd}%
  \BibitemOpen
  \bibfield  {author} {\bibinfo {author} {\bibfnamefont {S.}~\bibnamefont {Alam}} \emph {et~al.} (\bibinfo {collaboration} {eBOSS}),\ }\href {\doibase 10.1103/PhysRevD.103.083533} {\bibfield  {journal} {\bibinfo  {journal} {Phys. Rev. D}\ }\textbf {\bibinfo {volume} {103}},\ \bibinfo {pages} {083533} (\bibinfo {year} {2021})},\ \Eprint {http://arxiv.org/abs/2007.08991} {arXiv:2007.08991 [astro-ph.CO]} \BibitemShut {NoStop}%
\bibitem [{\citenamefont {Abbott}\ \emph {et~al.}(2022)\citenamefont {Abbott} \emph {et~al.}}]{DES:2021wwk}%
  \BibitemOpen
  \bibfield  {author} {\bibinfo {author} {\bibfnamefont {T.~M.~C.}\ \bibnamefont {Abbott}} \emph {et~al.} (\bibinfo {collaboration} {DES}),\ }\href {\doibase 10.1103/PhysRevD.105.023520} {\bibfield  {journal} {\bibinfo  {journal} {Phys. Rev. D}\ }\textbf {\bibinfo {volume} {105}},\ \bibinfo {pages} {023520} (\bibinfo {year} {2022})},\ \Eprint {http://arxiv.org/abs/2105.13549} {arXiv:2105.13549 [astro-ph.CO]} \BibitemShut {NoStop}%
\bibitem [{\citenamefont {Asgari}\ \emph {et~al.}(2021)\citenamefont {Asgari} \emph {et~al.}}]{KiDS:2020suj}%
  \BibitemOpen
  \bibfield  {author} {\bibinfo {author} {\bibfnamefont {M.}~\bibnamefont {Asgari}} \emph {et~al.} (\bibinfo {collaboration} {KiDS}),\ }\href {\doibase 10.1051/0004-6361/202039070} {\bibfield  {journal} {\bibinfo  {journal} {Astron. Astrophys.}\ }\textbf {\bibinfo {volume} {645}},\ \bibinfo {pages} {A104} (\bibinfo {year} {2021})},\ \Eprint {http://arxiv.org/abs/2007.15633} {arXiv:2007.15633 [astro-ph.CO]} \BibitemShut {NoStop}%
\bibitem [{\citenamefont {Freedman}\ and\ \citenamefont {Madore}(2023)}]{freedman2023progress}%
  \BibitemOpen
  \bibfield  {author} {\bibinfo {author} {\bibfnamefont {W.~L.}\ \bibnamefont {Freedman}}\ and\ \bibinfo {author} {\bibfnamefont {B.~F.}\ \bibnamefont {Madore}},\ }\href {\doibase 10.1088/1475-7516/2023/11/050} {\enquote {\bibinfo {title} {{Progress in direct measurements of the Hubble constant}},}\ } (\bibinfo {year} {2023}),\ \Eprint {http://arxiv.org/abs/2309.05618} {arXiv:2309.05618 [astro-ph.CO]} \BibitemShut {NoStop}%
\bibitem [{\citenamefont {Escamilla}\ \emph {et~al.}(2023{\natexlab{a}})\citenamefont {Escamilla}, \citenamefont {Giar\`e}, \citenamefont {Di~Valentino}, \citenamefont {Nunes},\ and\ \citenamefont {Vagnozzi}}]{Escamilla:2023oce}%
  \BibitemOpen
  \bibfield  {author} {\bibinfo {author} {\bibfnamefont {L.~A.}\ \bibnamefont {Escamilla}}, \bibinfo {author} {\bibfnamefont {W.}~\bibnamefont {Giar\`e}}, \bibinfo {author} {\bibfnamefont {E.}~\bibnamefont {Di~Valentino}}, \bibinfo {author} {\bibfnamefont {R.~C.}\ \bibnamefont {Nunes}}, \ and\ \bibinfo {author} {\bibfnamefont {S.}~\bibnamefont {Vagnozzi}},\ }\href@noop {} {\  (\bibinfo {year} {2023}{\natexlab{a}})},\ \Eprint {http://arxiv.org/abs/2307.14802} {arXiv:2307.14802 [astro-ph.CO]} \BibitemShut {NoStop}%
\bibitem [{\citenamefont {Di~Valentino}\ \emph {et~al.}(2021{\natexlab{a}})\citenamefont {Di~Valentino} \emph {et~al.}}]{DiValentino:2020vvd}%
  \BibitemOpen
  \bibfield  {author} {\bibinfo {author} {\bibfnamefont {E.}~\bibnamefont {Di~Valentino}} \emph {et~al.},\ }\href {\doibase 10.1016/j.astropartphys.2021.102604} {\bibfield  {journal} {\bibinfo  {journal} {Astropart. Phys.}\ }\textbf {\bibinfo {volume} {131}},\ \bibinfo {pages} {102604} (\bibinfo {year} {2021}{\natexlab{a}})},\ \Eprint {http://arxiv.org/abs/2008.11285} {arXiv:2008.11285 [astro-ph.CO]} \BibitemShut {NoStop}%
\bibitem [{\citenamefont {Verde}\ \emph {et~al.}(2019)\citenamefont {Verde}, \citenamefont {Treu},\ and\ \citenamefont {Riess}}]{Verde:2019ivm}%
  \BibitemOpen
  \bibfield  {author} {\bibinfo {author} {\bibfnamefont {L.}~\bibnamefont {Verde}}, \bibinfo {author} {\bibfnamefont {T.}~\bibnamefont {Treu}}, \ and\ \bibinfo {author} {\bibfnamefont {A.~G.}\ \bibnamefont {Riess}},\ }\href {\doibase 10.1038/s41550-019-0902-0} {\bibfield  {journal} {\bibinfo  {journal} {Nature Astron.}\ }\textbf {\bibinfo {volume} {3}},\ \bibinfo {pages} {891} (\bibinfo {year} {2019})},\ \Eprint {http://arxiv.org/abs/1907.10625} {arXiv:1907.10625 [astro-ph.CO]} \BibitemShut {NoStop}%
\bibitem [{\citenamefont {Knox}\ and\ \citenamefont {Millea}(2020)}]{Knox:2019rjx}%
  \BibitemOpen
  \bibfield  {author} {\bibinfo {author} {\bibfnamefont {L.}~\bibnamefont {Knox}}\ and\ \bibinfo {author} {\bibfnamefont {M.}~\bibnamefont {Millea}},\ }\href {\doibase 10.1103/PhysRevD.101.043533} {\bibfield  {journal} {\bibinfo  {journal} {Phys. Rev. D}\ }\textbf {\bibinfo {volume} {101}},\ \bibinfo {pages} {043533} (\bibinfo {year} {2020})},\ \Eprint {http://arxiv.org/abs/1908.03663} {arXiv:1908.03663 [astro-ph.CO]} \BibitemShut {NoStop}%
\bibitem [{\citenamefont {Riess}(2019)}]{Riess:2019qba}%
  \BibitemOpen
  \bibfield  {author} {\bibinfo {author} {\bibfnamefont {A.~G.}\ \bibnamefont {Riess}},\ }\href {\doibase 10.1038/s42254-019-0137-0} {\bibfield  {journal} {\bibinfo  {journal} {Nature Rev. Phys.}\ }\textbf {\bibinfo {volume} {2}},\ \bibinfo {pages} {10} (\bibinfo {year} {2019})},\ \Eprint {http://arxiv.org/abs/2001.03624} {arXiv:2001.03624 [astro-ph.CO]} \BibitemShut {NoStop}%
\bibitem [{\citenamefont {Di~Valentino}\ \emph {et~al.}(2021{\natexlab{b}})\citenamefont {Di~Valentino} \emph {et~al.}}]{DiValentino:2020zio}%
  \BibitemOpen
  \bibfield  {author} {\bibinfo {author} {\bibfnamefont {E.}~\bibnamefont {Di~Valentino}} \emph {et~al.},\ }\href {\doibase 10.1016/j.astropartphys.2021.102605} {\bibfield  {journal} {\bibinfo  {journal} {Astropart. Phys.}\ }\textbf {\bibinfo {volume} {131}},\ \bibinfo {pages} {102605} (\bibinfo {year} {2021}{\natexlab{b}})},\ \Eprint {http://arxiv.org/abs/2008.11284} {arXiv:2008.11284 [astro-ph.CO]} \BibitemShut {NoStop}%
\bibitem [{\citenamefont {Kamionkowski}\ and\ \citenamefont {Riess}(2023)}]{Kamionkowski:2022pkx}%
  \BibitemOpen
  \bibfield  {author} {\bibinfo {author} {\bibfnamefont {M.}~\bibnamefont {Kamionkowski}}\ and\ \bibinfo {author} {\bibfnamefont {A.~G.}\ \bibnamefont {Riess}},\ }\href@noop {} {\bibfield  {journal} {\bibinfo  {journal} {Ann. Rev. Nucl. Part. Sci.}\ }\textbf {\bibinfo {volume} {73}},\ \bibinfo {pages} {153} (\bibinfo {year} {2023})},\ \Eprint {http://arxiv.org/abs/2211.04492} {arXiv:2211.04492 [astro-ph.CO]} \BibitemShut {NoStop}%
\bibitem [{\citenamefont {Riess}\ \emph {et~al.}(2021)\citenamefont {Riess} \emph {et~al.}}]{Riess:2021jrx}%
  \BibitemOpen
  \bibfield  {author} {\bibinfo {author} {\bibfnamefont {A.~G.}\ \bibnamefont {Riess}} \emph {et~al.},\ }\href@noop {} {\  (\bibinfo {year} {2021})},\ \Eprint {http://arxiv.org/abs/2112.04510} {arXiv:2112.04510 [astro-ph.CO]} \BibitemShut {NoStop}%
\bibitem [{\citenamefont {Riess}\ \emph {et~al.}(2022)\citenamefont {Riess}, \citenamefont {Breuval}, \citenamefont {Yuan}, \citenamefont {Casertano}, \citenamefont {Macri}, \citenamefont {Bowers}, \citenamefont {Scolnic}, \citenamefont {Cantat-Gaudin}, \citenamefont {Anderson},\ and\ \citenamefont {Reyes}}]{Riess_2022}%
  \BibitemOpen
  \bibfield  {author} {\bibinfo {author} {\bibfnamefont {A.~G.}\ \bibnamefont {Riess}}, \bibinfo {author} {\bibfnamefont {L.}~\bibnamefont {Breuval}}, \bibinfo {author} {\bibfnamefont {W.}~\bibnamefont {Yuan}}, \bibinfo {author} {\bibfnamefont {S.}~\bibnamefont {Casertano}}, \bibinfo {author} {\bibfnamefont {L.~M.}\ \bibnamefont {Macri}}, \bibinfo {author} {\bibfnamefont {J.~B.}\ \bibnamefont {Bowers}}, \bibinfo {author} {\bibfnamefont {D.}~\bibnamefont {Scolnic}}, \bibinfo {author} {\bibfnamefont {T.}~\bibnamefont {Cantat-Gaudin}}, \bibinfo {author} {\bibfnamefont {R.~I.}\ \bibnamefont {Anderson}}, \ and\ \bibinfo {author} {\bibfnamefont {M.~C.}\ \bibnamefont {Reyes}},\ }\href {\doibase 10.3847/1538-4357/ac8f24} {\bibfield  {journal} {\bibinfo  {journal} {The Astrophysical Journal}\ }\textbf {\bibinfo {volume} {938}},\ \bibinfo {pages} {36} (\bibinfo {year} {2022})}\BibitemShut {NoStop}%
\bibitem [{\citenamefont {Murakami}\ \emph {et~al.}(2023)\citenamefont {Murakami}, \citenamefont {Riess}, \citenamefont {Stahl}, \citenamefont {Kenworthy}, \citenamefont {Pluck}, \citenamefont {Macoretta}, \citenamefont {Brout}, \citenamefont {Jones}, \citenamefont {Scolnic},\ and\ \citenamefont {Filippenko}}]{murakami2023leveraging}%
  \BibitemOpen
  \bibfield  {author} {\bibinfo {author} {\bibfnamefont {Y.~S.}\ \bibnamefont {Murakami}}, \bibinfo {author} {\bibfnamefont {A.~G.}\ \bibnamefont {Riess}}, \bibinfo {author} {\bibfnamefont {B.~E.}\ \bibnamefont {Stahl}}, \bibinfo {author} {\bibfnamefont {W.~D.}\ \bibnamefont {Kenworthy}}, \bibinfo {author} {\bibfnamefont {D.-M.~A.}\ \bibnamefont {Pluck}}, \bibinfo {author} {\bibfnamefont {A.}~\bibnamefont {Macoretta}}, \bibinfo {author} {\bibfnamefont {D.}~\bibnamefont {Brout}}, \bibinfo {author} {\bibfnamefont {D.~O.}\ \bibnamefont {Jones}}, \bibinfo {author} {\bibfnamefont {D.~M.}\ \bibnamefont {Scolnic}}, \ and\ \bibinfo {author} {\bibfnamefont {A.~V.}\ \bibnamefont {Filippenko}},\ }\href@noop {} {\enquote {\bibinfo {title} {Leveraging sn ia spectroscopic similarity to improve the measurement of $h_0$},}\ } (\bibinfo {year} {2023}),\ \Eprint {http://arxiv.org/abs/2306.00070} {arXiv:2306.00070 [astro-ph.CO]} \BibitemShut {NoStop}%
\bibitem [{\citenamefont {Di~Valentino}\ \emph {et~al.}(2021{\natexlab{c}})\citenamefont {Di~Valentino}, \citenamefont {Mena}, \citenamefont {Pan}, \citenamefont {Visinelli}, \citenamefont {Yang}, \citenamefont {Melchiorri}, \citenamefont {Mota}, \citenamefont {Riess},\ and\ \citenamefont {Silk}}]{DiValentino:2021izs}%
  \BibitemOpen
  \bibfield  {author} {\bibinfo {author} {\bibfnamefont {E.}~\bibnamefont {Di~Valentino}}, \bibinfo {author} {\bibfnamefont {O.}~\bibnamefont {Mena}}, \bibinfo {author} {\bibfnamefont {S.}~\bibnamefont {Pan}}, \bibinfo {author} {\bibfnamefont {L.}~\bibnamefont {Visinelli}}, \bibinfo {author} {\bibfnamefont {W.}~\bibnamefont {Yang}}, \bibinfo {author} {\bibfnamefont {A.}~\bibnamefont {Melchiorri}}, \bibinfo {author} {\bibfnamefont {D.~F.}\ \bibnamefont {Mota}}, \bibinfo {author} {\bibfnamefont {A.~G.}\ \bibnamefont {Riess}}, \ and\ \bibinfo {author} {\bibfnamefont {J.}~\bibnamefont {Silk}},\ }\href {\doibase 10.1088/1361-6382/ac086d} {\bibfield  {journal} {\bibinfo  {journal} {Class. Quant. Grav.}\ }\textbf {\bibinfo {volume} {38}},\ \bibinfo {pages} {153001} (\bibinfo {year} {2021}{\natexlab{c}})},\ \Eprint {http://arxiv.org/abs/2103.01183} {arXiv:2103.01183 [astro-ph.CO]} \BibitemShut {NoStop}%
\bibitem [{\citenamefont {Abdalla}\ \emph {et~al.}(2022)\citenamefont {Abdalla} \emph {et~al.}}]{Abdalla:2022yfr}%
  \BibitemOpen
  \bibfield  {author} {\bibinfo {author} {\bibfnamefont {E.}~\bibnamefont {Abdalla}} \emph {et~al.},\ }\href {\doibase 10.1016/j.jheap.2022.04.002} {\bibfield  {journal} {\bibinfo  {journal} {JHEAp}\ }\textbf {\bibinfo {volume} {34}},\ \bibinfo {pages} {49} (\bibinfo {year} {2022})},\ \Eprint {http://arxiv.org/abs/2203.06142} {arXiv:2203.06142 [astro-ph.CO]} \BibitemShut {NoStop}%
\bibitem [{\citenamefont {Perivolaropoulos}\ and\ \citenamefont {Skara}(2022)}]{Perivolaropoulos_2022}%
  \BibitemOpen
  \bibfield  {author} {\bibinfo {author} {\bibfnamefont {L.}~\bibnamefont {Perivolaropoulos}}\ and\ \bibinfo {author} {\bibfnamefont {F.}~\bibnamefont {Skara}},\ }\href {\doibase 10.1016/j.newar.2022.101659} {\bibfield  {journal} {\bibinfo  {journal} {New Astronomy Reviews}\ }\textbf {\bibinfo {volume} {95}},\ \bibinfo {pages} {101659} (\bibinfo {year} {2022})}\BibitemShut {NoStop}%
\bibitem [{\citenamefont {Akarsu}\ \emph {et~al.}(2024)\citenamefont {Akarsu}, \citenamefont {Colg{\'a}in}, \citenamefont {Sen},\ and\ \citenamefont {Sheikh-Jabbari}}]{akarsu2024lambda}%
  \BibitemOpen
  \bibfield  {author} {\bibinfo {author} {\bibfnamefont {{\"O}.}~\bibnamefont {Akarsu}}, \bibinfo {author} {\bibfnamefont {E.~{\'O}.}\ \bibnamefont {Colg{\'a}in}}, \bibinfo {author} {\bibfnamefont {A.~A.}\ \bibnamefont {Sen}}, \ and\ \bibinfo {author} {\bibfnamefont {M.}~\bibnamefont {Sheikh-Jabbari}},\ }\href {https://doi.org/10.48550/arXiv.2402.04767} {\bibfield  {journal} {\bibinfo  {journal} {arXiv preprint arXiv:2402.04767}\ } (\bibinfo {year} {2024})}\BibitemShut {NoStop}%
\bibitem [{\citenamefont {Cattoën}\ and\ \citenamefont {Visser}(2008)}]{Catto_n_2008}%
  \BibitemOpen
  \bibfield  {author} {\bibinfo {author} {\bibfnamefont {C.}~\bibnamefont {Cattoën}}\ and\ \bibinfo {author} {\bibfnamefont {M.}~\bibnamefont {Visser}},\ }\href {\doibase 10.1103/physrevd.78.063501} {\bibfield  {journal} {\bibinfo  {journal} {Physical Review D}\ }\textbf {\bibinfo {volume} {78}} (\bibinfo {year} {2008}),\ 10.1103/physrevd.78.063501}\BibitemShut {NoStop}%
\bibitem [{\citenamefont {Capozziello}\ \emph {et~al.}(2018)\citenamefont {Capozziello}, \citenamefont {D{\textquotesingle}Agostino},\ and\ \citenamefont {Luongo}}]{Capozziello_2018}%
  \BibitemOpen
  \bibfield  {author} {\bibinfo {author} {\bibfnamefont {S.}~\bibnamefont {Capozziello}}, \bibinfo {author} {\bibfnamefont {R.}~\bibnamefont {D{\textquotesingle}Agostino}}, \ and\ \bibinfo {author} {\bibfnamefont {O.}~\bibnamefont {Luongo}},\ }\href {\doibase 10.1093/mnras/sty422} {\bibfield  {journal} {\bibinfo  {journal} {Monthly Notices of the Royal Astronomical Society}\ }\textbf {\bibinfo {volume} {476}},\ \bibinfo {pages} {3924} (\bibinfo {year} {2018})}\BibitemShut {NoStop}%
\bibitem [{\citenamefont {D'Agostino}\ and\ \citenamefont {Nunes}(2023)}]{D_Agostino_2023}%
  \BibitemOpen
  \bibfield  {author} {\bibinfo {author} {\bibfnamefont {R.}~\bibnamefont {D'Agostino}}\ and\ \bibinfo {author} {\bibfnamefont {R.~C.}\ \bibnamefont {Nunes}},\ }\href {\doibase 10.1103/physrevd.108.023523} {\bibfield  {journal} {\bibinfo  {journal} {Physical Review D}\ }\textbf {\bibinfo {volume} {108}} (\bibinfo {year} {2023}),\ 10.1103/physrevd.108.023523}\BibitemShut {NoStop}%
\bibitem [{\citenamefont {Rasmussen}\ \emph {et~al.}(2006)\citenamefont {Rasmussen}, \citenamefont {Williams} \emph {et~al.}}]{rasmussen2006gaussian}%
  \BibitemOpen
  \bibfield  {author} {\bibinfo {author} {\bibfnamefont {C.~E.}\ \bibnamefont {Rasmussen}}, \bibinfo {author} {\bibfnamefont {C.~K.}\ \bibnamefont {Williams}},  \emph {et~al.},\ }\href@noop {} {\emph {\bibinfo {title} {Gaussian processes for machine learning}}},\ Vol.~\bibinfo {volume} {1}\ (\bibinfo  {publisher} {Springer},\ \bibinfo {year} {2006})\BibitemShut {NoStop}%
\bibitem [{\citenamefont {Williams}(2005)}]{williams2005gaussian}%
  \BibitemOpen
  \bibfield  {author} {\bibinfo {author} {\bibfnamefont {C.~K.}\ \bibnamefont {Williams}},\ }\href@noop {} {\emph {\bibinfo {title} {Gaussian processes for machine learning}}}\ (\bibinfo  {publisher} {MIT press},\ \bibinfo {year} {2005})\BibitemShut {NoStop}%
\bibitem [{\citenamefont {{Holsclaw}}\ \emph {et~al.}(2010{\natexlab{a}})\citenamefont {{Holsclaw}}, \citenamefont {{Alam}}, \citenamefont {{Sans{\'o}}}, \citenamefont {{Lee}}, \citenamefont {{Heitmann}}, \citenamefont {{Habib}},\ and\ \citenamefont {{Higdon}}}]{2010PhRvD..82j3502H}%
  \BibitemOpen
  \bibfield  {author} {\bibinfo {author} {\bibfnamefont {T.}~\bibnamefont {{Holsclaw}}}, \bibinfo {author} {\bibfnamefont {U.}~\bibnamefont {{Alam}}}, \bibinfo {author} {\bibfnamefont {B.}~\bibnamefont {{Sans{\'o}}}}, \bibinfo {author} {\bibfnamefont {H.}~\bibnamefont {{Lee}}}, \bibinfo {author} {\bibfnamefont {K.}~\bibnamefont {{Heitmann}}}, \bibinfo {author} {\bibfnamefont {S.}~\bibnamefont {{Habib}}}, \ and\ \bibinfo {author} {\bibfnamefont {D.}~\bibnamefont {{Higdon}}},\ }\href {\doibase 10.1103/PhysRevD.82.103502} {\bibfield  {journal} {\bibinfo  {journal} {\prd}\ }\textbf {\bibinfo {volume} {82}},\ \bibinfo {eid} {103502} (\bibinfo {year} {2010}{\natexlab{a}})},\ \Eprint {http://arxiv.org/abs/1009.5443} {arXiv:1009.5443 [astro-ph.CO]} \BibitemShut {NoStop}%
\bibitem [{\citenamefont {{Holsclaw}}\ \emph {et~al.}(2010{\natexlab{b}})\citenamefont {{Holsclaw}}, \citenamefont {{Alam}}, \citenamefont {{Sans{\'o}}}, \citenamefont {{Lee}}, \citenamefont {{Heitmann}}, \citenamefont {{Habib}},\ and\ \citenamefont {{Higdon}}}]{2010PhRvL.105x1302H}%
  \BibitemOpen
  \bibfield  {author} {\bibinfo {author} {\bibfnamefont {T.}~\bibnamefont {{Holsclaw}}}, \bibinfo {author} {\bibfnamefont {U.}~\bibnamefont {{Alam}}}, \bibinfo {author} {\bibfnamefont {B.}~\bibnamefont {{Sans{\'o}}}}, \bibinfo {author} {\bibfnamefont {H.}~\bibnamefont {{Lee}}}, \bibinfo {author} {\bibfnamefont {K.}~\bibnamefont {{Heitmann}}}, \bibinfo {author} {\bibfnamefont {S.}~\bibnamefont {{Habib}}}, \ and\ \bibinfo {author} {\bibfnamefont {D.}~\bibnamefont {{Higdon}}},\ }\href {\doibase 10.1103/PhysRevLett.105.241302} {\bibfield  {journal} {\bibinfo  {journal} {\prl}\ }\textbf {\bibinfo {volume} {105}},\ \bibinfo {eid} {241302} (\bibinfo {year} {2010}{\natexlab{b}})},\ \Eprint {http://arxiv.org/abs/1011.3079} {arXiv:1011.3079 [astro-ph.CO]} \BibitemShut {NoStop}%
\bibitem [{\citenamefont {Seikel}\ \emph {et~al.}(2012)\citenamefont {Seikel}, \citenamefont {Clarkson},\ and\ \citenamefont {Smith}}]{seikel2012reconstruction}%
  \BibitemOpen
  \bibfield  {author} {\bibinfo {author} {\bibfnamefont {M.}~\bibnamefont {Seikel}}, \bibinfo {author} {\bibfnamefont {C.}~\bibnamefont {Clarkson}}, \ and\ \bibinfo {author} {\bibfnamefont {M.}~\bibnamefont {Smith}},\ }\href {\doibase 10.1088/1475-7516/2012/06/036} {\bibfield  {journal} {\bibinfo  {journal} {Journal of Cosmology and Astroparticle Physics}\ }\textbf {\bibinfo {volume} {2012}},\ \bibinfo {pages} {036} (\bibinfo {year} {2012})}\BibitemShut {NoStop}%
\bibitem [{\citenamefont {Shafieloo}\ \emph {et~al.}(2012)\citenamefont {Shafieloo}, \citenamefont {Kim},\ and\ \citenamefont {Linder}}]{Shafieloo_2012}%
  \BibitemOpen
  \bibfield  {author} {\bibinfo {author} {\bibfnamefont {A.}~\bibnamefont {Shafieloo}}, \bibinfo {author} {\bibfnamefont {A.~G.}\ \bibnamefont {Kim}}, \ and\ \bibinfo {author} {\bibfnamefont {E.~V.}\ \bibnamefont {Linder}},\ }\href {\doibase 10.1103/physrevd.85.123530} {\bibfield  {journal} {\bibinfo  {journal} {Physical Review D}\ }\textbf {\bibinfo {volume} {85}} (\bibinfo {year} {2012}),\ 10.1103/physrevd.85.123530}\BibitemShut {NoStop}%
\bibitem [{\citenamefont {Sahni}\ \emph {et~al.}(2014)\citenamefont {Sahni}, \citenamefont {Shafieloo},\ and\ \citenamefont {Starobinsky}}]{Sahni_2014}%
  \BibitemOpen
  \bibfield  {author} {\bibinfo {author} {\bibfnamefont {V.}~\bibnamefont {Sahni}}, \bibinfo {author} {\bibfnamefont {A.}~\bibnamefont {Shafieloo}}, \ and\ \bibinfo {author} {\bibfnamefont {A.~A.}\ \bibnamefont {Starobinsky}},\ }\href {\doibase 10.1088/2041-8205/793/2/l40} {\bibfield  {journal} {\bibinfo  {journal} {The Astrophysical Journal}\ }\textbf {\bibinfo {volume} {793}},\ \bibinfo {pages} {L40} (\bibinfo {year} {2014})}\BibitemShut {NoStop}%
\bibitem [{\citenamefont {Belgacem}\ \emph {et~al.}(2020)\citenamefont {Belgacem}, \citenamefont {Foffa}, \citenamefont {Maggiore},\ and\ \citenamefont {Yang}}]{Belgacem_2020}%
  \BibitemOpen
  \bibfield  {author} {\bibinfo {author} {\bibfnamefont {E.}~\bibnamefont {Belgacem}}, \bibinfo {author} {\bibfnamefont {S.}~\bibnamefont {Foffa}}, \bibinfo {author} {\bibfnamefont {M.}~\bibnamefont {Maggiore}}, \ and\ \bibinfo {author} {\bibfnamefont {T.}~\bibnamefont {Yang}},\ }\href {\doibase 10.1103/physrevd.101.063505} {\bibfield  {journal} {\bibinfo  {journal} {Physical Review D}\ }\textbf {\bibinfo {volume} {101}} (\bibinfo {year} {2020}),\ 10.1103/physrevd.101.063505}\BibitemShut {NoStop}%
\bibitem [{\citenamefont {Bonilla}\ \emph {et~al.}(2021)\citenamefont {Bonilla}, \citenamefont {Kumar},\ and\ \citenamefont {Nunes}}]{bonilla2021measurements}%
  \BibitemOpen
  \bibfield  {author} {\bibinfo {author} {\bibfnamefont {A.}~\bibnamefont {Bonilla}}, \bibinfo {author} {\bibfnamefont {S.}~\bibnamefont {Kumar}}, \ and\ \bibinfo {author} {\bibfnamefont {R.~C.}\ \bibnamefont {Nunes}},\ }\href {https://doi.org/10.1140/epjc/s10052-021-08925-z} {\bibfield  {journal} {\bibinfo  {journal} {The European Physical Journal C}\ }\textbf {\bibinfo {volume} {81}},\ \bibinfo {pages} {1} (\bibinfo {year} {2021})}\BibitemShut {NoStop}%
\bibitem [{\citenamefont {Wang}\ and\ \citenamefont {Meng}(2017)}]{Wang_2017}%
  \BibitemOpen
  \bibfield  {author} {\bibinfo {author} {\bibfnamefont {D.}~\bibnamefont {Wang}}\ and\ \bibinfo {author} {\bibfnamefont {X.-H.}\ \bibnamefont {Meng}},\ }\href {\doibase 10.1103/physrevd.95.023508} {\bibfield  {journal} {\bibinfo  {journal} {Physical Review D}\ }\textbf {\bibinfo {volume} {95}} (\bibinfo {year} {2017}),\ 10.1103/physrevd.95.023508}\BibitemShut {NoStop}%
\bibitem [{\citenamefont {Bengaly}(2020)}]{Bengaly_2020}%
  \BibitemOpen
  \bibfield  {author} {\bibinfo {author} {\bibfnamefont {C.~A.~P.}\ \bibnamefont {Bengaly}},\ }\href {\doibase 10.1093/mnrasl/slaa040} {\bibfield  {journal} {\bibinfo  {journal} {Monthly Notices of the Royal Astronomical Society: Letters}\ }\textbf {\bibinfo {volume} {499}},\ \bibinfo {pages} {L6} (\bibinfo {year} {2020})}\BibitemShut {NoStop}%
\bibitem [{\citenamefont {G{\'{o}}mez-Valent}\ and\ \citenamefont {Amendola}(2018)}]{G_mez_Valent_2018}%
  \BibitemOpen
  \bibfield  {author} {\bibinfo {author} {\bibfnamefont {A.}~\bibnamefont {G{\'{o}}mez-Valent}}\ and\ \bibinfo {author} {\bibfnamefont {L.}~\bibnamefont {Amendola}},\ }\href {\doibase 10.1088/1475-7516/2018/04/051} {\bibfield  {journal} {\bibinfo  {journal} {Journal of Cosmology and Astroparticle Physics}\ }\textbf {\bibinfo {volume} {2018}},\ \bibinfo {pages} {051} (\bibinfo {year} {2018})}\BibitemShut {NoStop}%
\bibitem [{\citenamefont {Bonilla}\ \emph {et~al.}(2022)\citenamefont {Bonilla}, \citenamefont {Kumar}, \citenamefont {Nunes},\ and\ \citenamefont {Pan}}]{Bonilla_2022}%
  \BibitemOpen
  \bibfield  {author} {\bibinfo {author} {\bibfnamefont {A.}~\bibnamefont {Bonilla}}, \bibinfo {author} {\bibfnamefont {S.}~\bibnamefont {Kumar}}, \bibinfo {author} {\bibfnamefont {R.~C.}\ \bibnamefont {Nunes}}, \ and\ \bibinfo {author} {\bibfnamefont {S.}~\bibnamefont {Pan}},\ }\href {\doibase 10.1093/mnras/stac687} {\bibfield  {journal} {\bibinfo  {journal} {Monthly Notices of the Royal Astronomical Society}\ }\textbf {\bibinfo {volume} {512}},\ \bibinfo {pages} {4231} (\bibinfo {year} {2022})}\BibitemShut {NoStop}%
\bibitem [{\citenamefont {Dinda}\ and\ \citenamefont {Banerjee}(2023)}]{Dinda_2023}%
  \BibitemOpen
  \bibfield  {author} {\bibinfo {author} {\bibfnamefont {B.~R.}\ \bibnamefont {Dinda}}\ and\ \bibinfo {author} {\bibfnamefont {N.}~\bibnamefont {Banerjee}},\ }\href {\doibase 10.1103/physrevd.107.063513} {\bibfield  {journal} {\bibinfo  {journal} {Physical Review D}\ }\textbf {\bibinfo {volume} {107}} (\bibinfo {year} {2023}),\ 10.1103/physrevd.107.063513}\BibitemShut {NoStop}%
\bibitem [{\citenamefont {Avila}\ \emph {et~al.}(2022)\citenamefont {Avila}, \citenamefont {Bernui}, \citenamefont {Bonilla},\ and\ \citenamefont {Nunes}}]{Avila_2022}%
  \BibitemOpen
  \bibfield  {author} {\bibinfo {author} {\bibfnamefont {F.}~\bibnamefont {Avila}}, \bibinfo {author} {\bibfnamefont {A.}~\bibnamefont {Bernui}}, \bibinfo {author} {\bibfnamefont {A.}~\bibnamefont {Bonilla}}, \ and\ \bibinfo {author} {\bibfnamefont {R.~C.}\ \bibnamefont {Nunes}},\ }\href {\doibase 10.1140/epjc/s10052-022-10561-0} {\bibfield  {journal} {\bibinfo  {journal} {The European Physical Journal C}\ }\textbf {\bibinfo {volume} {82}} (\bibinfo {year} {2022}),\ 10.1140/epjc/s10052-022-10561-0}\BibitemShut {NoStop}%
\bibitem [{\citenamefont {Renzi}\ \emph {et~al.}(2022)\citenamefont {Renzi}, \citenamefont {Hogg},\ and\ \citenamefont {Giar{\`{e} }}}]{Renzi_2022}%
  \BibitemOpen
  \bibfield  {author} {\bibinfo {author} {\bibfnamefont {F.}~\bibnamefont {Renzi}}, \bibinfo {author} {\bibfnamefont {N.~B.}\ \bibnamefont {Hogg}}, \ and\ \bibinfo {author} {\bibfnamefont {W.}~\bibnamefont {Giar{\`{e} }}},\ }\href {\doibase 10.1093/mnras/stac1030} {\bibfield  {journal} {\bibinfo  {journal} {Monthly Notices of the Royal Astronomical Society}\ }\textbf {\bibinfo {volume} {513}},\ \bibinfo {pages} {4004} (\bibinfo {year} {2022})}\BibitemShut {NoStop}%
\bibitem [{\citenamefont {Rodrigues}\ and\ \citenamefont {Bengaly}(2022)}]{Rodrigues_2022}%
  \BibitemOpen
  \bibfield  {author} {\bibinfo {author} {\bibfnamefont {G.}~\bibnamefont {Rodrigues}}\ and\ \bibinfo {author} {\bibfnamefont {C.}~\bibnamefont {Bengaly}},\ }\href {\doibase 10.1088/1475-7516/2022/07/029} {\bibfield  {journal} {\bibinfo  {journal} {Journal of Cosmology and Astroparticle Physics}\ }\textbf {\bibinfo {volume} {2022}},\ \bibinfo {pages} {029} (\bibinfo {year} {2022})}\BibitemShut {NoStop}%
\bibitem [{\citenamefont {Sun}\ \emph {et~al.}(2021)\citenamefont {Sun}, \citenamefont {Jiao},\ and\ \citenamefont {Zhang}}]{Sun_2021}%
  \BibitemOpen
  \bibfield  {author} {\bibinfo {author} {\bibfnamefont {W.}~\bibnamefont {Sun}}, \bibinfo {author} {\bibfnamefont {K.}~\bibnamefont {Jiao}}, \ and\ \bibinfo {author} {\bibfnamefont {T.-J.}\ \bibnamefont {Zhang}},\ }\href {\doibase 10.3847/1538-4357/ac05b8} {\bibfield  {journal} {\bibinfo  {journal} {The Astrophysical Journal}\ }\textbf {\bibinfo {volume} {915}},\ \bibinfo {pages} {123} (\bibinfo {year} {2021})}\BibitemShut {NoStop}%
\bibitem [{\citenamefont {Keeley}\ \emph {et~al.}(2021)\citenamefont {Keeley}, \citenamefont {Shafieloo}, \citenamefont {Zhao}, \citenamefont {Vazquez},\ and\ \citenamefont {Koo}}]{Keeley:2020aym}%
  \BibitemOpen
  \bibfield  {author} {\bibinfo {author} {\bibfnamefont {R.~E.}\ \bibnamefont {Keeley}}, \bibinfo {author} {\bibfnamefont {A.}~\bibnamefont {Shafieloo}}, \bibinfo {author} {\bibfnamefont {G.-B.}\ \bibnamefont {Zhao}}, \bibinfo {author} {\bibfnamefont {J.~A.}\ \bibnamefont {Vazquez}}, \ and\ \bibinfo {author} {\bibfnamefont {H.}~\bibnamefont {Koo}},\ }\href {\doibase 10.3847/1538-3881/abdd2a} {\bibfield  {journal} {\bibinfo  {journal} {Astron. J.}\ }\textbf {\bibinfo {volume} {161}},\ \bibinfo {pages} {151} (\bibinfo {year} {2021})},\ \Eprint {http://arxiv.org/abs/2010.03234} {arXiv:2010.03234 [astro-ph.CO]} \BibitemShut {NoStop}%
\bibitem [{\citenamefont {Escamilla}\ \emph {et~al.}(2023{\natexlab{b}})\citenamefont {Escamilla}, \citenamefont {Akarsu}, \citenamefont {Di~Valentino},\ and\ \citenamefont {Vazquez}}]{Escamilla:2023shf}%
  \BibitemOpen
  \bibfield  {author} {\bibinfo {author} {\bibfnamefont {L.~A.}\ \bibnamefont {Escamilla}}, \bibinfo {author} {\bibfnamefont {O.}~\bibnamefont {Akarsu}}, \bibinfo {author} {\bibfnamefont {E.}~\bibnamefont {Di~Valentino}}, \ and\ \bibinfo {author} {\bibfnamefont {J.~A.}\ \bibnamefont {Vazquez}},\ }\href {\doibase 10.1088/1475-7516/2023/11/051} {\bibfield  {journal} {\bibinfo  {journal} {JCAP}\ }\textbf {\bibinfo {volume} {11}},\ \bibinfo {pages} {051} (\bibinfo {year} {2023}{\natexlab{b}})},\ \Eprint {http://arxiv.org/abs/2305.16290} {arXiv:2305.16290 [astro-ph.CO]} \BibitemShut {NoStop}%
\bibitem [{\citenamefont {Wang}\ \emph {et~al.}(2018)\citenamefont {Wang}, \citenamefont {Pogosian}, \citenamefont {Zhao},\ and\ \citenamefont {Zucca}}]{Wang:2018fng}%
  \BibitemOpen
  \bibfield  {author} {\bibinfo {author} {\bibfnamefont {Y.}~\bibnamefont {Wang}}, \bibinfo {author} {\bibfnamefont {L.}~\bibnamefont {Pogosian}}, \bibinfo {author} {\bibfnamefont {G.-B.}\ \bibnamefont {Zhao}}, \ and\ \bibinfo {author} {\bibfnamefont {A.}~\bibnamefont {Zucca}},\ }\href {\doibase 10.3847/2041-8213/aaf238} {\bibfield  {journal} {\bibinfo  {journal} {Astrophys. J. Lett.}\ }\textbf {\bibinfo {volume} {869}},\ \bibinfo {pages} {L8} (\bibinfo {year} {2018})},\ \Eprint {http://arxiv.org/abs/1807.03772} {arXiv:1807.03772 [astro-ph.CO]} \BibitemShut {NoStop}%
\bibitem [{\citenamefont {Escamilla}\ and\ \citenamefont {Vazquez}(2023)}]{Escamilla:2021uoj}%
  \BibitemOpen
  \bibfield  {author} {\bibinfo {author} {\bibfnamefont {L.~A.}\ \bibnamefont {Escamilla}}\ and\ \bibinfo {author} {\bibfnamefont {J.~A.}\ \bibnamefont {Vazquez}},\ }\href {\doibase 10.1140/epjc/s10052-023-11404-2} {\bibfield  {journal} {\bibinfo  {journal} {Eur. Phys. J. C}\ }\textbf {\bibinfo {volume} {83}},\ \bibinfo {pages} {251} (\bibinfo {year} {2023})},\ \Eprint {http://arxiv.org/abs/2111.10457} {arXiv:2111.10457 [astro-ph.CO]} \BibitemShut {NoStop}%
\bibitem [{\citenamefont {Dutta}\ \emph {et~al.}(2020)\citenamefont {Dutta}, \citenamefont {Ruchika}, \citenamefont {Roy}, \citenamefont {Sen},\ and\ \citenamefont {Sheikh-Jabbari}}]{Dutta:2018vmq}%
  \BibitemOpen
  \bibfield  {author} {\bibinfo {author} {\bibfnamefont {K.}~\bibnamefont {Dutta}}, \bibinfo {author} {\bibnamefont {Ruchika}}, \bibinfo {author} {\bibfnamefont {A.}~\bibnamefont {Roy}}, \bibinfo {author} {\bibfnamefont {A.~A.}\ \bibnamefont {Sen}}, \ and\ \bibinfo {author} {\bibfnamefont {M.~M.}\ \bibnamefont {Sheikh-Jabbari}},\ }\href {\doibase 10.1007/s10714-020-2665-4} {\bibfield  {journal} {\bibinfo  {journal} {Gen. Rel. Grav.}\ }\textbf {\bibinfo {volume} {52}},\ \bibinfo {pages} {15} (\bibinfo {year} {2020})},\ \Eprint {http://arxiv.org/abs/1808.06623} {arXiv:1808.06623 [astro-ph.CO]} \BibitemShut {NoStop}%
\bibitem [{\citenamefont {Visinelli}\ \emph {et~al.}(2019)\citenamefont {Visinelli}, \citenamefont {Vagnozzi},\ and\ \citenamefont {Danielsson}}]{Visinelli:2019qqu}%
  \BibitemOpen
  \bibfield  {author} {\bibinfo {author} {\bibfnamefont {L.}~\bibnamefont {Visinelli}}, \bibinfo {author} {\bibfnamefont {S.}~\bibnamefont {Vagnozzi}}, \ and\ \bibinfo {author} {\bibfnamefont {U.}~\bibnamefont {Danielsson}},\ }\href {\doibase 10.3390/sym11081035} {\bibfield  {journal} {\bibinfo  {journal} {Symmetry}\ }\textbf {\bibinfo {volume} {11}},\ \bibinfo {pages} {1035} (\bibinfo {year} {2019})},\ \Eprint {http://arxiv.org/abs/1907.07953} {arXiv:1907.07953 [astro-ph.CO]} \BibitemShut {NoStop}%
\bibitem [{\citenamefont {Akarsu}\ \emph {et~al.}(2020)\citenamefont {Akarsu}, \citenamefont {Barrow}, \citenamefont {Escamilla},\ and\ \citenamefont {Vazquez}}]{Akarsu:2019hmw}%
  \BibitemOpen
  \bibfield  {author} {\bibinfo {author} {\bibfnamefont {O.}~\bibnamefont {Akarsu}}, \bibinfo {author} {\bibfnamefont {J.~D.}\ \bibnamefont {Barrow}}, \bibinfo {author} {\bibfnamefont {L.~A.}\ \bibnamefont {Escamilla}}, \ and\ \bibinfo {author} {\bibfnamefont {J.~A.}\ \bibnamefont {Vazquez}},\ }\href {\doibase 10.1103/PhysRevD.101.063528} {\bibfield  {journal} {\bibinfo  {journal} {Phys. Rev. D}\ }\textbf {\bibinfo {volume} {101}},\ \bibinfo {pages} {063528} (\bibinfo {year} {2020})},\ \Eprint {http://arxiv.org/abs/1912.08751} {arXiv:1912.08751 [astro-ph.CO]} \BibitemShut {NoStop}%
\bibitem [{\citenamefont {Akarsu}\ \emph {et~al.}(2021)\citenamefont {Akarsu}, \citenamefont {Kumar}, \citenamefont {\"Oz\"ulker},\ and\ \citenamefont {Vazquez}}]{Akarsu:2021fol}%
  \BibitemOpen
  \bibfield  {author} {\bibinfo {author} {\bibfnamefont {O.}~\bibnamefont {Akarsu}}, \bibinfo {author} {\bibfnamefont {S.}~\bibnamefont {Kumar}}, \bibinfo {author} {\bibfnamefont {E.}~\bibnamefont {\"Oz\"ulker}}, \ and\ \bibinfo {author} {\bibfnamefont {J.~A.}\ \bibnamefont {Vazquez}},\ }\href {\doibase 10.1103/PhysRevD.104.123512} {\bibfield  {journal} {\bibinfo  {journal} {Phys. Rev. D}\ }\textbf {\bibinfo {volume} {104}},\ \bibinfo {pages} {123512} (\bibinfo {year} {2021})},\ \Eprint {http://arxiv.org/abs/2108.09239} {arXiv:2108.09239 [astro-ph.CO]} \BibitemShut {NoStop}%
\bibitem [{\citenamefont {Akarsu}\ \emph {et~al.}(2023{\natexlab{a}})\citenamefont {Akarsu}, \citenamefont {Kumar}, \citenamefont {\"Oz\"ulker}, \citenamefont {Vazquez},\ and\ \citenamefont {Yadav}}]{Akarsu:2022typ}%
  \BibitemOpen
  \bibfield  {author} {\bibinfo {author} {\bibfnamefont {O.}~\bibnamefont {Akarsu}}, \bibinfo {author} {\bibfnamefont {S.}~\bibnamefont {Kumar}}, \bibinfo {author} {\bibfnamefont {E.}~\bibnamefont {\"Oz\"ulker}}, \bibinfo {author} {\bibfnamefont {J.~A.}\ \bibnamefont {Vazquez}}, \ and\ \bibinfo {author} {\bibfnamefont {A.}~\bibnamefont {Yadav}},\ }\href {\doibase 10.1103/PhysRevD.108.023513} {\bibfield  {journal} {\bibinfo  {journal} {Phys. Rev. D}\ }\textbf {\bibinfo {volume} {108}},\ \bibinfo {pages} {023513} (\bibinfo {year} {2023}{\natexlab{a}})},\ \Eprint {http://arxiv.org/abs/2211.05742} {arXiv:2211.05742 [astro-ph.CO]} \BibitemShut {NoStop}%
\bibitem [{\citenamefont {Akarsu}\ \emph {et~al.}(2023{\natexlab{b}})\citenamefont {Akarsu}, \citenamefont {Di~Valentino}, \citenamefont {Kumar}, \citenamefont {Nunes}, \citenamefont {Vazquez},\ and\ \citenamefont {Yadav}}]{Akarsu:2023mfb}%
  \BibitemOpen
  \bibfield  {author} {\bibinfo {author} {\bibfnamefont {O.}~\bibnamefont {Akarsu}}, \bibinfo {author} {\bibfnamefont {E.}~\bibnamefont {Di~Valentino}}, \bibinfo {author} {\bibfnamefont {S.}~\bibnamefont {Kumar}}, \bibinfo {author} {\bibfnamefont {R.~C.}\ \bibnamefont {Nunes}}, \bibinfo {author} {\bibfnamefont {J.~A.}\ \bibnamefont {Vazquez}}, \ and\ \bibinfo {author} {\bibfnamefont {A.}~\bibnamefont {Yadav}},\ }\href@noop {} {\  (\bibinfo {year} {2023}{\natexlab{b}})},\ \Eprint {http://arxiv.org/abs/2307.10899} {arXiv:2307.10899 [astro-ph.CO]} \BibitemShut {NoStop}%
\bibitem [{\citenamefont {Sen}\ \emph {et~al.}(2022)\citenamefont {Sen}, \citenamefont {Adil},\ and\ \citenamefont {Sen}}]{Sen:2021wld}%
  \BibitemOpen
  \bibfield  {author} {\bibinfo {author} {\bibfnamefont {A.~A.}\ \bibnamefont {Sen}}, \bibinfo {author} {\bibfnamefont {S.~A.}\ \bibnamefont {Adil}}, \ and\ \bibinfo {author} {\bibfnamefont {S.}~\bibnamefont {Sen}},\ }\href {\doibase 10.1093/mnras/stac2796} {\bibfield  {journal} {\bibinfo  {journal} {Mon. Not. Roy. Astron. Soc.}\ }\textbf {\bibinfo {volume} {518}},\ \bibinfo {pages} {1098} (\bibinfo {year} {2022})},\ \Eprint {http://arxiv.org/abs/2112.10641} {arXiv:2112.10641 [astro-ph.CO]} \BibitemShut {NoStop}%
\bibitem [{\citenamefont {Adil}\ \emph {et~al.}(2024)\citenamefont {Adil}, \citenamefont {Akarsu}, \citenamefont {Di~Valentino}, \citenamefont {Nunes}, \citenamefont {\"Oz\"ulker}, \citenamefont {Sen},\ and\ \citenamefont {Specogna}}]{Adil:2023exv}%
  \BibitemOpen
  \bibfield  {author} {\bibinfo {author} {\bibfnamefont {S.~A.}\ \bibnamefont {Adil}}, \bibinfo {author} {\bibfnamefont {O.}~\bibnamefont {Akarsu}}, \bibinfo {author} {\bibfnamefont {E.}~\bibnamefont {Di~Valentino}}, \bibinfo {author} {\bibfnamefont {R.~C.}\ \bibnamefont {Nunes}}, \bibinfo {author} {\bibfnamefont {E.}~\bibnamefont {\"Oz\"ulker}}, \bibinfo {author} {\bibfnamefont {A.~A.}\ \bibnamefont {Sen}}, \ and\ \bibinfo {author} {\bibfnamefont {E.}~\bibnamefont {Specogna}},\ }\href {\doibase 10.1103/PhysRevD.109.023527} {\bibfield  {journal} {\bibinfo  {journal} {Phys. Rev. D}\ }\textbf {\bibinfo {volume} {109}},\ \bibinfo {pages} {023527} (\bibinfo {year} {2024})},\ \Eprint {http://arxiv.org/abs/2306.08046} {arXiv:2306.08046 [astro-ph.CO]} \BibitemShut {NoStop}%
\bibitem [{\citenamefont {\'O~Colg\'ain}\ \emph {et~al.}(2021)\citenamefont {\'O~Colg\'ain}, \citenamefont {Sheikh-Jabbari},\ and\ \citenamefont {Yin}}]{Colgain:2021pmf}%
  \BibitemOpen
  \bibfield  {author} {\bibinfo {author} {\bibfnamefont {E.}~\bibnamefont {\'O~Colg\'ain}}, \bibinfo {author} {\bibfnamefont {M.~M.}\ \bibnamefont {Sheikh-Jabbari}}, \ and\ \bibinfo {author} {\bibfnamefont {L.}~\bibnamefont {Yin}},\ }\href {\doibase 10.1103/PhysRevD.104.023510} {\bibfield  {journal} {\bibinfo  {journal} {Phys. Rev. D}\ }\textbf {\bibinfo {volume} {104}},\ \bibinfo {pages} {023510} (\bibinfo {year} {2021})}\BibitemShut {NoStop}%
\bibitem [{\citenamefont {Raveri}\ \emph {et~al.}(2023)\citenamefont {Raveri} \emph {et~al.}}]{Raveri_2023}%
  \BibitemOpen
  \bibfield  {author} {\bibinfo {author} {\bibfnamefont {M.}~\bibnamefont {Raveri}} \emph {et~al.},\ }\href {https://doi.org/10.1088/1475-7516/2023/02/061} {\bibfield  {journal} {\bibinfo  {journal} {Journal of Cosmology and Astroparticle Physics}\ }\textbf {\bibinfo {volume} {2023}},\ \bibinfo {pages} {061} (\bibinfo {year} {2023})}\BibitemShut {NoStop}%
\bibitem [{\citenamefont {Pogosian}\ \emph {et~al.}(2022)\citenamefont {Pogosian} \emph {et~al.}}]{pogosian2022imprints}%
  \BibitemOpen
  \bibfield  {author} {\bibinfo {author} {\bibfnamefont {L.}~\bibnamefont {Pogosian}} \emph {et~al.},\ }\href {https://doi.org/10.1038/s41550-022-01808-7} {\bibfield  {journal} {\bibinfo  {journal} {Nature Astronomy}\ }\textbf {\bibinfo {volume} {6}},\ \bibinfo {pages} {1484} (\bibinfo {year} {2022})}\BibitemShut {NoStop}%
\bibitem [{\citenamefont {Colg{\'a}in}\ \emph {et~al.}(2023)\citenamefont {Colg{\'a}in}, \citenamefont {Pourojaghi}, \citenamefont {Sheikh-Jabbari},\ and\ \citenamefont {Sherwin}}]{colgain2023mcmc}%
  \BibitemOpen
  \bibfield  {author} {\bibinfo {author} {\bibfnamefont {E.~{\'O}.}\ \bibnamefont {Colg{\'a}in}}, \bibinfo {author} {\bibfnamefont {S.}~\bibnamefont {Pourojaghi}}, \bibinfo {author} {\bibfnamefont {M.}~\bibnamefont {Sheikh-Jabbari}}, \ and\ \bibinfo {author} {\bibfnamefont {D.}~\bibnamefont {Sherwin}},\ }\href {https://doi.org/10.48550/arXiv.2307.16349} {\bibfield  {journal} {\bibinfo  {journal} {arXiv preprint arXiv:2307.16349}\ } (\bibinfo {year} {2023})}\BibitemShut {NoStop}%
\bibitem [{\citenamefont {Moresco.}\ \emph {et~al.}(2022)\citenamefont {Moresco.} \emph {et~al.}}]{Moresco_2022}%
  \BibitemOpen
  \bibfield  {author} {\bibinfo {author} {\bibfnamefont {M.}~\bibnamefont {Moresco.}} \emph {et~al.},\ }\href {\doibase 10.1007/s41114-022-00040-z} {\bibfield  {journal} {\bibinfo  {journal} {Living Reviews in Relativity}\ }\textbf {\bibinfo {volume} {25}} (\bibinfo {year} {2022}),\ 10.1007/s41114-022-00040-z}\BibitemShut {NoStop}%
\bibitem [{\citenamefont {et~al}(2018{\natexlab{a}})}]{Scolnic_2018}%
  \BibitemOpen
  \bibfield  {author} {\bibinfo {author} {\bibfnamefont {D.~M.~S.}\ \bibnamefont {et~al}},\ }\href {\doibase 10.3847/1538-4357/aab9bb} {\bibfield  {journal} {\bibinfo  {journal} {The Astrophysical Journal}\ }\textbf {\bibinfo {volume} {859}},\ \bibinfo {pages} {101} (\bibinfo {year} {2018}{\natexlab{a}})}\BibitemShut {NoStop}%
\bibitem [{\citenamefont {et~al}(2018{\natexlab{b}})}]{Riess_2018}%
  \BibitemOpen
  \bibfield  {author} {\bibinfo {author} {\bibfnamefont {A.~G.~R.}\ \bibnamefont {et~al}},\ }\href {\doibase 10.3847/1538-4357/aaa5a9} {\bibfield  {journal} {\bibinfo  {journal} {The Astrophysical Journal}\ }\textbf {\bibinfo {volume} {853}},\ \bibinfo {pages} {126} (\bibinfo {year} {2018}{\natexlab{b}})}\BibitemShut {NoStop}%
\bibitem [{\citenamefont {Haridasu}\ \emph {et~al.}(2018)\citenamefont {Haridasu}, \citenamefont {Lukovi{\'c}}, \citenamefont {Moresco},\ and\ \citenamefont {Vittorio}}]{haridasu2018improved}%
  \BibitemOpen
  \bibfield  {author} {\bibinfo {author} {\bibfnamefont {B.~S.}\ \bibnamefont {Haridasu}}, \bibinfo {author} {\bibfnamefont {V.~V.}\ \bibnamefont {Lukovi{\'c}}}, \bibinfo {author} {\bibfnamefont {M.}~\bibnamefont {Moresco}}, \ and\ \bibinfo {author} {\bibfnamefont {N.}~\bibnamefont {Vittorio}},\ }\href@noop {} {\bibfield  {journal} {\bibinfo  {journal} {Journal of Cosmology and Astroparticle Physics}\ }\textbf {\bibinfo {volume} {2018}},\ \bibinfo {pages} {015} (\bibinfo {year} {2018})}\BibitemShut {NoStop}%
\bibitem [{\citenamefont {Mukherjee}\ and\ \citenamefont {Banerjee}(2021)}]{Mukherjee_2021}%
  \BibitemOpen
  \bibfield  {author} {\bibinfo {author} {\bibfnamefont {P.}~\bibnamefont {Mukherjee}}\ and\ \bibinfo {author} {\bibfnamefont {N.}~\bibnamefont {Banerjee}},\ }\href {\doibase 10.1103/physrevd.103.123530} {\bibfield  {journal} {\bibinfo  {journal} {Physical Review D}\ }\textbf {\bibinfo {volume} {103}} (\bibinfo {year} {2021}),\ 10.1103/physrevd.103.123530}\BibitemShut {NoStop}%
\bibitem [{\citenamefont {Akarsu}\ \emph {et~al.}(2023{\natexlab{c}})\citenamefont {Akarsu}, \citenamefont {Colgain}, \citenamefont {\"Ozulker}, \citenamefont {Thakur},\ and\ \citenamefont {Yin}}]{Akarsu:2022lhx}%
  \BibitemOpen
  \bibfield  {author} {\bibinfo {author} {\bibfnamefont {O.}~\bibnamefont {Akarsu}}, \bibinfo {author} {\bibfnamefont {E.~O.}\ \bibnamefont {Colgain}}, \bibinfo {author} {\bibfnamefont {E.}~\bibnamefont {\"Ozulker}}, \bibinfo {author} {\bibfnamefont {S.}~\bibnamefont {Thakur}}, \ and\ \bibinfo {author} {\bibfnamefont {L.}~\bibnamefont {Yin}},\ }\href {\doibase 10.1103/PhysRevD.107.123526} {\bibfield  {journal} {\bibinfo  {journal} {Phys. Rev. D}\ }\textbf {\bibinfo {volume} {107}},\ \bibinfo {pages} {123526} (\bibinfo {year} {2023}{\natexlab{c}})},\ \Eprint {http://arxiv.org/abs/2207.10609} {arXiv:2207.10609 [astro-ph.CO]} \BibitemShut {NoStop}%
\bibitem [{\citenamefont {Weinberg}(1989)}]{Weinberg:1988cp}%
  \BibitemOpen
  \bibfield  {author} {\bibinfo {author} {\bibfnamefont {S.}~\bibnamefont {Weinberg}},\ }\href {\doibase 10.1103/RevModPhys.61.1} {\bibfield  {journal} {\bibinfo  {journal} {Rev. Mod. Phys.}\ }\textbf {\bibinfo {volume} {61}},\ \bibinfo {pages} {1} (\bibinfo {year} {1989})}\BibitemShut {NoStop}%
\bibitem [{\citenamefont {Weinberg}(2000)}]{Weinberg:2000yb}%
  \BibitemOpen
  \bibfield  {author} {\bibinfo {author} {\bibfnamefont {S.}~\bibnamefont {Weinberg}},\ }in\ \href@noop {} {\emph {\bibinfo {booktitle} {{4th International Symposium on Sources and Detection of Dark Matter in the Universe (DM 2000)}}}}\ (\bibinfo {year} {2000})\ pp.\ \bibinfo {pages} {18--26},\ \Eprint {http://arxiv.org/abs/astro-ph/0005265} {arXiv:astro-ph/0005265} \BibitemShut {NoStop}%
\bibitem [{\citenamefont {Peebles}\ and\ \citenamefont {Ratra}(2003)}]{Peebles:2002gy}%
  \BibitemOpen
  \bibfield  {author} {\bibinfo {author} {\bibfnamefont {P.~J.~E.}\ \bibnamefont {Peebles}}\ and\ \bibinfo {author} {\bibfnamefont {B.}~\bibnamefont {Ratra}},\ }\href {\doibase 10.1103/RevModPhys.75.559} {\bibfield  {journal} {\bibinfo  {journal} {Rev. Mod. Phys.}\ }\textbf {\bibinfo {volume} {75}},\ \bibinfo {pages} {559} (\bibinfo {year} {2003})},\ \Eprint {http://arxiv.org/abs/astro-ph/0207347} {arXiv:astro-ph/0207347} \BibitemShut {NoStop}%
\bibitem [{\citenamefont {Ozulker}(2022)}]{Ozulker:2022slu}%
  \BibitemOpen
  \bibfield  {author} {\bibinfo {author} {\bibfnamefont {E.}~\bibnamefont {Ozulker}},\ }\href {\doibase 10.1103/PhysRevD.106.063509} {\bibfield  {journal} {\bibinfo  {journal} {Phys. Rev. D}\ }\textbf {\bibinfo {volume} {106}},\ \bibinfo {pages} {063509} (\bibinfo {year} {2022})},\ \Eprint {http://arxiv.org/abs/2203.04167} {arXiv:2203.04167 [astro-ph.CO]} \BibitemShut {NoStop}%
\bibitem [{\citenamefont {Copeland}\ \emph {et~al.}(2006)\citenamefont {Copeland}, \citenamefont {Sami},\ and\ \citenamefont {Tsujikawa}}]{Copeland:2006wr}%
  \BibitemOpen
  \bibfield  {author} {\bibinfo {author} {\bibfnamefont {E.~J.}\ \bibnamefont {Copeland}}, \bibinfo {author} {\bibfnamefont {M.}~\bibnamefont {Sami}}, \ and\ \bibinfo {author} {\bibfnamefont {S.}~\bibnamefont {Tsujikawa}},\ }\href {\doibase 10.1142/S021827180600942X} {\bibfield  {journal} {\bibinfo  {journal} {Int. J. Mod. Phys. D}\ }\textbf {\bibinfo {volume} {15}},\ \bibinfo {pages} {1753} (\bibinfo {year} {2006})},\ \Eprint {http://arxiv.org/abs/hep-th/0603057} {arXiv:hep-th/0603057} \BibitemShut {NoStop}%
\bibitem [{\citenamefont {Di~Valentino}\ \emph {et~al.}(2021{\natexlab{d}})\citenamefont {Di~Valentino}, \citenamefont {Mukherjee},\ and\ \citenamefont {Sen}}]{DiValentino:2020naf}%
  \BibitemOpen
  \bibfield  {author} {\bibinfo {author} {\bibfnamefont {E.}~\bibnamefont {Di~Valentino}}, \bibinfo {author} {\bibfnamefont {A.}~\bibnamefont {Mukherjee}}, \ and\ \bibinfo {author} {\bibfnamefont {A.~A.}\ \bibnamefont {Sen}},\ }\href {\doibase 10.3390/e23040404} {\bibfield  {journal} {\bibinfo  {journal} {Entropy}\ }\textbf {\bibinfo {volume} {23}},\ \bibinfo {pages} {404} (\bibinfo {year} {2021}{\natexlab{d}})},\ \Eprint {http://arxiv.org/abs/2005.12587} {arXiv:2005.12587 [astro-ph.CO]} \BibitemShut {NoStop}%
\bibitem [{\citenamefont {Kumar}\ and\ \citenamefont {Nunes}(2017)}]{Kumar:2017dnp}%
  \BibitemOpen
  \bibfield  {author} {\bibinfo {author} {\bibfnamefont {S.}~\bibnamefont {Kumar}}\ and\ \bibinfo {author} {\bibfnamefont {R.~C.}\ \bibnamefont {Nunes}},\ }\href {\doibase 10.1103/PhysRevD.96.103511} {\bibfield  {journal} {\bibinfo  {journal} {Phys. Rev. D}\ }\textbf {\bibinfo {volume} {96}},\ \bibinfo {pages} {103511} (\bibinfo {year} {2017})},\ \Eprint {http://arxiv.org/abs/1702.02143} {arXiv:1702.02143 [astro-ph.CO]} \BibitemShut {NoStop}%
\bibitem [{\citenamefont {Di~Valentino}\ \emph {et~al.}(2017)\citenamefont {Di~Valentino}, \citenamefont {Melchiorri},\ and\ \citenamefont {Mena}}]{DiValentino:2017iww}%
  \BibitemOpen
  \bibfield  {author} {\bibinfo {author} {\bibfnamefont {E.}~\bibnamefont {Di~Valentino}}, \bibinfo {author} {\bibfnamefont {A.}~\bibnamefont {Melchiorri}}, \ and\ \bibinfo {author} {\bibfnamefont {O.}~\bibnamefont {Mena}},\ }\href {\doibase 10.1103/PhysRevD.96.043503} {\bibfield  {journal} {\bibinfo  {journal} {Phys. Rev. D}\ }\textbf {\bibinfo {volume} {96}},\ \bibinfo {pages} {043503} (\bibinfo {year} {2017})},\ \Eprint {http://arxiv.org/abs/1704.08342} {arXiv:1704.08342 [astro-ph.CO]} \BibitemShut {NoStop}%
\bibitem [{\citenamefont {Yang}\ \emph {et~al.}(2018)\citenamefont {Yang}, \citenamefont {Mukherjee}, \citenamefont {Di~Valentino},\ and\ \citenamefont {Pan}}]{Yang:2018uae}%
  \BibitemOpen
  \bibfield  {author} {\bibinfo {author} {\bibfnamefont {W.}~\bibnamefont {Yang}}, \bibinfo {author} {\bibfnamefont {A.}~\bibnamefont {Mukherjee}}, \bibinfo {author} {\bibfnamefont {E.}~\bibnamefont {Di~Valentino}}, \ and\ \bibinfo {author} {\bibfnamefont {S.}~\bibnamefont {Pan}},\ }\href {\doibase 10.1103/PhysRevD.98.123527} {\bibfield  {journal} {\bibinfo  {journal} {Phys. Rev. D}\ }\textbf {\bibinfo {volume} {98}},\ \bibinfo {pages} {123527} (\bibinfo {year} {2018})},\ \Eprint {http://arxiv.org/abs/1809.06883} {arXiv:1809.06883 [astro-ph.CO]} \BibitemShut {NoStop}%
\bibitem [{\citenamefont {Pan}\ \emph {et~al.}(2019)\citenamefont {Pan}, \citenamefont {Yang}, \citenamefont {Di~Valentino}, \citenamefont {Saridakis},\ and\ \citenamefont {Chakraborty}}]{Pan:2019gop}%
  \BibitemOpen
  \bibfield  {author} {\bibinfo {author} {\bibfnamefont {S.}~\bibnamefont {Pan}}, \bibinfo {author} {\bibfnamefont {W.}~\bibnamefont {Yang}}, \bibinfo {author} {\bibfnamefont {E.}~\bibnamefont {Di~Valentino}}, \bibinfo {author} {\bibfnamefont {E.~N.}\ \bibnamefont {Saridakis}}, \ and\ \bibinfo {author} {\bibfnamefont {S.}~\bibnamefont {Chakraborty}},\ }\href {\doibase 10.1103/PhysRevD.100.103520} {\bibfield  {journal} {\bibinfo  {journal} {Phys. Rev. D}\ }\textbf {\bibinfo {volume} {100}},\ \bibinfo {pages} {103520} (\bibinfo {year} {2019})},\ \Eprint {http://arxiv.org/abs/1907.07540} {arXiv:1907.07540 [astro-ph.CO]} \BibitemShut {NoStop}%
\bibitem [{\citenamefont {Kumar}\ \emph {et~al.}(2019)\citenamefont {Kumar}, \citenamefont {Nunes},\ and\ \citenamefont {Yadav}}]{Kumar:2019wfs}%
  \BibitemOpen
  \bibfield  {author} {\bibinfo {author} {\bibfnamefont {S.}~\bibnamefont {Kumar}}, \bibinfo {author} {\bibfnamefont {R.~C.}\ \bibnamefont {Nunes}}, \ and\ \bibinfo {author} {\bibfnamefont {S.~K.}\ \bibnamefont {Yadav}},\ }\href {\doibase 10.1140/epjc/s10052-019-7087-7} {\bibfield  {journal} {\bibinfo  {journal} {Eur. Phys. J. C}\ }\textbf {\bibinfo {volume} {79}},\ \bibinfo {pages} {576} (\bibinfo {year} {2019})},\ \Eprint {http://arxiv.org/abs/1903.04865} {arXiv:1903.04865 [astro-ph.CO]} \BibitemShut {NoStop}%
\bibitem [{\citenamefont {Di~Valentino}\ \emph {et~al.}(2020{\natexlab{a}})\citenamefont {Di~Valentino}, \citenamefont {Melchiorri}, \citenamefont {Mena},\ and\ \citenamefont {Vagnozzi}}]{DiValentino:2019ffd}%
  \BibitemOpen
  \bibfield  {author} {\bibinfo {author} {\bibfnamefont {E.}~\bibnamefont {Di~Valentino}}, \bibinfo {author} {\bibfnamefont {A.}~\bibnamefont {Melchiorri}}, \bibinfo {author} {\bibfnamefont {O.}~\bibnamefont {Mena}}, \ and\ \bibinfo {author} {\bibfnamefont {S.}~\bibnamefont {Vagnozzi}},\ }\href {\doibase 10.1016/j.dark.2020.100666} {\bibfield  {journal} {\bibinfo  {journal} {Phys. Dark Univ.}\ }\textbf {\bibinfo {volume} {30}},\ \bibinfo {pages} {100666} (\bibinfo {year} {2020}{\natexlab{a}})},\ \Eprint {http://arxiv.org/abs/1908.04281} {arXiv:1908.04281 [astro-ph.CO]} \BibitemShut {NoStop}%
\bibitem [{\citenamefont {Di~Valentino}\ \emph {et~al.}(2020{\natexlab{b}})\citenamefont {Di~Valentino}, \citenamefont {Melchiorri}, \citenamefont {Mena},\ and\ \citenamefont {Vagnozzi}}]{DiValentino:2019jae}%
  \BibitemOpen
  \bibfield  {author} {\bibinfo {author} {\bibfnamefont {E.}~\bibnamefont {Di~Valentino}}, \bibinfo {author} {\bibfnamefont {A.}~\bibnamefont {Melchiorri}}, \bibinfo {author} {\bibfnamefont {O.}~\bibnamefont {Mena}}, \ and\ \bibinfo {author} {\bibfnamefont {S.}~\bibnamefont {Vagnozzi}},\ }\href {\doibase 10.1103/PhysRevD.101.063502} {\bibfield  {journal} {\bibinfo  {journal} {Phys. Rev. D}\ }\textbf {\bibinfo {volume} {101}},\ \bibinfo {pages} {063502} (\bibinfo {year} {2020}{\natexlab{b}})},\ \Eprint {http://arxiv.org/abs/1910.09853} {arXiv:1910.09853 [astro-ph.CO]} \BibitemShut {NoStop}%
\bibitem [{\citenamefont {Lucca}\ and\ \citenamefont {Hooper}(2020)}]{Lucca:2020zjb}%
  \BibitemOpen
  \bibfield  {author} {\bibinfo {author} {\bibfnamefont {M.}~\bibnamefont {Lucca}}\ and\ \bibinfo {author} {\bibfnamefont {D.~C.}\ \bibnamefont {Hooper}},\ }\href {\doibase 10.1103/PhysRevD.102.123502} {\bibfield  {journal} {\bibinfo  {journal} {Phys. Rev. D}\ }\textbf {\bibinfo {volume} {102}},\ \bibinfo {pages} {123502} (\bibinfo {year} {2020})},\ \Eprint {http://arxiv.org/abs/2002.06127} {arXiv:2002.06127 [astro-ph.CO]} \BibitemShut {NoStop}%
\bibitem [{\citenamefont {G\'omez-Valent}\ \emph {et~al.}(2020)\citenamefont {G\'omez-Valent}, \citenamefont {Pettorino},\ and\ \citenamefont {Amendola}}]{Gomez-Valent:2020mqn}%
  \BibitemOpen
  \bibfield  {author} {\bibinfo {author} {\bibfnamefont {A.}~\bibnamefont {G\'omez-Valent}}, \bibinfo {author} {\bibfnamefont {V.}~\bibnamefont {Pettorino}}, \ and\ \bibinfo {author} {\bibfnamefont {L.}~\bibnamefont {Amendola}},\ }\href {\doibase 10.1103/PhysRevD.101.123513} {\bibfield  {journal} {\bibinfo  {journal} {Phys. Rev. D}\ }\textbf {\bibinfo {volume} {101}},\ \bibinfo {pages} {123513} (\bibinfo {year} {2020})},\ \Eprint {http://arxiv.org/abs/2004.00610} {arXiv:2004.00610 [astro-ph.CO]} \BibitemShut {NoStop}%
\bibitem [{\citenamefont {Kumar}(2021)}]{Kumar:2021eev}%
  \BibitemOpen
  \bibfield  {author} {\bibinfo {author} {\bibfnamefont {S.}~\bibnamefont {Kumar}},\ }\href {\doibase 10.1016/j.dark.2021.100862} {\bibfield  {journal} {\bibinfo  {journal} {Phys. Dark Univ.}\ }\textbf {\bibinfo {volume} {33}},\ \bibinfo {pages} {100862} (\bibinfo {year} {2021})},\ \Eprint {http://arxiv.org/abs/2102.12902} {arXiv:2102.12902 [astro-ph.CO]} \BibitemShut {NoStop}%
\bibitem [{\citenamefont {Nunes}\ \emph {et~al.}(2022)\citenamefont {Nunes}, \citenamefont {Vagnozzi}, \citenamefont {Kumar}, \citenamefont {Di~Valentino},\ and\ \citenamefont {Mena}}]{Nunes:2022bhn}%
  \BibitemOpen
  \bibfield  {author} {\bibinfo {author} {\bibfnamefont {R.~C.}\ \bibnamefont {Nunes}}, \bibinfo {author} {\bibfnamefont {S.}~\bibnamefont {Vagnozzi}}, \bibinfo {author} {\bibfnamefont {S.}~\bibnamefont {Kumar}}, \bibinfo {author} {\bibfnamefont {E.}~\bibnamefont {Di~Valentino}}, \ and\ \bibinfo {author} {\bibfnamefont {O.}~\bibnamefont {Mena}},\ }\href {\doibase 10.1103/PhysRevD.105.123506} {\bibfield  {journal} {\bibinfo  {journal} {Phys. Rev. D}\ }\textbf {\bibinfo {volume} {105}},\ \bibinfo {pages} {123506} (\bibinfo {year} {2022})},\ \Eprint {http://arxiv.org/abs/2203.08093} {arXiv:2203.08093 [astro-ph.CO]} \BibitemShut {NoStop}%
\bibitem [{\citenamefont {Bernui}\ \emph {et~al.}(2023)\citenamefont {Bernui}, \citenamefont {Di~Valentino}, \citenamefont {Giar\`e}, \citenamefont {Kumar},\ and\ \citenamefont {Nunes}}]{Bernui:2023byc}%
  \BibitemOpen
  \bibfield  {author} {\bibinfo {author} {\bibfnamefont {A.}~\bibnamefont {Bernui}}, \bibinfo {author} {\bibfnamefont {E.}~\bibnamefont {Di~Valentino}}, \bibinfo {author} {\bibfnamefont {W.}~\bibnamefont {Giar\`e}}, \bibinfo {author} {\bibfnamefont {S.}~\bibnamefont {Kumar}}, \ and\ \bibinfo {author} {\bibfnamefont {R.~C.}\ \bibnamefont {Nunes}},\ }\href {\doibase 10.1103/PhysRevD.107.103531} {\bibfield  {journal} {\bibinfo  {journal} {Phys. Rev. D}\ }\textbf {\bibinfo {volume} {107}},\ \bibinfo {pages} {103531} (\bibinfo {year} {2023})},\ \Eprint {http://arxiv.org/abs/2301.06097} {arXiv:2301.06097 [astro-ph.CO]} \BibitemShut {NoStop}%
\bibitem [{\citenamefont {Malekjani}\ \emph {et~al.}(2024)\citenamefont {Malekjani}, \citenamefont {Conville}, \citenamefont {Colg\'ain}, \citenamefont {Pourojaghi},\ and\ \citenamefont {Sheikh-Jabbari}}]{Malekjani:2023ple}%
  \BibitemOpen
  \bibfield  {author} {\bibinfo {author} {\bibfnamefont {M.}~\bibnamefont {Malekjani}}, \bibinfo {author} {\bibfnamefont {R.~M.}\ \bibnamefont {Conville}}, \bibinfo {author} {\bibfnamefont {E.~O.}\ \bibnamefont {Colg\'ain}}, \bibinfo {author} {\bibfnamefont {S.}~\bibnamefont {Pourojaghi}}, \ and\ \bibinfo {author} {\bibfnamefont {M.~M.}\ \bibnamefont {Sheikh-Jabbari}},\ }\href {\doibase 10.1140/epjc/s10052-024-12667-z} {\bibfield  {journal} {\bibinfo  {journal} {Eur. Phys. J. C}\ }\textbf {\bibinfo {volume} {84}},\ \bibinfo {pages} {317} (\bibinfo {year} {2024})},\ \Eprint {http://arxiv.org/abs/2301.12725} {arXiv:2301.12725 [astro-ph.CO]} \BibitemShut {NoStop}%
\bibitem [{\citenamefont {Wen}\ \emph {et~al.}(2024)\citenamefont {Wen}, \citenamefont {Hergt}, \citenamefont {Afshordi},\ and\ \citenamefont {Scott}}]{Wen:2023wes}%
  \BibitemOpen
  \bibfield  {author} {\bibinfo {author} {\bibfnamefont {R.~Y.}\ \bibnamefont {Wen}}, \bibinfo {author} {\bibfnamefont {L.~T.}\ \bibnamefont {Hergt}}, \bibinfo {author} {\bibfnamefont {N.}~\bibnamefont {Afshordi}}, \ and\ \bibinfo {author} {\bibfnamefont {D.}~\bibnamefont {Scott}},\ }\href {\doibase 10.1088/1475-7516/2024/03/045} {\bibfield  {journal} {\bibinfo  {journal} {JCAP}\ }\textbf {\bibinfo {volume} {03}},\ \bibinfo {pages} {045} (\bibinfo {year} {2024})},\ \Eprint {http://arxiv.org/abs/2311.03028} {arXiv:2311.03028 [astro-ph.CO]} \BibitemShut {NoStop}%
\bibitem [{\citenamefont {Brinckmann}\ and\ \citenamefont {Lesgourgues}(2018)}]{Brinckmann:2018cvx}%
  \BibitemOpen
  \bibfield  {author} {\bibinfo {author} {\bibfnamefont {T.}~\bibnamefont {Brinckmann}}\ and\ \bibinfo {author} {\bibfnamefont {J.}~\bibnamefont {Lesgourgues}},\ }\href@noop {} {\  (\bibinfo {year} {2018})},\ \Eprint {http://arxiv.org/abs/1804.07261} {arXiv:1804.07261 [astro-ph.CO]} \BibitemShut {NoStop}%
\bibitem [{\citenamefont {Audren}\ \emph {et~al.}(2013)\citenamefont {Audren}, \citenamefont {Lesgourgues}, \citenamefont {Benabed},\ and\ \citenamefont {Prunet}}]{Audren:2012wb}%
  \BibitemOpen
  \bibfield  {author} {\bibinfo {author} {\bibfnamefont {B.}~\bibnamefont {Audren}}, \bibinfo {author} {\bibfnamefont {J.}~\bibnamefont {Lesgourgues}}, \bibinfo {author} {\bibfnamefont {K.}~\bibnamefont {Benabed}}, \ and\ \bibinfo {author} {\bibfnamefont {S.}~\bibnamefont {Prunet}},\ }\href {\doibase 10.1088/1475-7516/2013/02/001} {\bibfield  {journal} {\bibinfo  {journal} {JCAP}\ }\textbf {\bibinfo {volume} {1302}},\ \bibinfo {pages} {001} (\bibinfo {year} {2013})},\ \Eprint {http://arxiv.org/abs/1210.7183} {arXiv:1210.7183 [astro-ph.CO]} \BibitemShut {NoStop}%
\bibitem [{\citenamefont {Lemos}\ \emph {et~al.}(2019)\citenamefont {Lemos}, \citenamefont {Lee}, \citenamefont {Efstathiou},\ and\ \citenamefont {Gratton}}]{lemos2019model}%
  \BibitemOpen
  \bibfield  {author} {\bibinfo {author} {\bibfnamefont {P.}~\bibnamefont {Lemos}}, \bibinfo {author} {\bibfnamefont {E.}~\bibnamefont {Lee}}, \bibinfo {author} {\bibfnamefont {G.}~\bibnamefont {Efstathiou}}, \ and\ \bibinfo {author} {\bibfnamefont {S.}~\bibnamefont {Gratton}},\ }\href {https://doi.org/10.1093/mnras/sty3082} {\bibfield  {journal} {\bibinfo  {journal} {Monthly Notices of the Royal Astronomical Society}\ }\textbf {\bibinfo {volume} {483}},\ \bibinfo {pages} {4803} (\bibinfo {year} {2019})}\BibitemShut {NoStop}%
\bibitem [{\citenamefont {Efstathiou}(2021)}]{Efstathiou_2021}%
  \BibitemOpen
  \bibfield  {author} {\bibinfo {author} {\bibfnamefont {G.}~\bibnamefont {Efstathiou}},\ }\href {\doibase 10.1093/mnras/stab1588} {\bibfield  {journal} {\bibinfo  {journal} {Monthly Notices of the Royal Astronomical Society}\ }\textbf {\bibinfo {volume} {505}},\ \bibinfo {pages} {3866} (\bibinfo {year} {2021})}\BibitemShut {NoStop}%
\bibitem [{\citenamefont {Moresco}(2023)}]{Moresco_2023}%
  \BibitemOpen
  \bibfield  {author} {\bibinfo {author} {\bibfnamefont {M.}~\bibnamefont {Moresco}},\ }\href {https://doi.org/10.48550/arXiv.2307.09501} {\bibfield  {journal} {\bibinfo  {journal} {arXiv preprint arXiv:2307.09501}\ } (\bibinfo {year} {2023})}\BibitemShut {NoStop}%
\bibitem [{\citenamefont {Holsclaw}\ \emph {et~al.}(2011)\citenamefont {Holsclaw}, \citenamefont {Alam}, \citenamefont {Sans\'o}, \citenamefont {Lee}, \citenamefont {Heitmann}, \citenamefont {Habib},\ and\ \citenamefont {Higdon}}]{Holsclaw_2011}%
  \BibitemOpen
  \bibfield  {author} {\bibinfo {author} {\bibfnamefont {T.}~\bibnamefont {Holsclaw}}, \bibinfo {author} {\bibfnamefont {U.}~\bibnamefont {Alam}}, \bibinfo {author} {\bibfnamefont {B.}~\bibnamefont {Sans\'o}}, \bibinfo {author} {\bibfnamefont {H.}~\bibnamefont {Lee}}, \bibinfo {author} {\bibfnamefont {K.}~\bibnamefont {Heitmann}}, \bibinfo {author} {\bibfnamefont {S.}~\bibnamefont {Habib}}, \ and\ \bibinfo {author} {\bibfnamefont {D.}~\bibnamefont {Higdon}},\ }\href {\doibase 10.1103/PhysRevD.84.083501} {\bibfield  {journal} {\bibinfo  {journal} {Phys. Rev. D}\ }\textbf {\bibinfo {volume} {84}},\ \bibinfo {pages} {083501} (\bibinfo {year} {2011})}\BibitemShut {NoStop}%
\bibitem [{\citenamefont {Hwang}\ \emph {et~al.}(2023)\citenamefont {Hwang}, \citenamefont {L’Huillier}, \citenamefont {Keeley}, \citenamefont {Jee},\ and\ \citenamefont {Shafieloo}}]{Hwang_2023}%
  \BibitemOpen
  \bibfield  {author} {\bibinfo {author} {\bibfnamefont {S.-g.}\ \bibnamefont {Hwang}}, \bibinfo {author} {\bibfnamefont {B.}~\bibnamefont {L’Huillier}}, \bibinfo {author} {\bibfnamefont {R.~E.}\ \bibnamefont {Keeley}}, \bibinfo {author} {\bibfnamefont {M.~J.}\ \bibnamefont {Jee}}, \ and\ \bibinfo {author} {\bibfnamefont {A.}~\bibnamefont {Shafieloo}},\ }\href {\doibase 10.1088/1475-7516/2023/02/014} {\bibfield  {journal} {\bibinfo  {journal} {Journal of Cosmology and Astroparticle Physics}\ }\textbf {\bibinfo {volume} {2023}},\ \bibinfo {pages} {014} (\bibinfo {year} {2023})}\BibitemShut {NoStop}%
\end{thebibliography}%

\end{document}